\documentclass[12pt, a4paper]{article}
\pdfoutput=1

\usepackage{amsmath}
\usepackage{amsfonts}
\usepackage{amssymb}
\usepackage{latexsym}
\usepackage{graphicx}
\usepackage{color}
\usepackage{mathrsfs}
\usepackage{slashed}
\usepackage{hyperref}
\usepackage{enumerate}
\usepackage{multirow}
\usepackage[table]{xcolor}
\usepackage{colortbl}
\usepackage{url}
\usepackage{enumitem}
\usepackage{comment}
\numberwithin{equation}{section}

\hypersetup{colorlinks, citecolor=bluscuro, linkcolor=magenta, urlcolor=bluscuro}
\definecolor{rossos}{rgb}{0.7,0,0.3}
\definecolor{violachiaro}{rgb}{1,0.6,1}
\definecolor{rossochiaro}{rgb}{1,0.6,0.6}
\definecolor{verdechiaro}{rgb}{0.6,1,0.6}
\definecolor{giallochiaro}{rgb}{1,1,0.6}
\definecolor{bluscuro}{rgb}{0.15, 0.2, 0.9}
\definecolor{verdes}{rgb}{0.1, 0.5, 0.1}
\makeatother   % Cancel the effect of \makeatletter
\newcommand{\GeV}{{\rm \,GeV}}

 \def\be   {\begin{equation}}   \def\ee   {\end{equation}}
 \def\ba   {\begin{array}}      \def\ea   {\end{array}}
 \def\bea  {\begin{eqnarray}}   \def\eea  {\end{eqnarray}}
 \def\bean {\begin{eqnarray*}}  \def\eean {\end{eqnarray*}}

\def\g{\gamma}

\def\s{\sigma}

\begin{document}
%\hfill\textit{\currenttime\qquad \dmyyyydate\today}

\vspace{7cm}

\begin{center}
\vspace{5cm}
{\Huge
Fermi Bubbles under Dark Matter Scrutiny\\ [0.9 cm]
Part I: Astrophysical Analysis
}
\\ [2.5cm]
{\large{\textsc{
Wei-Chih Huang$^{\,a,b,}$\footnote{\textsl{whuang@sissa.it}}, Alfredo Urbano$^{\,a,}$\footnote{\textsl{alfredo.urbano@sissa.it}}, Wei Xue$^{\,b,a,}$\footnote{\textsl{wxue@sissa.it}}}}}
\\[1cm]

\large{\textit{
$^{a}$~SISSA, via Bonomea 265, I-34136 Trieste, ITALY.\\ \vspace{1.5mm}
$^{b}$~INFN, sezione di Trieste, I-34136 Trieste, ITALY.
}}
\\ [2 cm]
{ \large{\textrm{
Abstract
}}}
\\ [1.5cm]
\end{center}

The quest for Dark Matter signals in the gamma-ray sky is one of the most intriguing and exciting challenges in astrophysics. 
%Recently, two giant gamma-ray structures known as \textit{Fermi bubbles}, extending above and below the center of the Milky Way galaxy, have been discovered.  
In this paper we perform the analysis of the energy spectrum of the \textit{Fermi bubbles} at different latitudes, making use of the gamma-ray data collected by the Fermi Large Area Telescope. 
By exploring various setups for the full-sky analysis we achieve stable
results in all the analyzed latitudes.
%, reporting a North-South symmetry with respect to the Galactic plane.
At high latitude, $|b|=20^{\circ}-50^{\circ}$, the \textit{Fermi bubbles} energy spectrum can be reproduced by gamma-ray photons generated by inverse Compton scattering processes, assuming the existence of a population of high-energy electrons. At low latitude, $|b|=10^{\circ}-20^{\circ}$, the presence 
of a bump at $E_{\gamma}\sim 1-4$ GeV, reveals the existence of an extra component compatible with Dark Matter annihilation. Our best-fit candidate corresponds to annihilation into $b\overline{b}$ with mass $M_{\rm DM}= 61.8^{+6.9}_{-4.9}$ GeV and cross section $\langle \sigma v\rangle =  3.30^{+0.69}_{-0.49}\times  10^{-26}$ cm$^{3}$s$^{-1}$. In addition, using the energy spectrum of the \textit{Fermi bubbles}, we derive new conservative but stringent upper limits on the Dark Matter annihilation cross section.

\def\thefootnote{\arabic{footnote}}
\setcounter{footnote}{0}
\pagestyle{empty}

\newpage
\pagestyle{plain}
\setcounter{page}{1}

%%%%%%%%%%%%%%%%%%%%%%%%%%%%%%%%%%%%%
\section{Introduction}
%%%%%%%%%%%%%%%%%%%%%%%%%%%%%%%%%%%%%

Since the dawn of civilization,
the desire to gaze, study and understand the mysteries hedged in the astonishing beauty of the sky has been an unavoidable and innate prerogative of human nature. In March $1610$ Galileo Galilei published the \emph{Sidereus Nuncius}, the first scientific work based on telescope observations. Through the eye of this revolutionary instrument Galileo was able to take the first steps in the exploration of a completely unknown world, describing the results of his studies about the 
mountainous surface of the Moon, a myriad of stars never seen before with the naked eye, and the discovery of four Erratic Stars that appeared to be orbiting around the planet Jupiter.

After more than four hundred years, telescopes are becoming the most important scientific instrument in astronomy
and astrophysics, reaching a degree of technical perfection that enables us to study in great detail the Universe.
Among them, the Fermi Large Area Telescope (LAT) \cite{LAT_website} is devoted to the study of photons in the high energy region of gamma-rays, 
and one of the most challenging goals of the mission is to shed light on the elusive nature of Dark Matter (DM).

Many efforts have been made, for instance, to study and understand the nature of a spatially extended excess, peaked at few GeV, found in the gamma-ray emission from the Galactic center~\cite{Hooper:2010mq,Boyarsky:2010dr,Hooper:2011ti,Abazajian:2012pn,Gordon:2013vta}. The signal can be explained by $\mathcal{O} (10)$ GeV DM annihilating into $\tau^+\tau^-$, $b\bar{b}$, or by model with dark forces~\cite{Hooper:2012cw}.

 In May $2010$,
analyzing $1.66$ years of data, two giant gamma-ray bubbles that extend 25,000 light-years north and south of the center of the Milky Way galaxy have been discovered \cite{Su:2010qj},
clarifying the morphology of the ``gamma-ray haze" previously found in Ref.~\cite{Dobler:2009xz} studying the first year of data.
The spatial extension of these \textit{Fermi bubbles} ($|b| < 50^{\circ}$, $|l| < 30^{\circ}$ in Galactic coordinates) gives rise to a majestic and unique structure.
Their origin is still shrouded in mystery but the analysis of the corresponding energy spectrum reveals the most important characteristics of the emission. In Ref.~\cite{Su:2010qj} this analysis has been performed in the region $|b|> 30^{\circ}$, where the observed gamma-ray spectrum turns out to be harder ($d\Phi/dE_{\gamma}d\Omega\sim E_{\gamma}^{-2}$) than those of the Galactic diffuse emission, e.g. photons from Inverse Compton Scattering (ICS) between cosmic ray electrons and the low-energy interstellar radiation field or from  the decay of neutral pions produced by the interaction of cosmic ray protons with the interstellar medium. 
The most proposed mechanism to account for these features postulates the existence of an extra population of electrons, accelerated in shocks or turbulence, producing ICS photons. Additionally these electrons, at the same time, generate synchrotron radiation spiraling in magnetic fields thus providing the possibility to correlate the Fermi bubbles with the WMAP haze observed in the microwave \cite{Finkbeiner:2003im, Dobler:2007wv}. The chance to reproduce all the spectral features of the Fermi bubbles considering the annihilation of DM particles in the Galactic halo, on the contrary, seems to be very unlikely with a standard spherical halo and isotropic cosmic-ray diffusion \cite{Dobler:2011mk}. Nevertheless it is worthwhile to put the spectrum of the Fermi bubbles through a more careful investigation, looking in particular for spectral variation with latitude. This approach has been recently pursued in Ref.~\cite{Hooper:2013rwa,Hooper:2013nhl}, where the Fermi bubbles region is sliced in ten stripes of different latitude. From this perspective the Fermi bubbles spectrum ( $E_{\gamma}^2d\Phi/dE_{\gamma}d\Omega$) shows the presence at low latitude ($|b|< 20^{\circ}$) of a bump at $E_{\gamma}\sim 1-4$ GeV, thus revealing the existence of a possible extra component in addition to the ICS photons that, on the contrary, dominate the spectrum at high latitudes. This extra component seems to be compatible with a $\mathcal{O}(10)$ GeV DM particle annihilating into leptons or quarks, with a thermal averaged cross section $\langle \sigma v\rangle \sim 10^{-27}~{\rm cm}^{3}\,{\rm s}^{-1}$, close to the value suggested by the WIMP-miracle paradigm, $\langle \sigma v\rangle \sim 3\times 10^{-26}~{\rm cm}^{3}\,{\rm s}^{-1}$.

In this paper we study the Fermi bubbles spectrum using the same latitude-dependent approach adopted in Ref.~\cite{Hooper:2013rwa}. The aim of our analysis is twofold. 
 On the one hand we perform our own analysis of the energy spectrum. We confirm, opting for an alternative subtraction method compared to the one used in  Ref.~\cite{Hooper:2013rwa},  the existence at low latitude of an extra component in addition to the ICS emission compatible with DM annihilation. On the other hand we use the spectrum of the Fermi bubbles in order to obtain new bounds on the DM annihilation cross section, comparing our results with the existing literature.
 
This work is organized as follows. In Section~\ref{fermi_s} we describe in detail our procedure to compute the energy spectrum of the Fermi bubbles at different latitudes. In Section~\ref{sec:ICS} we discuss the interplay between 
the ICS component and the DM contribution. Section~\ref{sec:bounds} is devoted to the computation of
the bounds on DM annihilation cross section. Finally, we conclude in Section~\ref{sec:Conclusions}. In
Appendix~\ref{app:A} we provide further details about data taking and the analysis procedure. In Appendix~\ref{App:B} we discuss alternative setups.

%%%%%%%%%%%%%%%%%%%%%%%%%%%%%%%%%%%%%
\section{The energy spectrum of the Fermi bubbles}\label{fermi_s}
%%%%%%%%%%%%%%%%%%%%%%%%%%%%%%%%%%%%%

In this Section we compute and discuss the energy spectrum of the Fermi bubbles as a function of the Galactic latitude.
In a nutshell the procedure to get this spectrum can be summarized as follows.

Analyzing the data collected by the LAT, it is possible to obtain two different maps of the sky.
On the one hand the \textit{counts map} contains - in a given energy range and in each point of the sky - the number of photons collected
 by the LAT. The \textit{exposure map}, on the other hand, measures in cm$^2$s the corresponding exposure.
  The differential flux measured by the experiment is given by the count map divided  by the exposure map times energy width and solid angle.  We briefly review in Section~\ref{sec:LATdata} the main points of the data analysis; for completeness, the interest reader can find in Appendix~\ref{app:A} a more detailed summary.

In order to obtain the energy spectrum of the Fermi bubbles,
it is necessary to subtract from the observed photons those originating from all the known gamma-ray sources.
In our analysis we take into account the point and extended sources as well as the Galactic diffuse emission and the isotropic extragalactic component. Point and extended sources are masked, while the diffuse model and the isotropic component are subtracted as a consequence of a fitting procedure. We illustrate this point in Section~\ref{sec:fitting}. In Section~\ref{sec:residual} we present and  discuss our results.

\subsection{Fermi-LAT gamma-ray full sky analysis: a quick outline}\label{sec:LATdata}

The Fermi Gamma Ray Space Telescope spacecraft \cite{Fermi_website}  - launched on 11 June 2008 - is a space observatory
devoted to the gamma-ray analysis of the Milky Way galaxy. The main instrument aboard is the
LAT \cite{LAT_website},  a pair-conversion telescope able to detect photons in the energy range from about $0.02$ GeV to more than $300$ GeV.
\begin{figure}[!htb!]
 \centering
  \begin{minipage}{0.4\textwidth}
   \centering
   \includegraphics[scale=0.32]{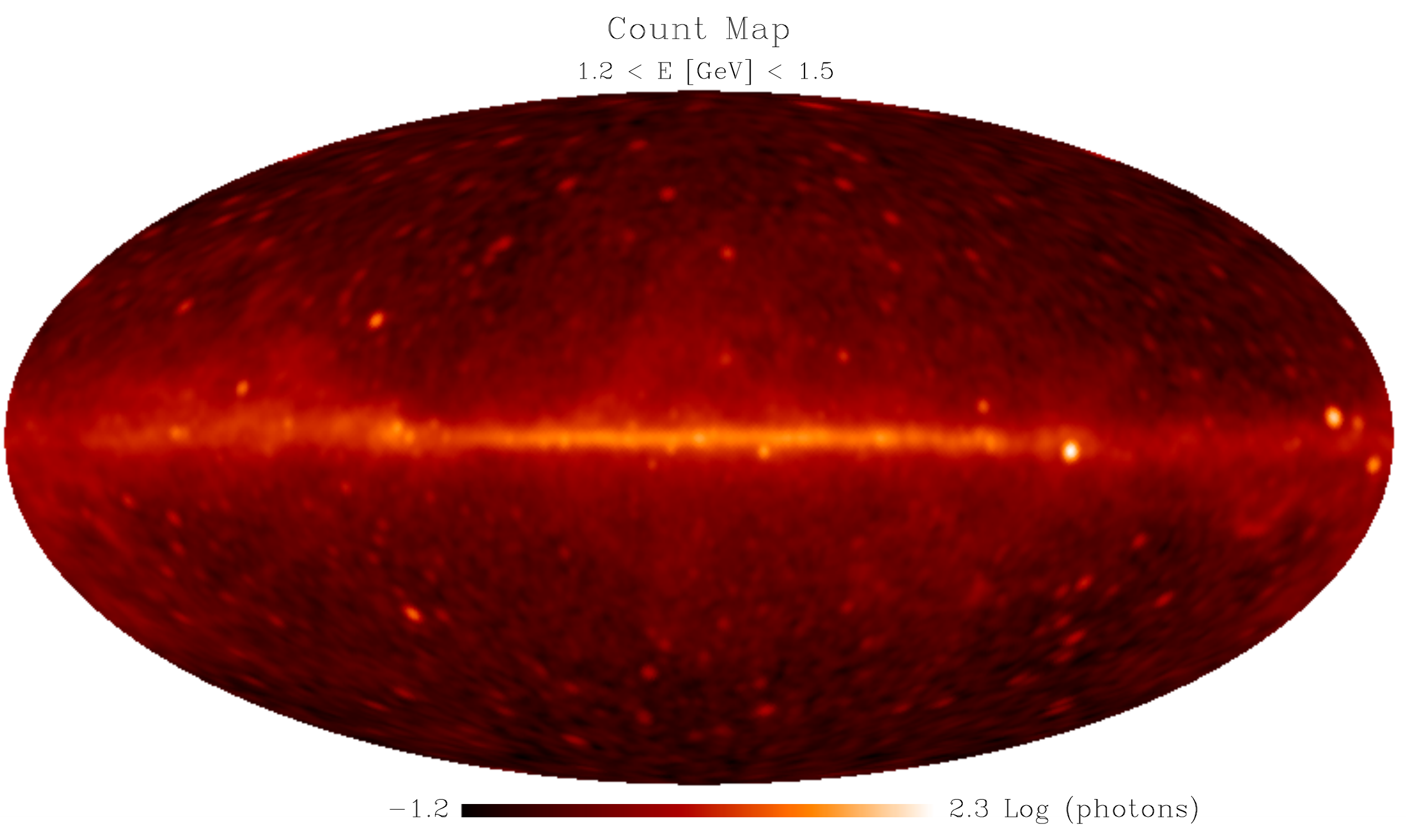}
   %\caption{\textit{Count Map}}    \label{fig:CountMap}
    \end{minipage}\hspace{1.8 cm}
   \begin{minipage}{0.4\textwidth}
    \centering
    \includegraphics[scale=0.32]{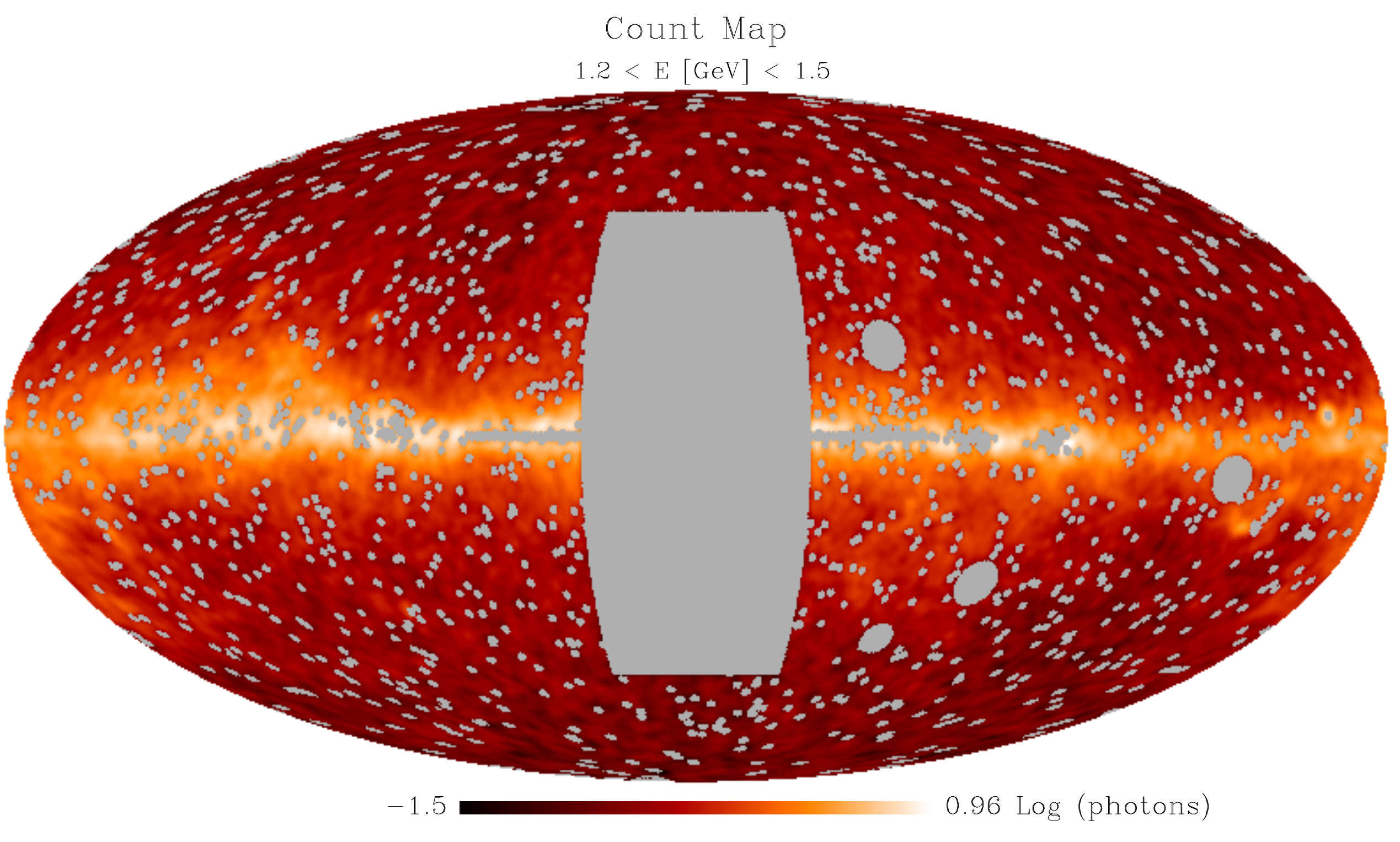}
    \end{minipage}\\
    \vspace{0.5 cm}
   \begin{minipage}{0.4\textwidth}
    \centering
   \includegraphics[scale=0.32]{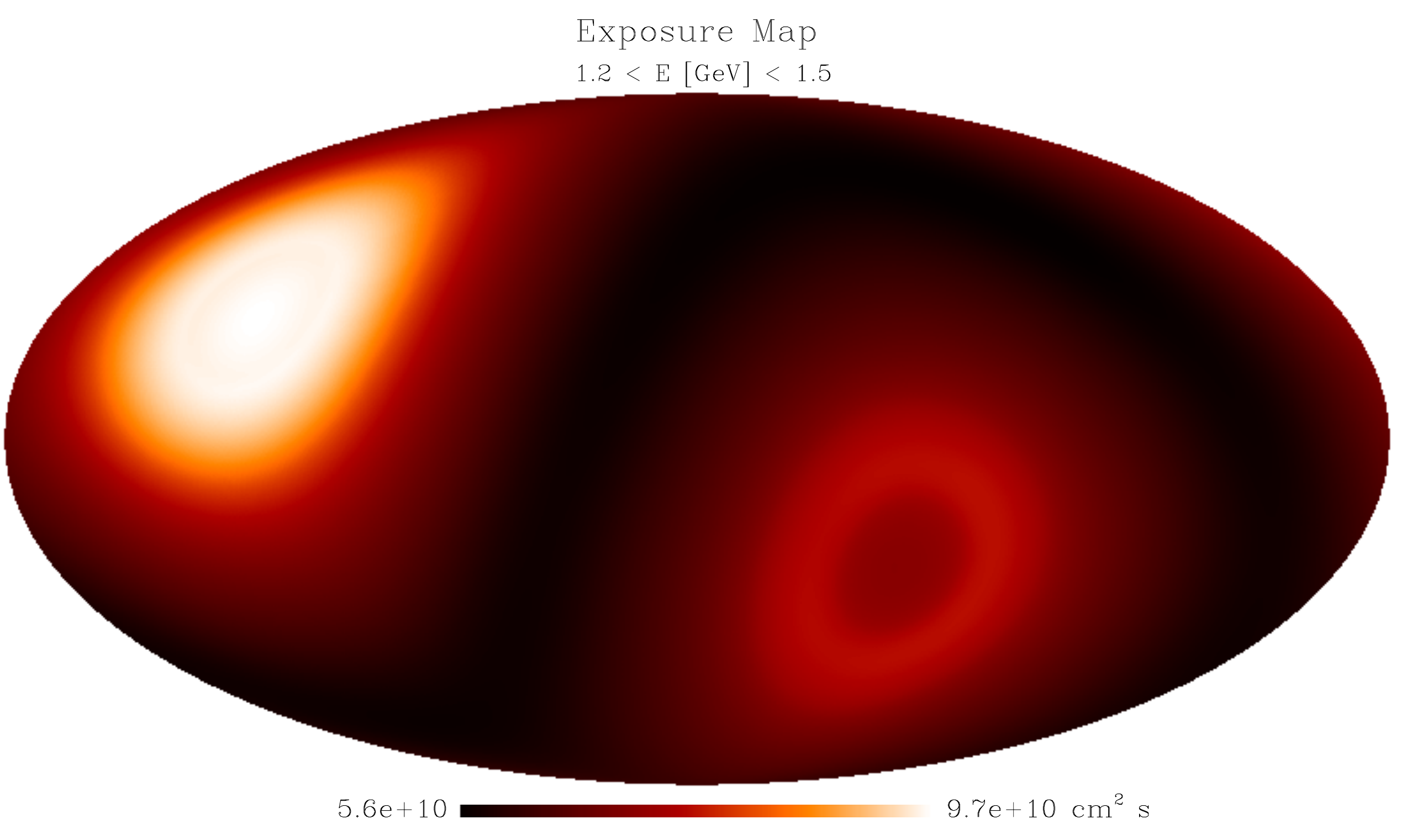}
   %\caption{\textit{Exposure Map}}\label{fig:ExposureMap}
    \end{minipage}\hspace{1.8 cm}
   \begin{minipage}{0.4\textwidth}
    \centering
    \includegraphics[scale=0.32]{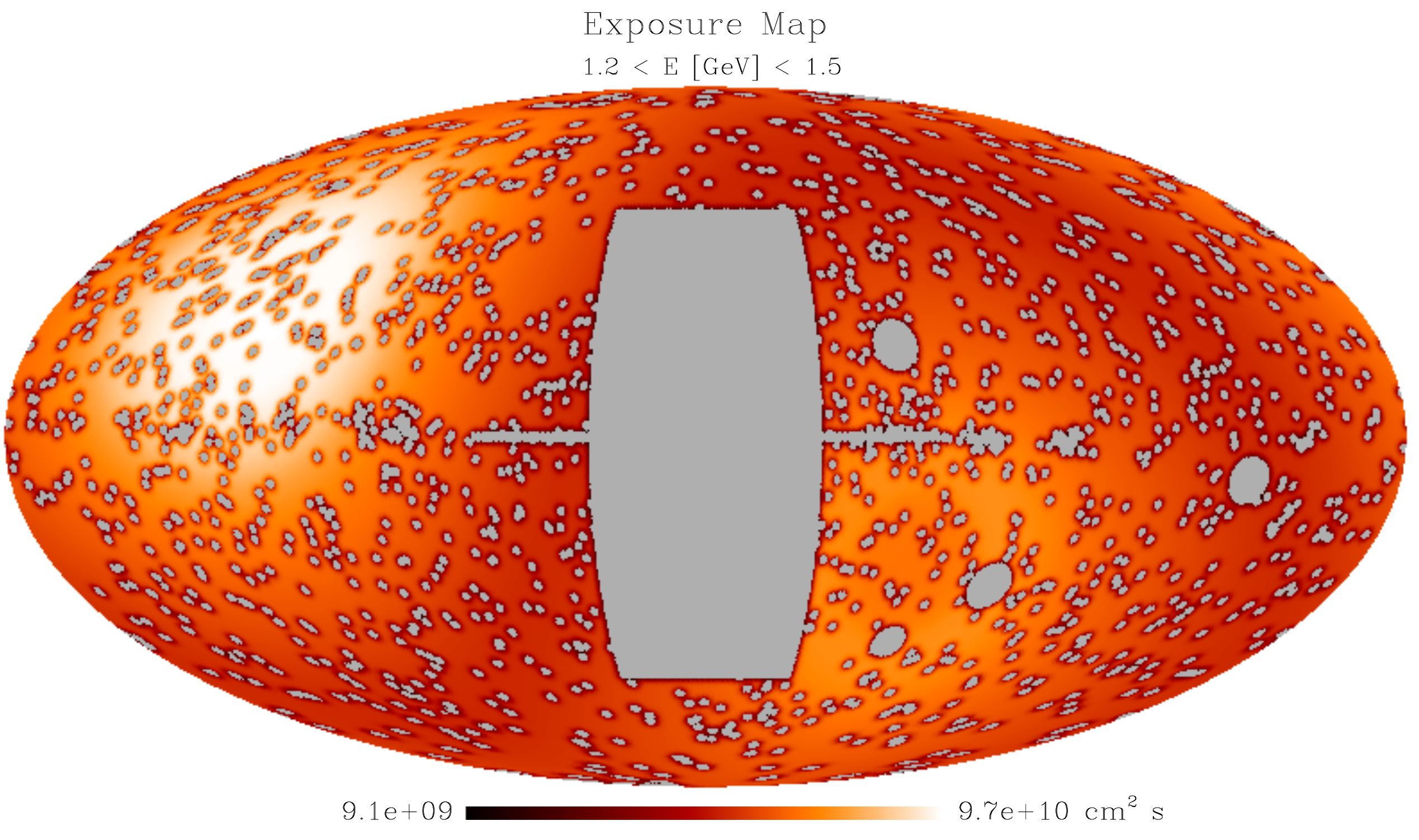}
    \end{minipage}\\
     \vspace{0.5 cm}
       \begin{minipage}{0.4\textwidth}
    \centering
   \includegraphics[scale=0.32]{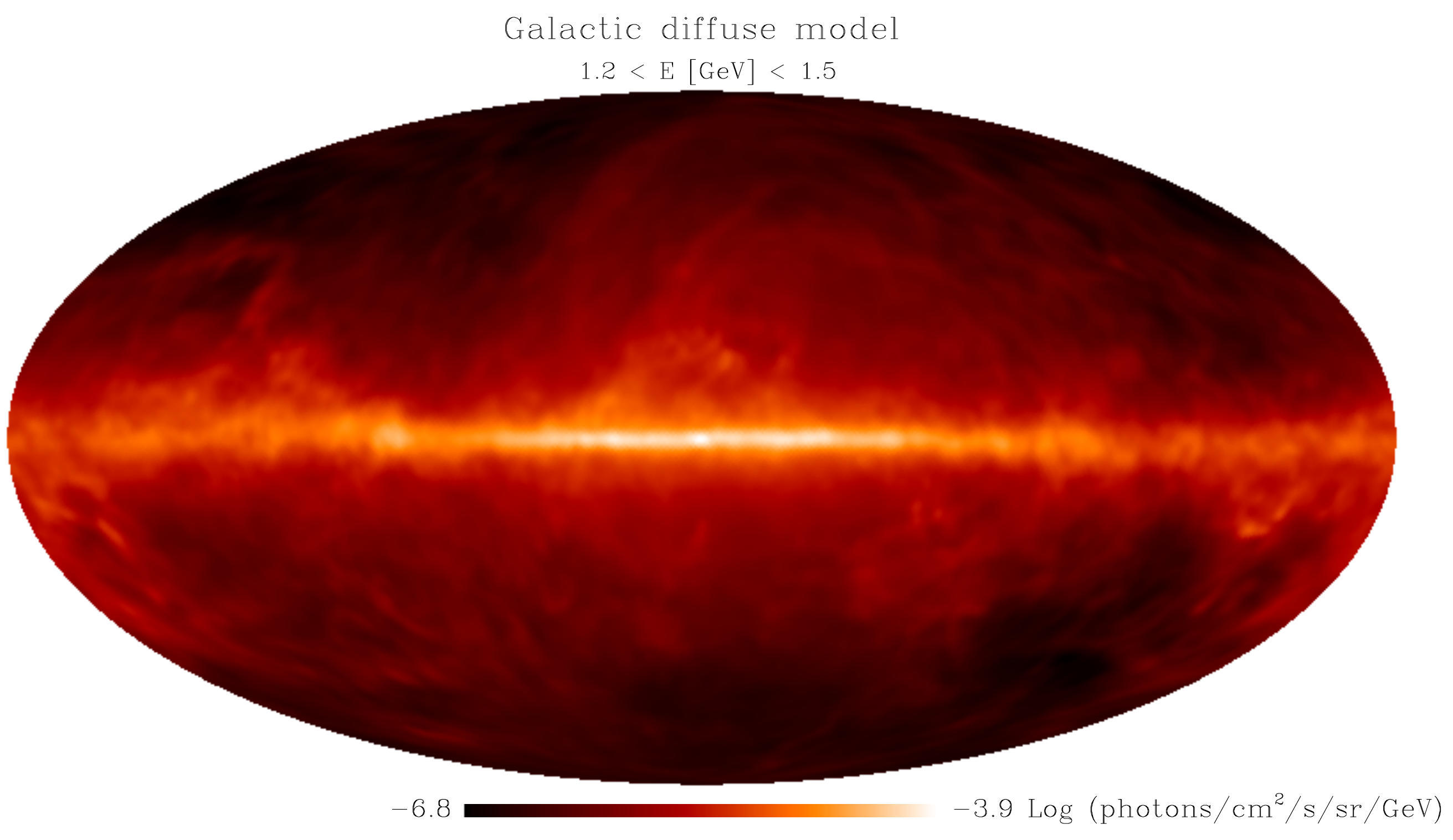}
  % \caption{\textit{Galactic diffuse model}}\label{fig:DiffuseMap}
    \end{minipage}\hspace{1.8 cm}
   \begin{minipage}{0.4\textwidth}
    \centering
    \includegraphics[scale=0.32]{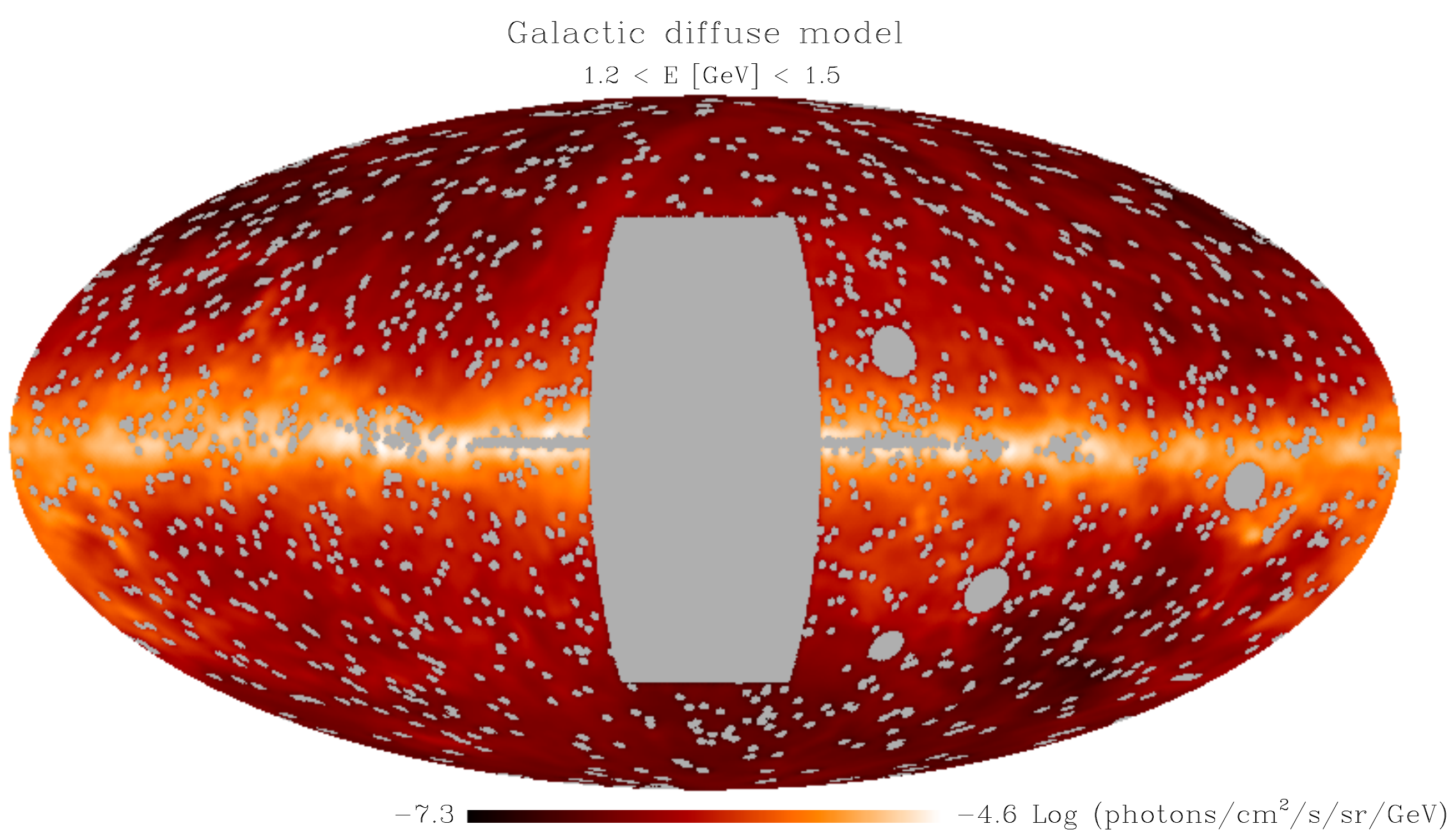}
    \end{minipage}\\
    \caption{\textit{Representative Fermi skymaps, obtained following the prescriptions outlined in Section~\ref{fermi_s}. For definiteness we show only front-converting events (see  Appendix~\ref{App:Count} for details), in a single energy bin centered in $E_{\gamma} = 1.34$ GeV. From top to bottom we show the corresponding counts map, the exposure map and the Galactic diffuse template. In the right panel we show
    the same skymaps after masking and smoothing.
    }}
\label{maps_f}
\end{figure}
In our analysis we employ the public available dataset from week $9$ to week $255$ (August $4$th $2008$ - April $25$th $2013$) \cite{Fermi_Data},
binning the data into $30$ log-spaced energy bins in the range from $0.3$ GeV to $300$ GeV.
  We use HEALPix~\cite{healpix_web} to bin the full skymap into iso-latitude equal-area pixels with NSIDE $= 256$. We use the event class denoted as \texttt{ULTRACLEAN}  and we apply the zenith angle cut $z_{\rm cut}=100^\circ$ to remove contributions from the Earth limb.\footnote{In Appendix~\ref{App:eventclass} we repeat our analysis using the \texttt{SOURCE} and \texttt{CLEAN} categories. We also tested different zenith angle cuts. In particular we find that both the zenith angle cut $z_{\rm cut}=90^\circ$ and the harder cut $z_{\rm cut}=80^\circ$ give results that are consistent if compared with $z_{\rm cut} = 100^\circ$.} With the information of the photon events and the exposure time of the LAT data, we generate the counts maps and exposure maps as shown in the first two rows of the left panel in Fig.~\ref{maps_f}. More information about data taking, counts map and exposure maps is given in Appendix~\ref{App:Count} and \ref{App:Exposure}.

  A large fraction of these observed photons comes from the Galactic diffuse gamma-ray emission and the isotropic extragalactic component. The diffuse emission is produced by interactions of cosmic rays
with interstellar gas and low-energy radiation fields.
Explicitly, cosmic ray electrons produce synchrotron radiation in the presence of magnetic fields due to their spiral motion. Furthermore they produce Bremsstrahlung radiation via interactions with the matter in the interstellar medium.
Another contribution is the ICS between cosmic ray electrons and photons of the low-energy
interstellar radiation field. Finally, cosmic ray protons interacting with the interstellar medium produce gamma rays via neutral pion decay. We use the \texttt{PASS6(V11)} diffuse model template provided by the Fermi collaboration in order to model the Galactic diffuse emission.\footnote{We use the \texttt{PASS6(V11)} diffuse model template instead of the more recent  \texttt{PASS7(V6)} because the latter already contains a template for the Fermi bubbles.} We show a representative skymap for the Galactic diffuse model in the third row of the left panel in Fig.~\ref{maps_f}. The isotropic  component, on the contrary, describes
diffuse gamma rays of extragalactic origin and the residual cosmic-ray contamination.
We model this component using a constant spectral template.

  The masking and smoothing procedures are performed on the noisy skymap before any statistical analysis. In particular we start masking the central disk in the region $|b| < 1^\circ$, $|l|< 60^\circ$ to exclude the region around the galactic center, notoriously plagued with large uncertainties. Point sources are masked as well, taking into account the energy dependence of the Point Spread Function (PSF) and the informations provided by the Fermi collaboration in the LAT 2-years Point Source Catalogue. For the extended sources, on the contrary, we use a fixed mask according to the corresponding templates. Notice that since the Galactic diffuse model does not include the Fermi bubbles we need to exclude this region in order to compare with the observed counts maps. Consequently, before performing the fitting procedure, a rectangular region overlapping the Fermi bubbles template  is masked. Finally, we smooth each map using an appropriate kernel in order to obtain
   a common Gaussian PSF with  $2^\circ$ full width at half maximum (FWHM) at $E_{\gamma} > 1$ GeV.
   Due to the large PSF at lower energy, a Gaussian PSF with $3^\circ$ FWHM is used at $E_{\gamma} < 1$ GeV.
 Representative masked and smoothed skymaps are shown in the right panel of Fig.~\ref{maps_f}. More information about masking and smoothing are given in Appendix~\ref{App:masking} and \ref{App:smoothing}.
 As an alternative approach to the masking method, it is possible to subtract the point sources from the counts maps. We discuss in detail the point source subtraction in Appendix~\ref{App:masking} and compare the results from the two different approaches.

  \subsection{Residual maps and fitting procedure}\label{sec:fitting}

\begin{figure}[!htb!]
   \centering
       \includegraphics[scale=0.45]{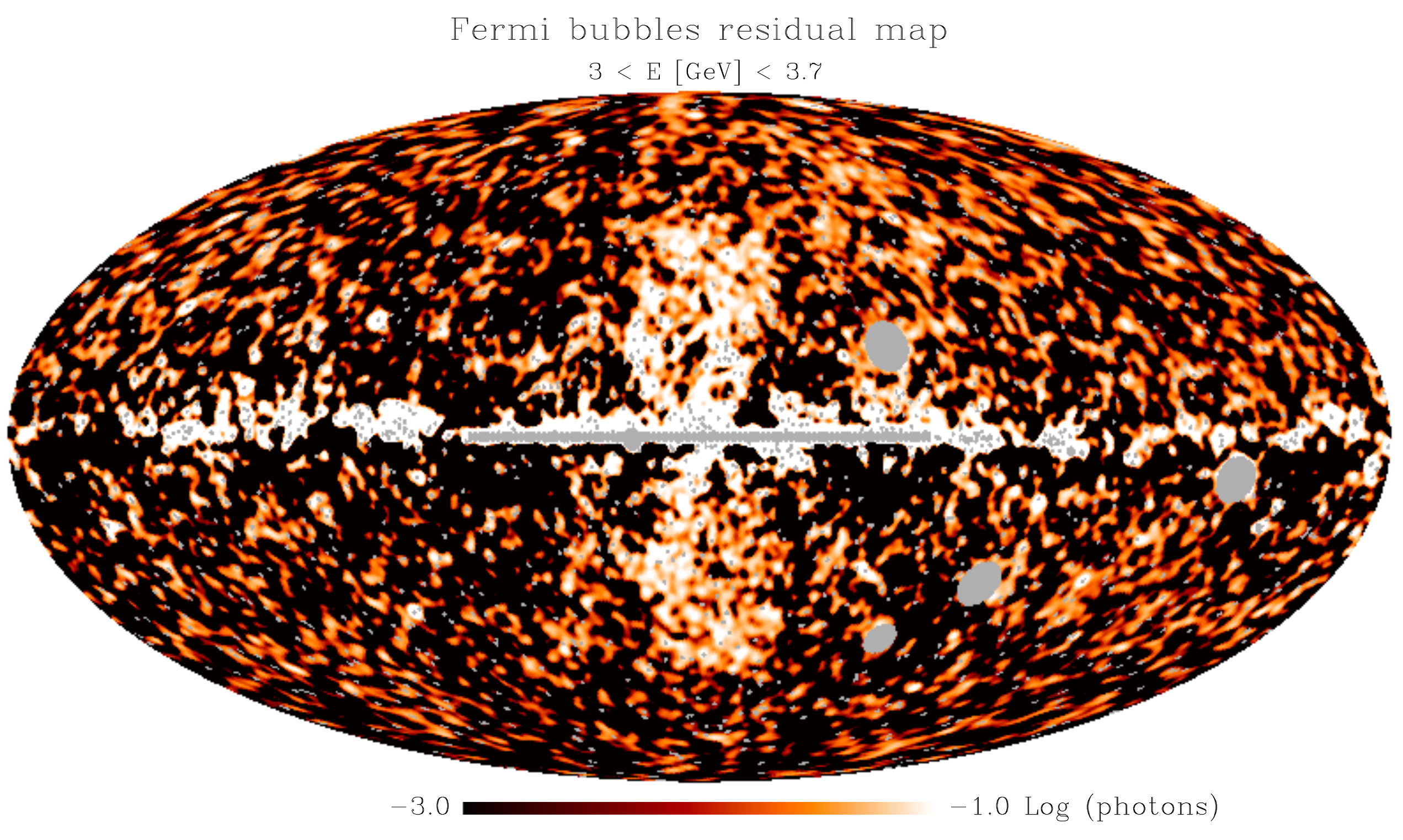}
 \caption{\emph{
 Observed gamma-ray sky after subtraction of the Galactic diffuse model and isotropic extragalactic component. We show front-converting events in the energy interval $E_{\gamma} =3-3.7$ GeV. The Fermi bubbles clearly stand out.
  }}
 \label{Fig:FermiBubblesSkymap}
\end{figure}

\begin{figure}[!htb!]
\begin{center}
\centering
  \begin{minipage}{0.4\textwidth}
   \centering
   \includegraphics[scale=0.65]{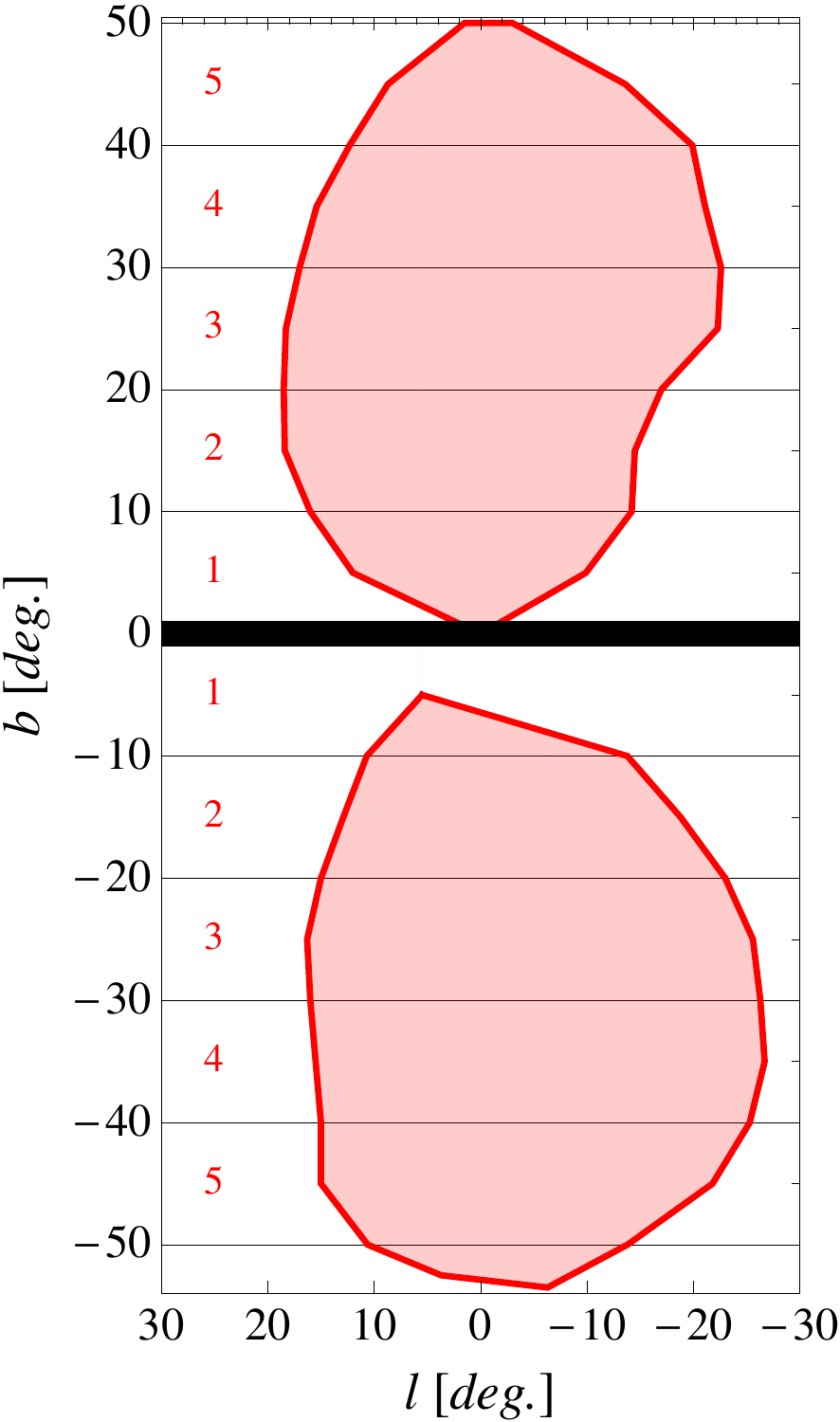}
   %\caption{\textit{Count Map}}\label{fig:CountMap}
    \end{minipage}\hspace{0.5 cm}
   \begin{minipage}{0.4\textwidth}
    \centering
    \includegraphics[scale=0.65]{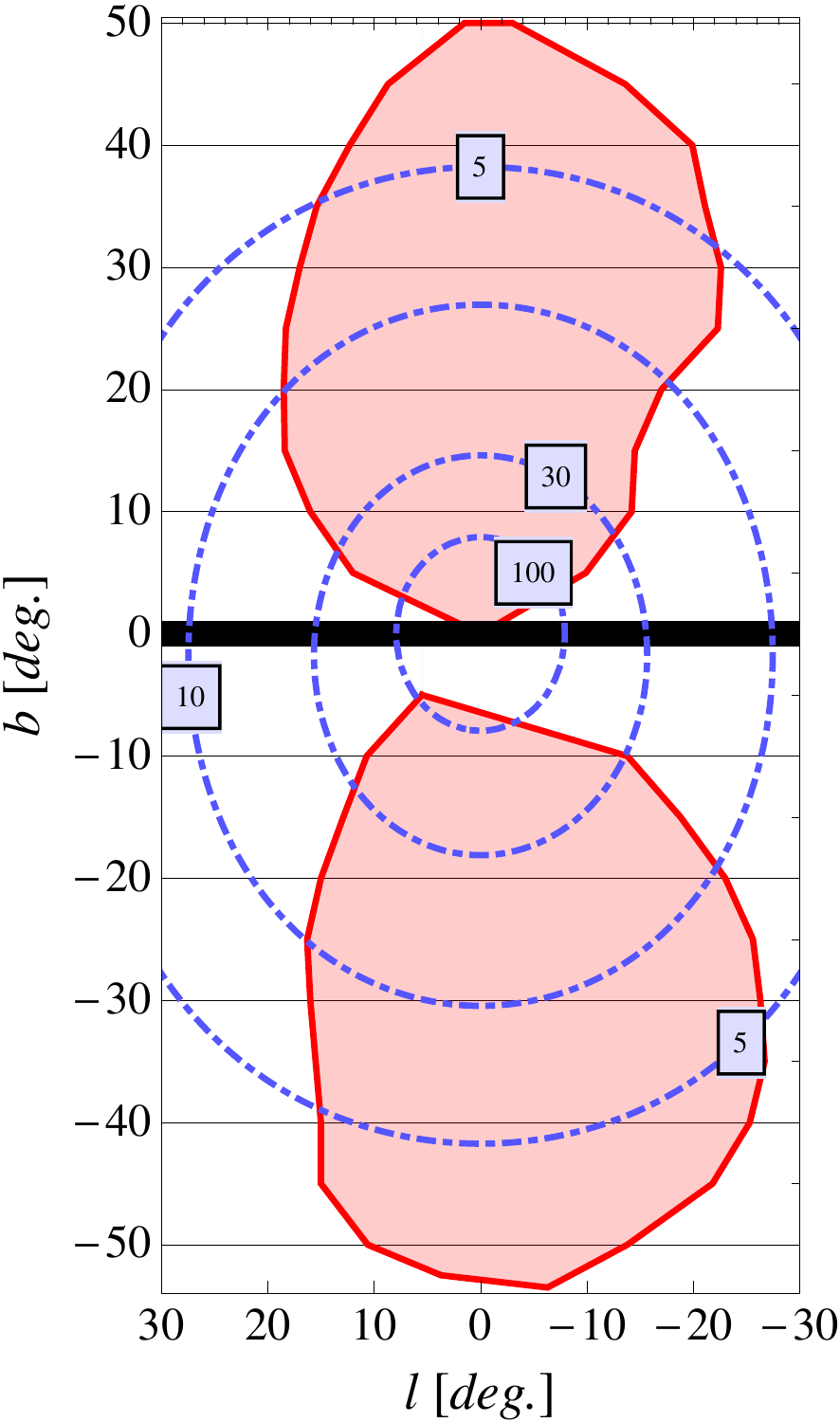}
    \end{minipage}\\
\caption{\textit{Fermi bubbles region. The edges of the bubbles follow the $(l,b)$
coordinates of the template defined in Ref.~\cite{Su:2010qj}. As in Ref.~\cite{Hooper:2013rwa}, we slice the bubbles in $5$ regions of different latitude, $|b|=1^{\circ}-10^{\circ},\dots, |b|=40^{\circ}-50^{\circ}$, labelled as $1,\dots, 5$ (left panel).
 In each slice the region inside the Fermi bubbles (red shadow) is used to get the residual energy spectra. In the right panel we also show the contours of constant $J$ factor, defined in Eq.~(\ref{eq:FSR}), using the generalized NFW profile in Eq.~(\ref{eq:genNFW}).
 % the outer region, on the contrary, is used to get the complementary energy spectra.
 }}
\label{fig:BubbleTemplate}
\end{center}
\end{figure}

Residual maps are obtained by subtracting from the observed counts maps a linear combination of the Galactic diffuse model and the isotropic template. As a representative example we show in
Fig.~\ref{Fig:FermiBubblesSkymap} the residual skymap obtained in the energy interval $E_{\gamma} =3-3.7$ GeV. To be more concrete, the subtraction procedure is as follows.
\begin{enumerate}
\item First we obtain the amplitudes of the templates performing a likelihood fit, focusing on the region outside the Fermi bubbles with the sky maps that include the rectangular mask. For each energy bin we define the following log-likelihood distribution\footnote{Notice that in principle the Poisson likelihood distribution in Eq.~(\ref{eq:logL}) should be computed on unsmoothed counts maps. However, as pointed out in Ref.~\cite{Dobler:2009xz}, the smoothing procedure does not entail any problems for the likelihood analysis. We explicitly checked that our results are stable if compared with those obtained using unsmoothed counts maps in the fitting procedure.}
\begin{equation}\label{eq:logL}
\ln \mathcal{L} = \sum_{i = {\rm pixels}}\left[ k_i \ln \mu_i -\mu_i -\ln (k_i !) \right]~,
\end{equation}
where $i$ runs over all the unmasked pixels and $ k_i$ ($\mu_i$) represents the observed (predicted) number of photons in the pixel $i$
\begin{eqnarray}
k_i &=& {\rm Count\,Map}|_i~,\label{eq:measured} \\
\mu_i &=& (a\cdot {\rm Diffuse\,Model}|_i + b)\cdot {\rm Exposure\,Map}|_i \cdot {\rm px}\cdot \Delta E_{\gamma}~.\label{eq:predicted}
\end{eqnarray}
In Eq.~(\ref{eq:predicted}) $\Delta E_{\gamma}$ is the width of the analyzed energy bin while ${\rm px}= \pi/(3\,{\rm NSIDE}^2)$ is the pixel solid angle. The predicted number of photons in Eq.~(\ref{eq:predicted}) takes into account the Galactic diffuse emission and the isotropic extragalactic
component. In each energy bin the log-likelihood distribution in Eq.~(\ref{eq:logL}) has two free parameters: the overall normalization of the diffuse emission, $a$, and the amplitude of the isotropic component,
$b$. We minimize the distribution $\mathscr{L}(a,b)\equiv -\ln\mathcal{L}$ w.r.t. these parameters to get the best-fit values $(a_{0}, b_{ 0})$. The likelihood fit of the templates takes only the statistical uncertainties into account. In order to find the $1$-$\sigma$ errors $\delta a$
 and $\delta b$ first we expand $\mathscr{L}(a,b)$ around the minimum $(a_{0}, b_{0})$, based on the Gaussian approximation, neglecting higher-order (i.e. $\geqslant 3$) derivatives; imposing the condition $\Delta \mathscr{L}=1/2$, the $1$-$\sigma$ errors are the square roots of the diagonals elements of the inverse of the Hessian matrix \cite{Su:2010qj}. 
 The values of the best fit coefficients $(a_0,b_0)$ 
 that we obtain, in each energy bin, from our likelihood analysis  are reasonable. In particular, we find values of $a_0$ close to $a_0=1$ while the values of $b_0$ are in good agreement with the typical order of magnitude describing the  isotropic component quoted by the Fermi collaboration
 for the \texttt{PASS6(V11)} diffuse model template.
 As an example, considering for definiteness front-converting events, at low energy in the second energy bin ($E_{\gamma}=424$ MeV) we find $a_0\simeq 1.18$, $b_0\simeq 7.5\times10^{-9}$; at high energy  ($E_{\gamma}=84$ GeV), 
 we find $a_0\simeq 0.94$, $b_0\simeq 10^{-14}$.

\item  We then  unmask the Fermi bubbles region keeping masked the
point
sources and the inner disk. Following  Ref.~\cite{Hooper:2013rwa}, we slice the Fermi bubbles in $5$ different regions, as shown in Fig.~\ref{fig:BubbleTemplate}.
In each one of these slices, and in each energy bin, we compute the difference
\begin{equation}\label{eq:ResidualFlux}
{\rm Res} =
\sum_i{\rm Count\, Map}|_i - \sum_i\left[(a_{0}
\cdot {\rm Diffuse\,Model}|_i + b_{0})\cdot {\rm Exposure\,Map}|_i
\cdot {\rm px}\cdot \Delta E_{\gamma}\right]~,
\end{equation}
where the sum runs over the unmasked pixels of the analyzed region. Eq.~(\ref{eq:ResidualFlux}) represents the residual number of photons after background subtraction. Dividing by the total exposure times pixel solid angle and energy width, we obtain the differential flux of the Fermi bubbles ($d\Phi/dE_{\gamma}d\Omega$, energy spectrum in units of photons GeV$^{-1}$ cm$^{-2}$ s$^{-1}$ sr$^{-1}$). The error bars on the residual value in Eq.~(\ref{eq:ResidualFlux})
are the statistical errors.

\end{enumerate}

%%%%%%%%%%%%%%%%%%%%%%%%%%%%%%%%%%%%%
\subsection{The latitude-dependent energy spectrum of the Fermi bubbles}\label{sec:residual}
%%%%%%%%%%%%%%%%%%%%%%%%%%%%%%%%%%%%%

\subsubsection{On the relevance of the latitude-dependent approach}

In this Subsection, we stress the importance of the latitude-dependent approach. The aim of our analysis is to look for the hint of a DM component in the
residual energy spectrum at the Fermi bubble region. The DM annihilation produces gamma-rays both by electromagnetic Final State Radiation (FSR) and by ICS on the ambient light. The differential flux of FSR photons from the angular direction $d\Omega$ is given by \cite{Cirelli:2010xx,Ciafaloni:2010ti}
\begin{equation}\label{eq:FSR}
\frac{d\Phi}{dE_{\gamma}d\Omega }=\frac{r_{\odot}}{8\pi}\left(\frac{\rho_{\odot}}{M_{\rm DM}}\right)^2 J
\sum_f \langle \sigma v\rangle_f\frac{dN_{\gamma}^{f}}{dE_{\gamma}},~~~~
J(\theta)=\int_{\rm l.o.s.}\frac{ds}{r_{\odot}}\left[\frac{\rho_{\rm DM}(r(s,\theta))}{\rho_{\odot}}\right]^{2}~,
\end{equation}
where $M_{\rm DM}$ is the DM mass,  $\rho_{\odot}=0.4$ GeV/cm$^3$ is the density of DM at the location of the Sun $r_{\odot}=8.33$ kpc, and $dN_{\gamma}^{f}/dE_{\gamma}$ is the number of photon per unit energy per DM annihilation with final state $f$ and thermal averaged cross section $\langle \sigma v\rangle_f$.\footnote{We consider here the annihilation of self-conjugate/Majorana DM particles.} The $J$ factor in Eq.~(\ref{eq:FSR}) is obtained by integrating the square of the normalized annihilating DM density over the line of sight (l.o.s.), where $s$ is the distance between the Earth and the point of interest and the spherical radial coordinate $r$, centered at the Galactic center, is given by $r(s,\theta)=(
r_{\odot}^2 + s^2 -2sr_{\odot}\cos \theta
)^{1/2}$ in which $\theta$ is the angle between the l.o.s. and the axis connecting the Earth with the Galactic center.
The $J$ factor clearly depends on the DM density distribution $\rho_{\rm DM}(r)$. On the other hand, the density profile of the DM in the Milky Way galaxy is not well understood. Even if numerical N-body simulation seems to favor a distribution peaking toward the center, the inclusion of baryons may overturn this conclusion in favor of a density distribution described by an isothermal sphere \cite{RomanoDiaz:2008wz, RomanoDiaz:2009yq}.
For illustration, we choose the generalized Navarro-Frenk-White (gNFW) profile \cite{Navarro:1995iw,Klypin:2001xu}
\begin{equation}\label{eq:genNFW}
\rho_{\rm gNFW}(r)=\rho_s\left(\frac{r}{R_s}\right)^{-\gamma}
\left(1+\frac{r}{R_s}\right)^{\gamma-3}~,
\end{equation}
with inner slope $\gamma =1.2$ and scale radius $R_s = 20$ kpc.\footnote{The standard NFW profile corresponds to $\gamma=1$ and $R_s = 24.42$ kpc.} The normalization $\rho_s$ is fixed by $\rho_{\rm gNFW}(r_{\odot})=\rho_{\odot}=0.4$ GeV/cm$^3$. In the right panel of Fig.~\ref{fig:BubbleTemplate} we show different contours of constant value for the $J$ factor in Eq.~(\ref{eq:FSR}) superimposed on the Fermi bubbles template. It is clear that the $J$ factor, and hence the DM photon flux, is larger near the Galactic center, i.e. in the low latitude region of the Fermi bubbles.

A similar argument is still valid also postulating the existence of a population of unresolved millisecond pulsars (MSP). In the so called baseline model \cite{FaucherGiguere:2009df}, for instance, the surface density of the MSP is described by $\rho_{\rm MSP}(r)\propto \exp(-r^2/2\sigma_r^2)$,
where $\sigma_r\sim 5$ kpc.\footnote{See Ref.~\cite{Hooper:2013nhl} for a recent analysis about the possible connection between the MSP and the spectrum of the Fermi bubbles.}

\subsubsection{Results and comments}\label{Sec:ResultsAndComments}

\begin{figure}[!htb!]
 \centering
  \begin{minipage}{0.4\textwidth}
   \centering
   \includegraphics[scale=0.42]{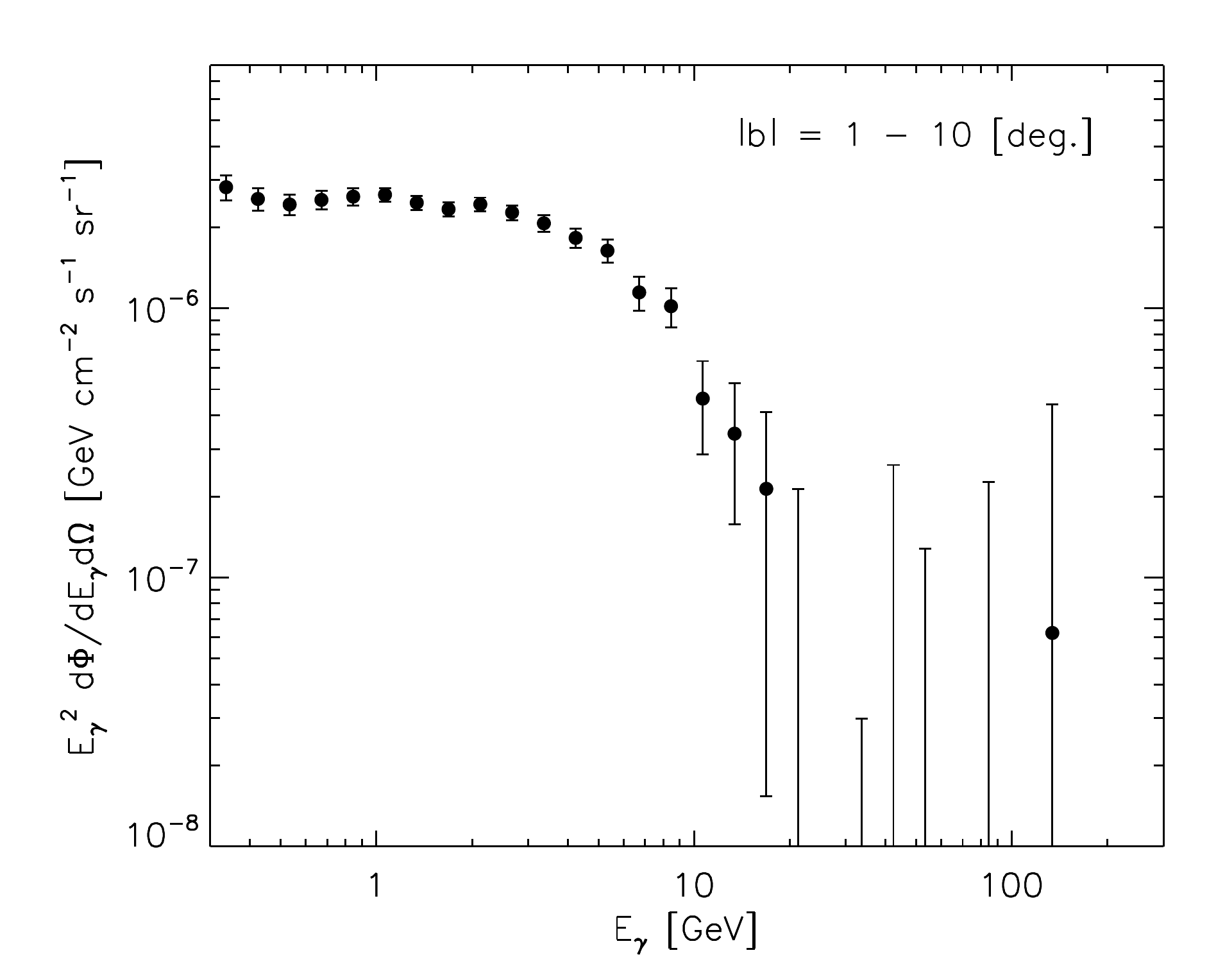}
   %\caption{\textit{Count Map}}\label{fig:CountMap}
    \end{minipage}\hspace{1.2 cm}
   \begin{minipage}{0.4\textwidth}
    \centering
    \includegraphics[scale=0.42]{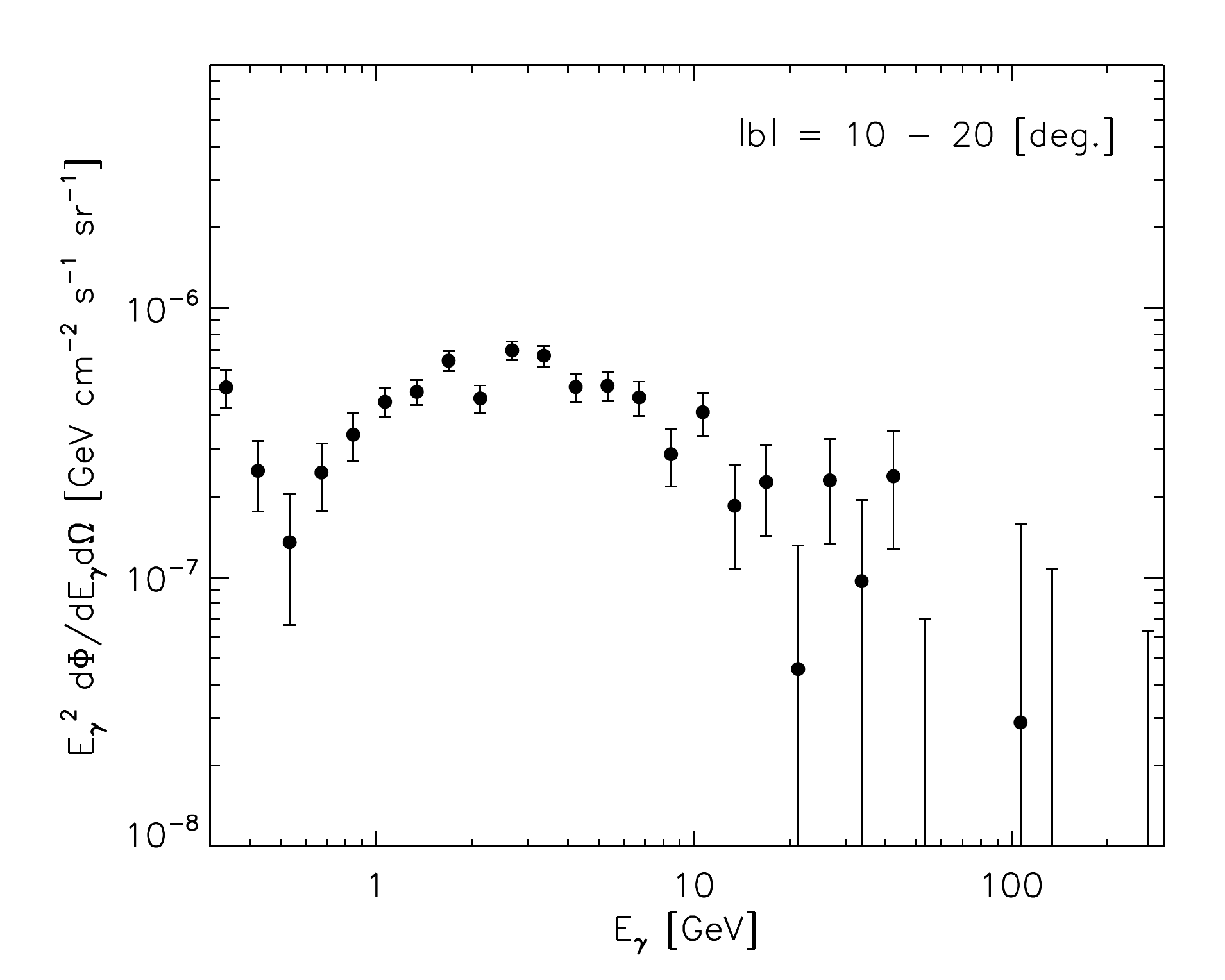}
    \end{minipage}\\
    \vspace{0.5 cm}
   \begin{minipage}{0.4\textwidth}
    \centering
   \includegraphics[scale=0.42]{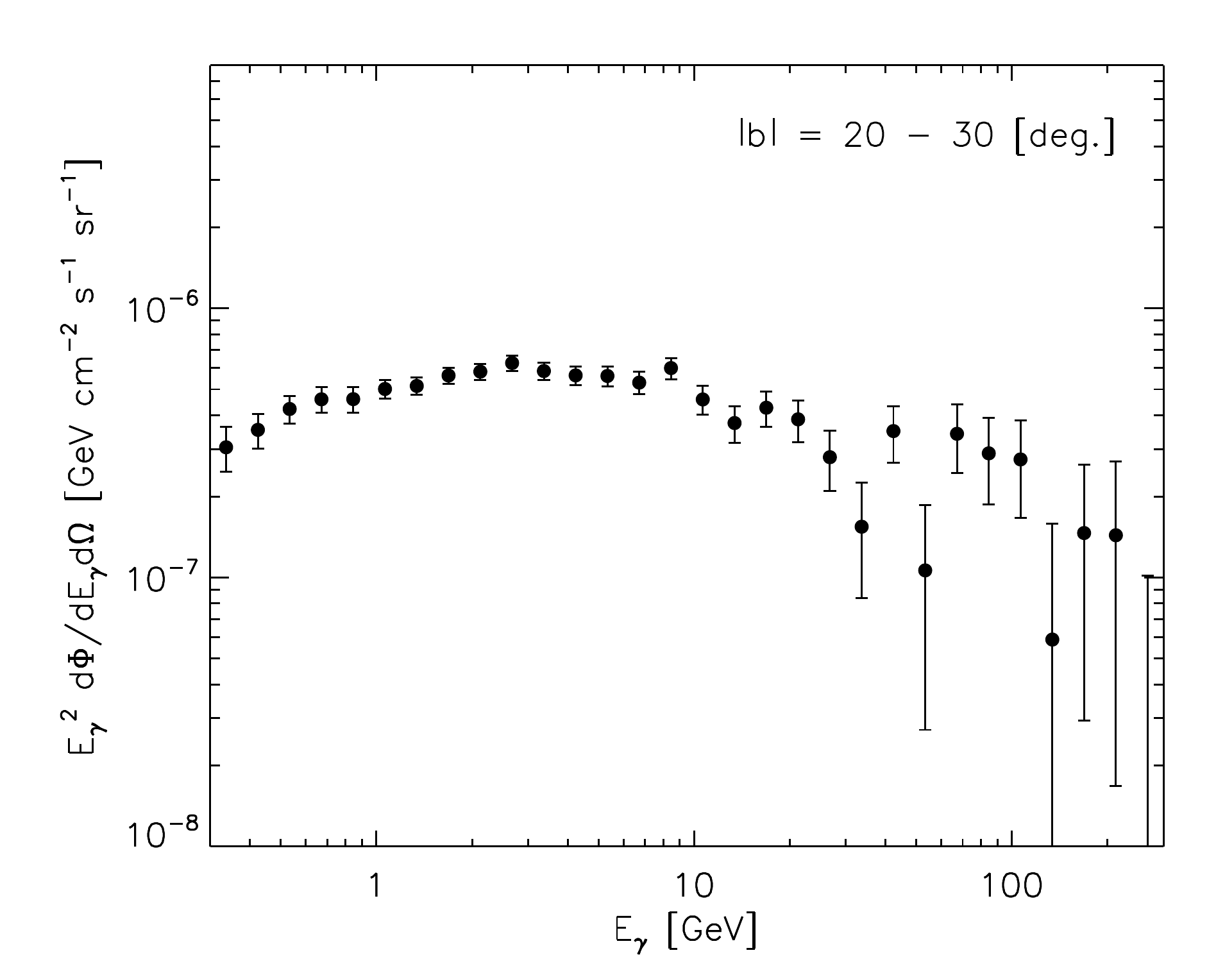}
   %\caption{\textit{Exposure Map}}\label{fig:ExposureMap}
    \end{minipage}\hspace{1.2 cm}
   \begin{minipage}{0.4\textwidth}
    \centering
    \includegraphics[scale=0.42]{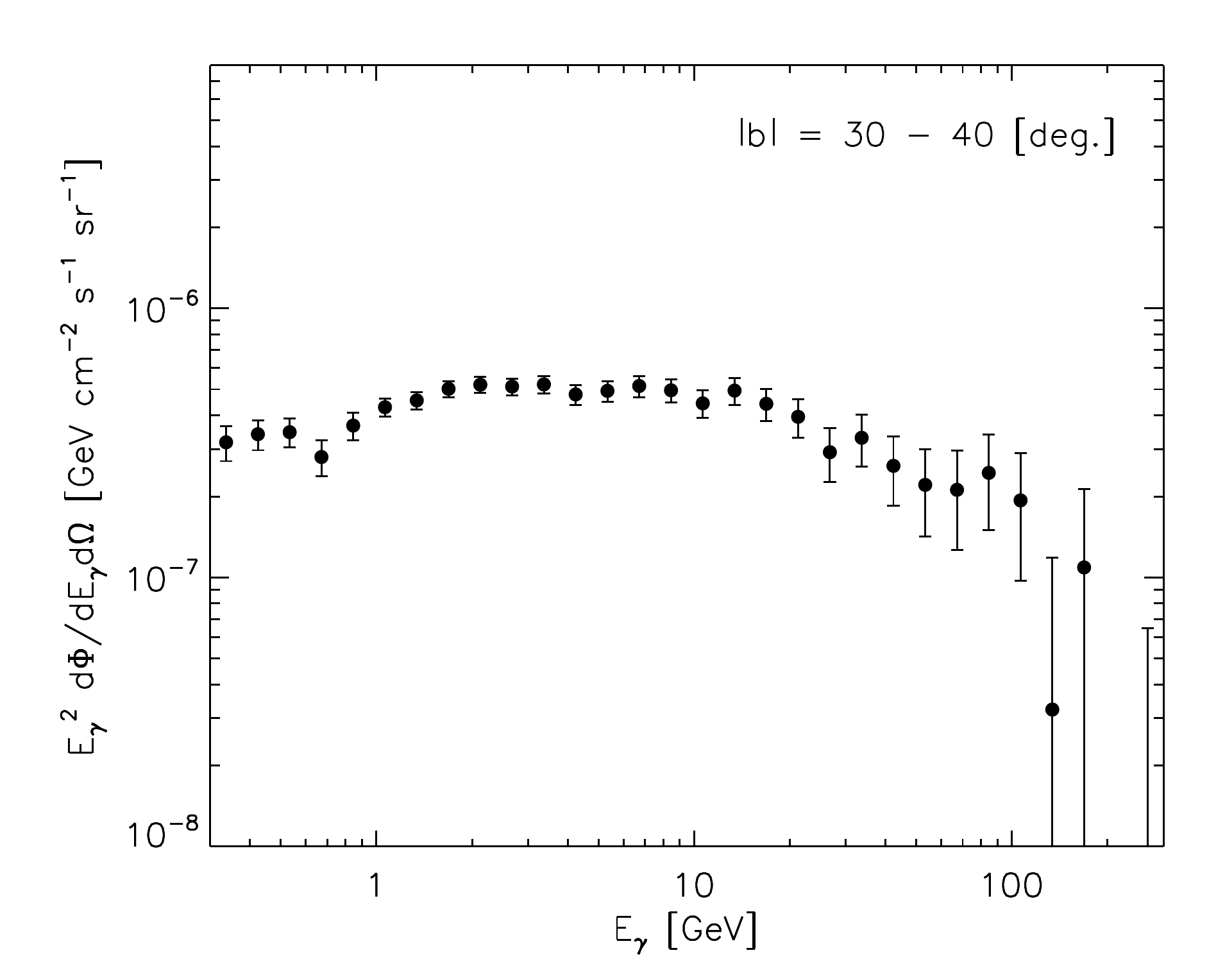}
    \end{minipage}\\
     \vspace{0.5 cm}
       \begin{minipage}{1\textwidth}
    \centering
   \includegraphics[scale=0.42]{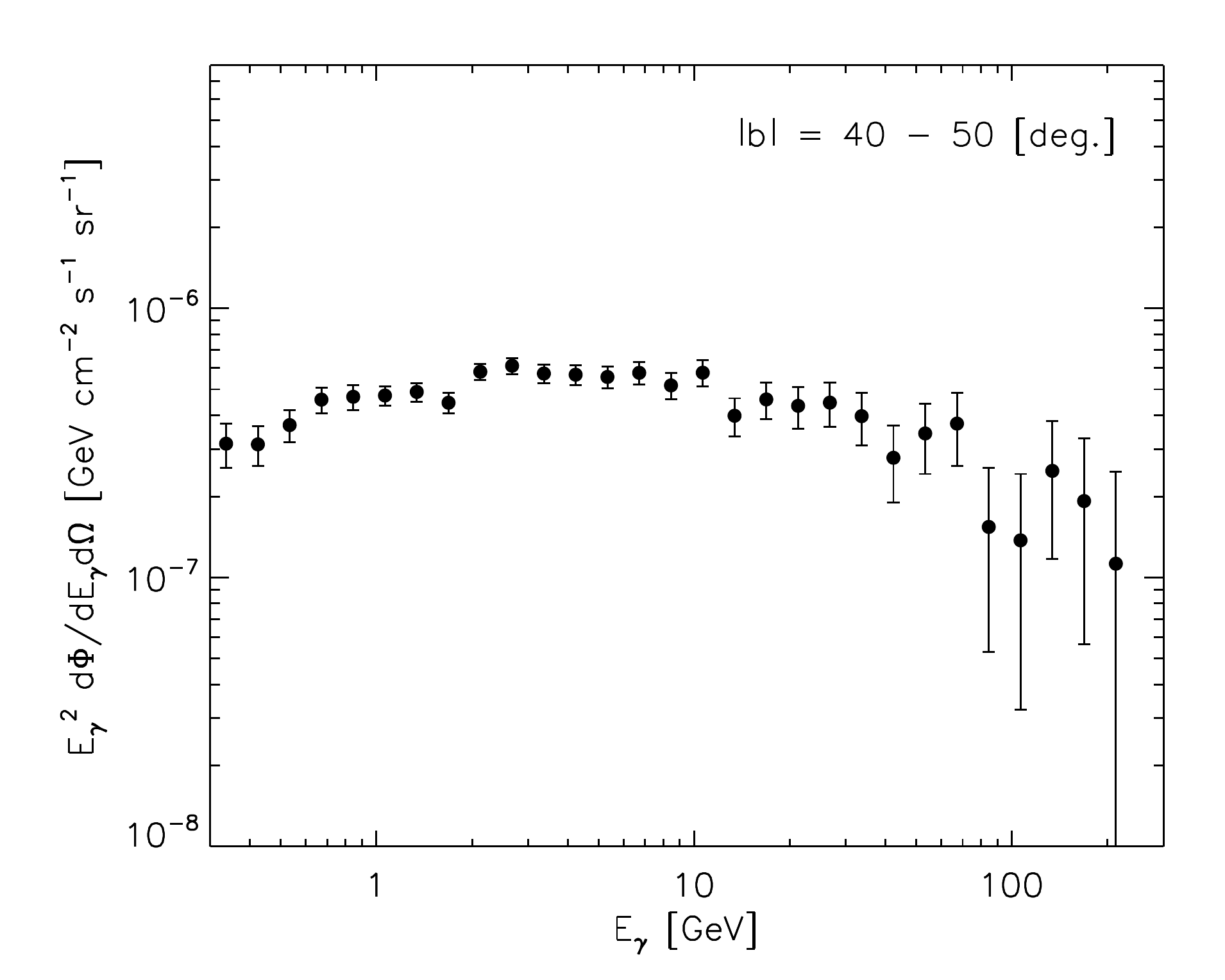}
  % \caption{\textit{Galactic diffuse model}}\label{fig:DiffuseMap}
    \end{minipage}
    \caption{\textit{
    Fermi bubbles energy spectrum broken into the five strips shown in Fig.~\ref{fig:BubbleTemplate}. We use \texttt{ULTRACLEAN} events, masking the inner disk in the region  $|b|<1^{\circ}$, $|l|<60^{\circ}$.
    }}\label{fig:MAIN}
\end{figure}

We show the energy spectrum (in $E_{\gamma}^2d\Phi/dE_{\gamma}d\Omega$) of the Fermi bubbles as a function of the latitude in Fig.~\ref{fig:MAIN}.
We can clearly discern three main spectral features.

\begin{figure}[!htb!]
 \centering
  \begin{minipage}{0.4\textwidth}
   \centering
   \includegraphics[scale=0.42]{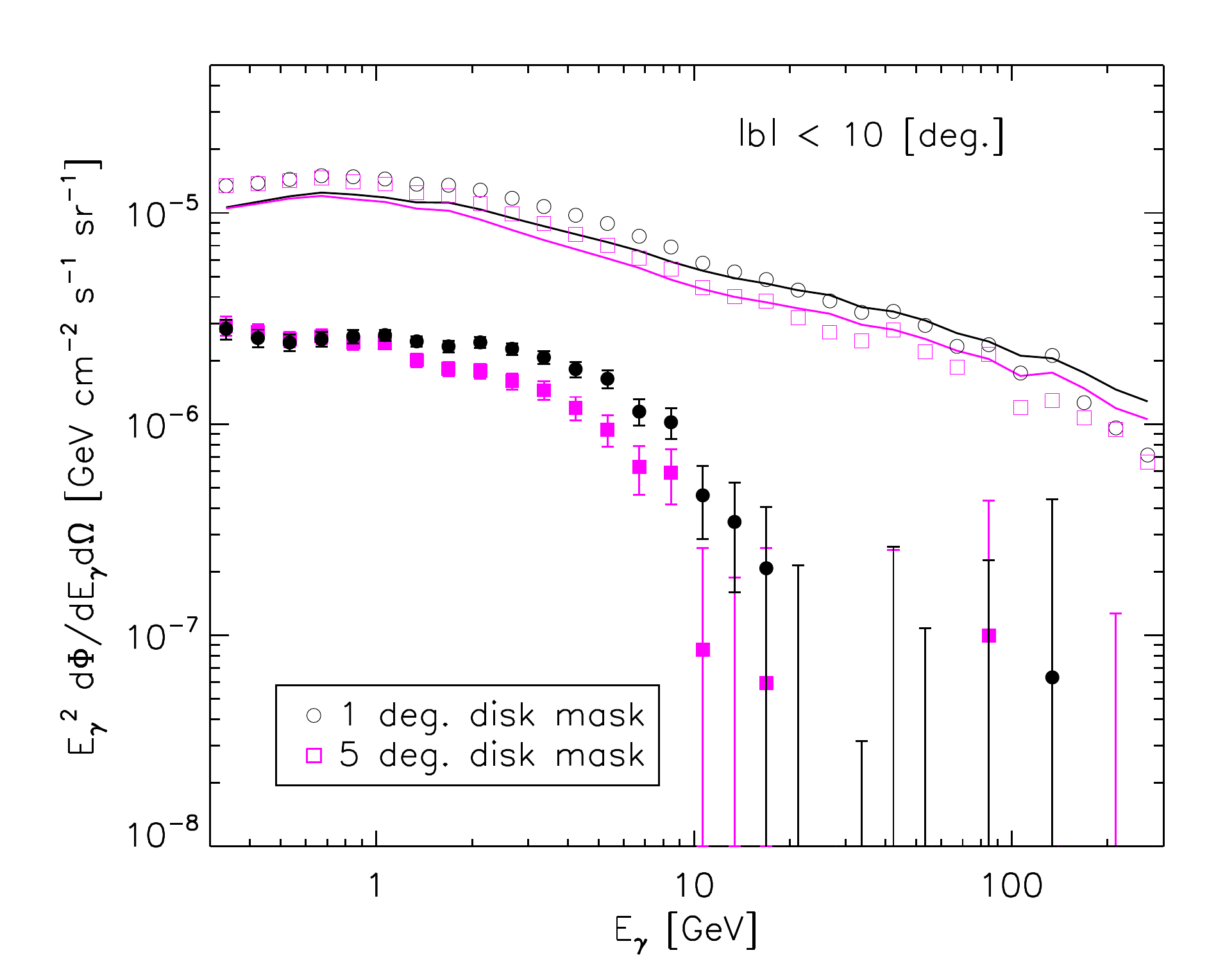}
   %\caption{\textit{Count Map}}\label{fig:CountMap}
    \end{minipage}\hspace{1.2 cm}
   \begin{minipage}{0.4\textwidth}
    \centering
    \includegraphics[scale=0.42]{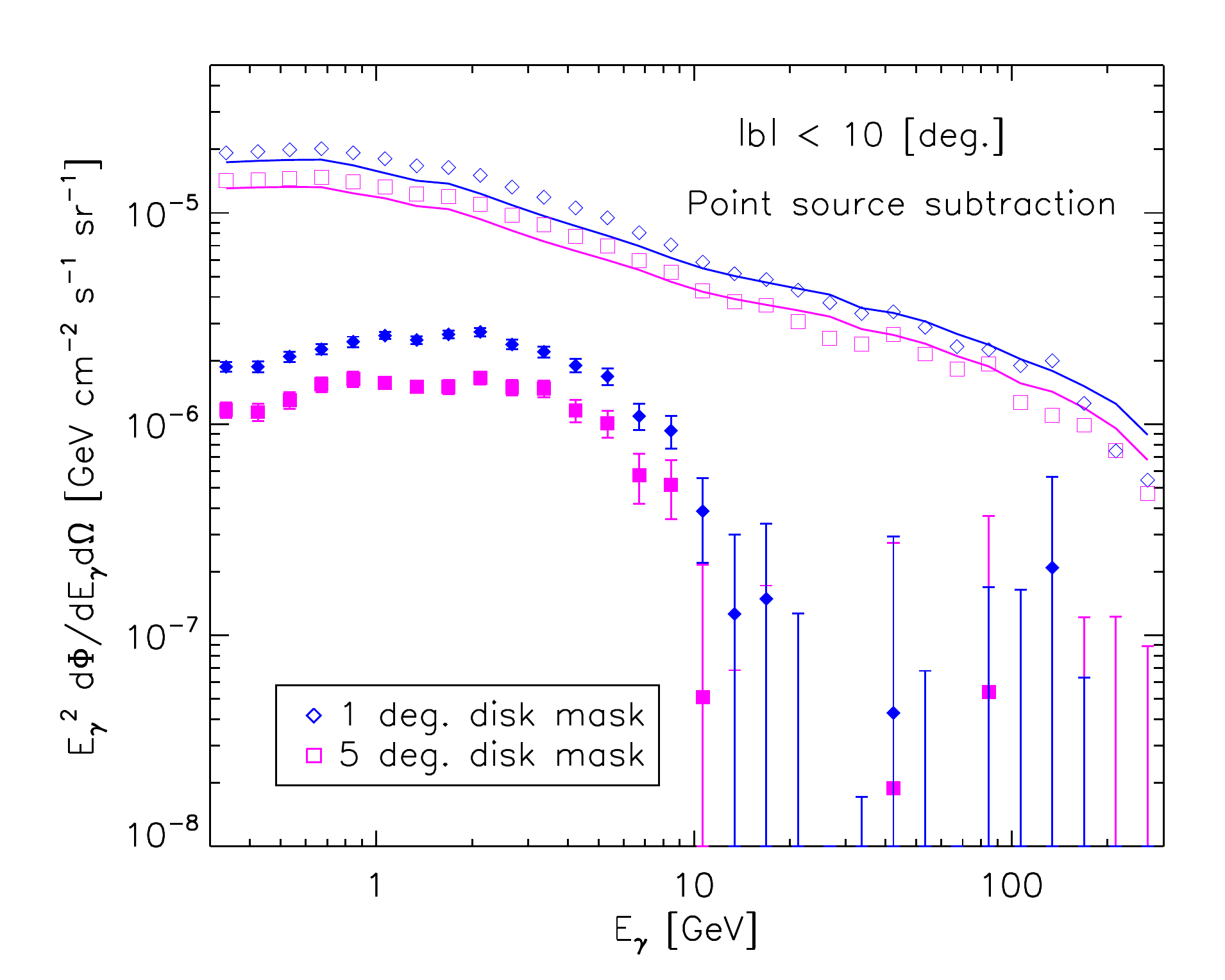}
    \end{minipage}\\
    \vspace{0.5 cm}
   \begin{minipage}{0.4\textwidth}
    \centering
   \includegraphics[scale=0.42]{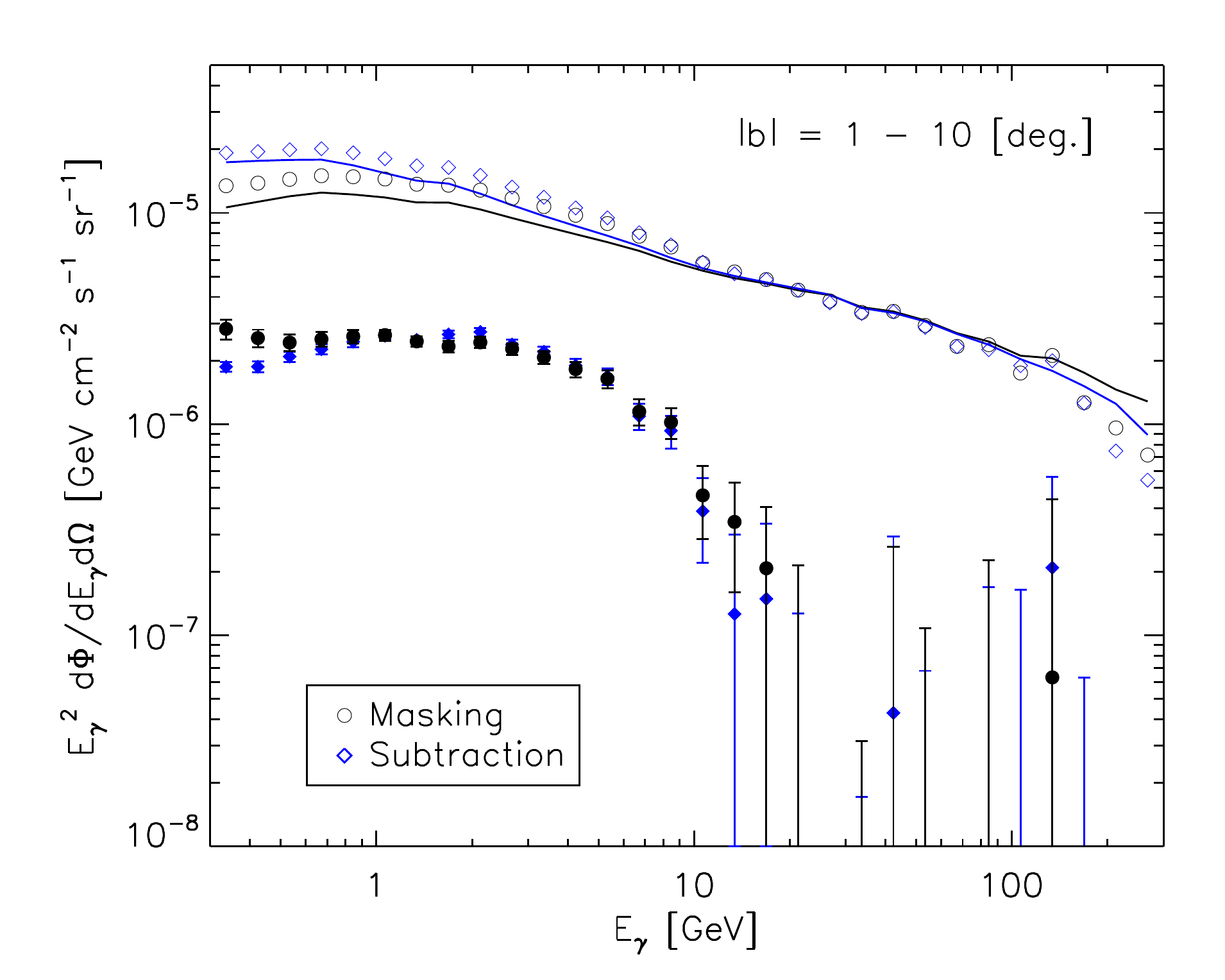}
   %\caption{\textit{Exposure Map}}\label{fig:ExposureMap}
    \end{minipage}\hspace{1.2 cm}
   \begin{minipage}{0.4\textwidth}
    \centering
    \includegraphics[scale=0.42]{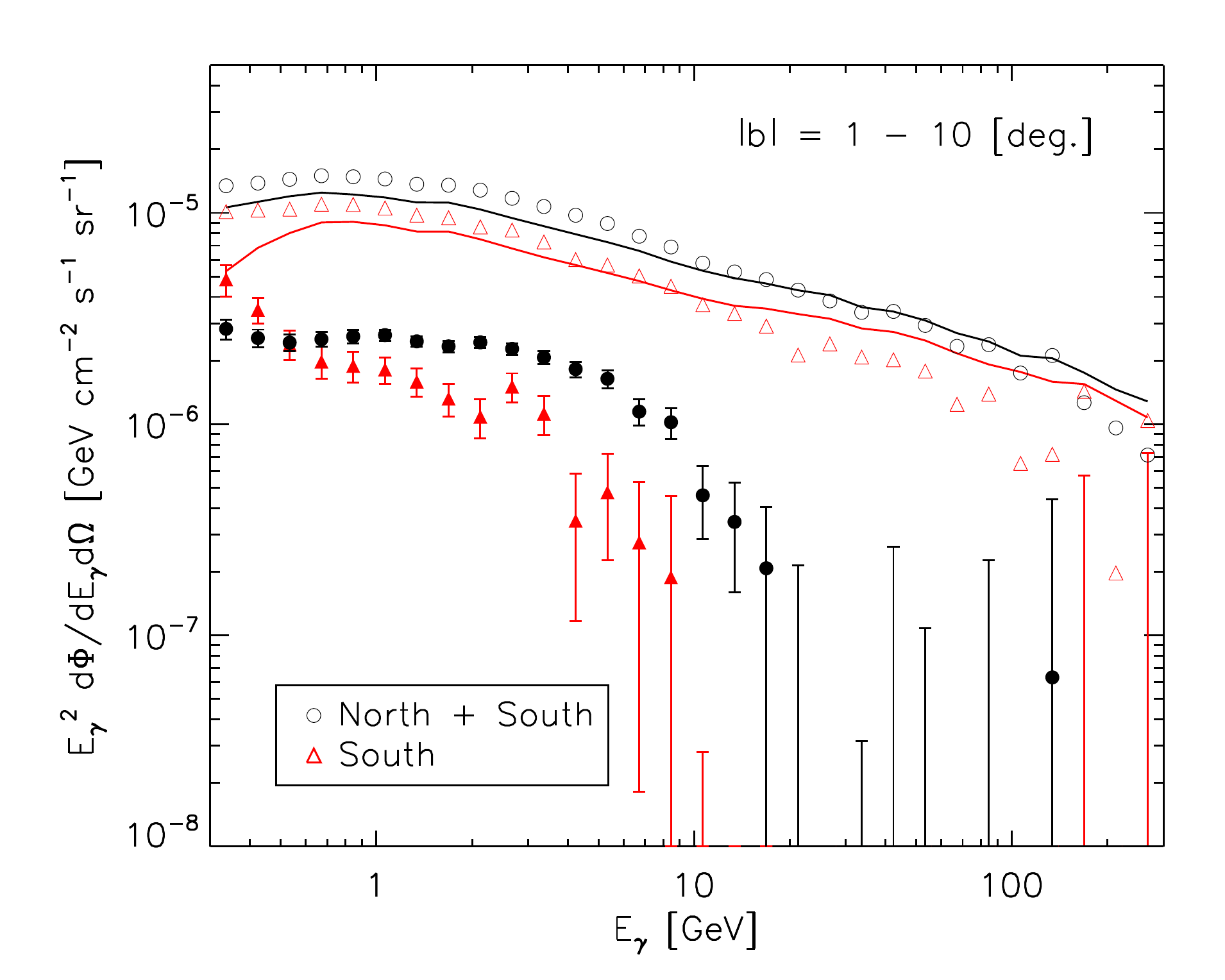}
    \end{minipage}
    \caption{\textit{
    Fermi bubbles energy spectrum in the first slice $|b|<10^{\circ}$ with the corresponding error bars. In addition, the observed flux and the best-fit theoretical prediction from the Galactic diffuse model and the isotropic extragalactic component are plotted (we use same color code w.r.t. the residual values but, respectively,  with empty symbols and solid lines).
    In the upper panel we compare the results obtained masking the inner disk in the region $|b|<1^\circ$, $|l|<60^{\circ}$ and $|b|< 5^\circ$, $|l|< 60^{\circ}$.  In the left one, the point sources are masked; in the right one, the point sources are subtracted.
In the bottom left panel we compare point source masking and subtraction with $1^\circ$ inner disk mask. In the bottom right panel we consider the Southern hemisphere and North+South hemispheres.
   % In the upper panel we show the results obtained masking the inner disk in the region $|b|<1^{\circ}$, $|l|<60^{\circ}$ (black dots) and $|b|<5^{\circ}$, $|l|<60^{\circ}$ (purple squares). In the lower panel we show
 %the  energy spectrum obtained considering the point source subtraction (blue filled diamonds) instead of the masking procedure (black filled dots). In Appendix~\ref{App:masking} we describe in detail the two methods. In both panels we also show the observed flux and the best-fit theoretical prediction from the Galactic diffuse model and the isotropic extragalactic component (we use same color code w.r.t. the residual values but, respectively,  with empty symbols and solid lines).
    }}\label{fig:MAIN2}
\end{figure}
\begin{figure}[!htb!]
 \centering
  \begin{minipage}{0.4\textwidth}
   \centering
   \includegraphics[scale=0.42]{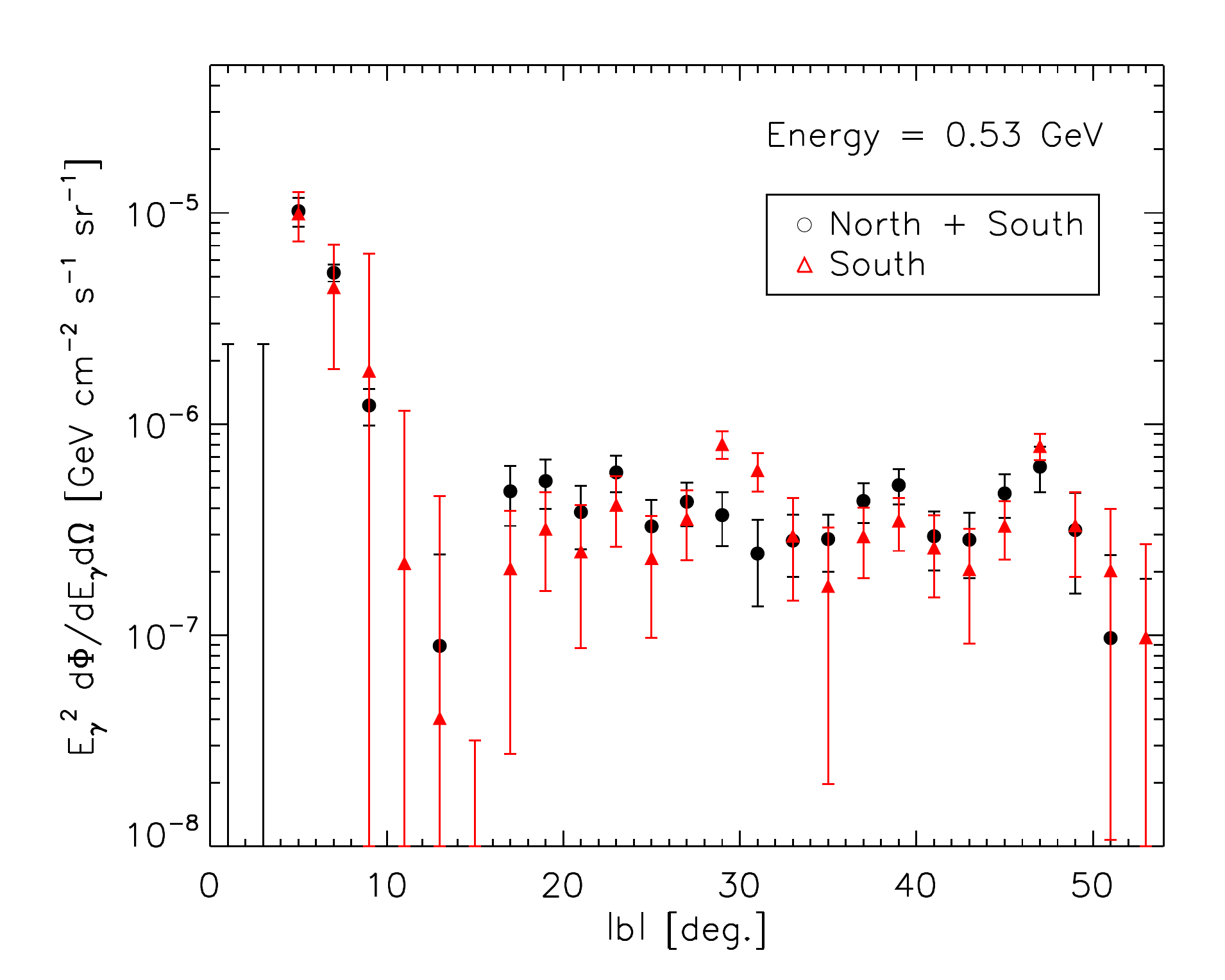}
   %\caption{\textit{Count Map}}\label{fig:CountMap}
    \end{minipage}\hspace{1.2 cm}
   \begin{minipage}{0.4\textwidth}
    \centering
    \includegraphics[scale=0.42]{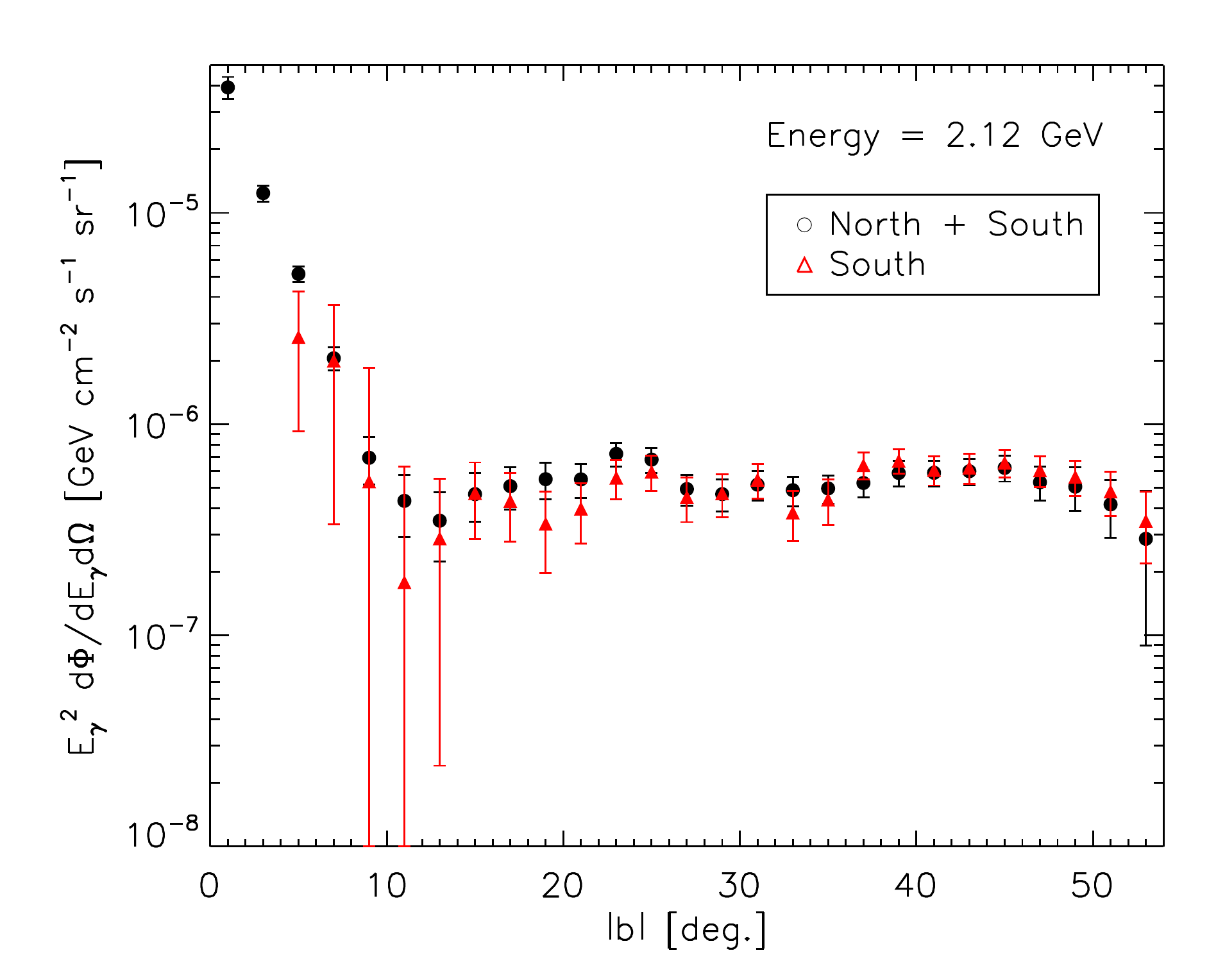}
    \end{minipage}
    \caption{\textit{
    Fermi bubbles energy spectrum as a function of the latitude. We bin the data in intervals $\Delta b = 2^{\circ}$, masking the inner disk for $|b|< 1^{\circ}$. We show two representative energy bins, $E_{\gamma}=0.53$ GeV (left panel) and $E_{\gamma} = 2.12$ GeV (right panel). Moreover, we compare the energy spectrum of the whole Fermi bubbles region (black dots) with the part lying in the Southern hemisphere of the sky (red triangles).
    }}\label{fig:LAT}
\end{figure}
\begin{figure}[!htb!]
 \centering
  \begin{minipage}{0.4\textwidth}
   \centering
   \includegraphics[scale=0.42]{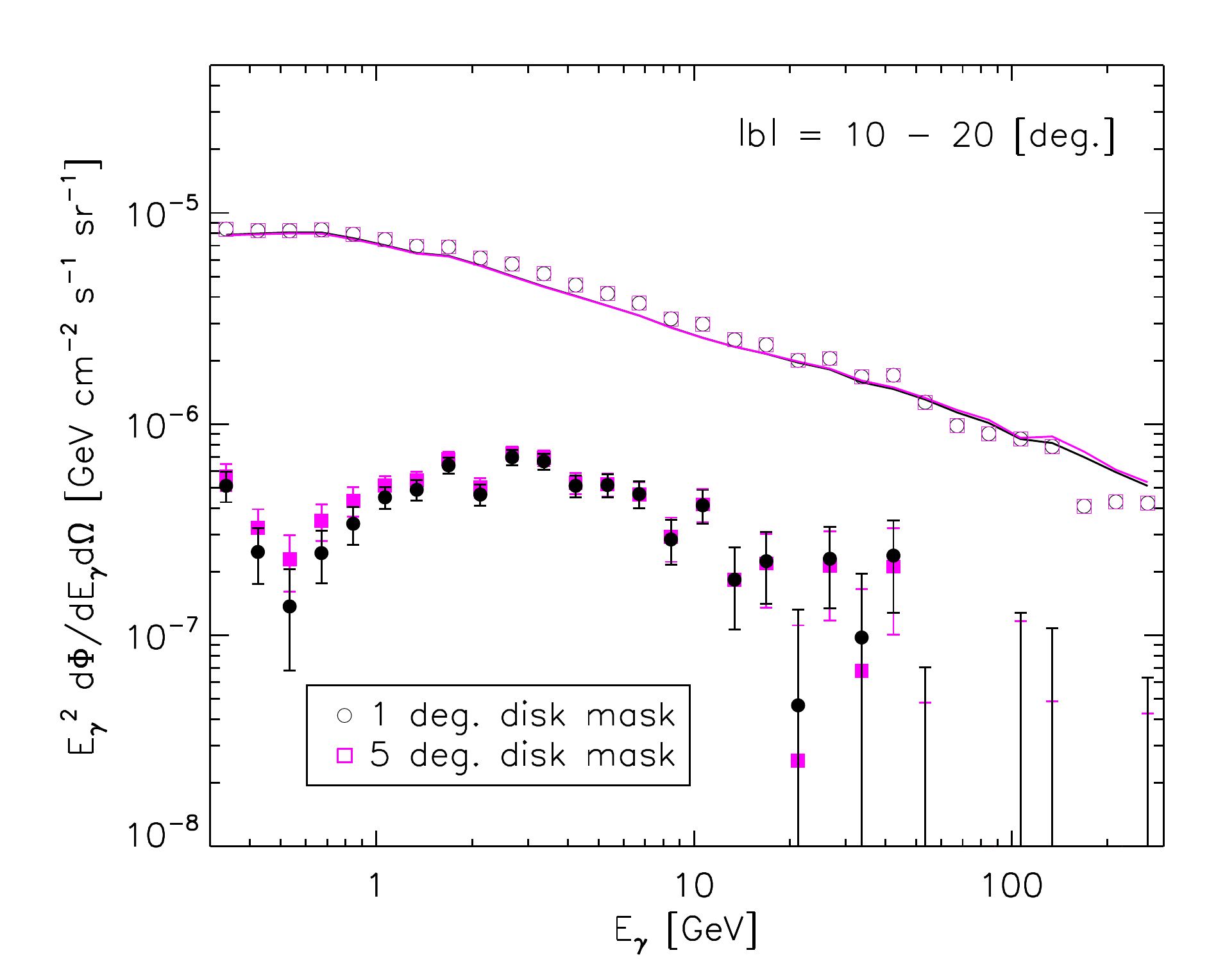}
   %\caption{\textit{Count Map}}\label{fig:CountMap}
    \end{minipage}\hspace{1.2 cm}
   \begin{minipage}{0.4\textwidth}
    \centering
    \includegraphics[scale=0.42]{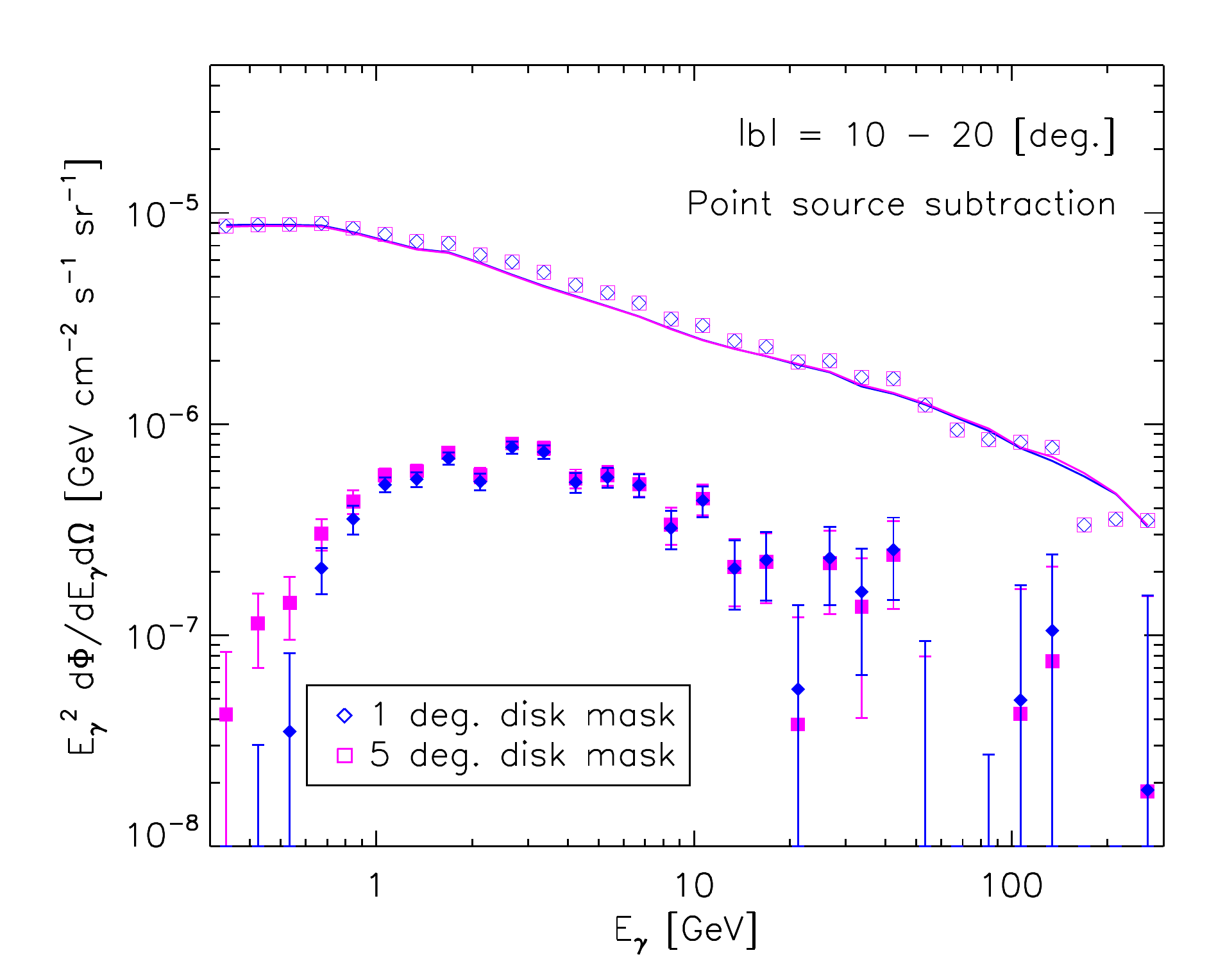}
    \end{minipage}\\
    \vspace{0.5 cm}
   \begin{minipage}{0.4\textwidth}
    \centering
   \includegraphics[scale=0.42]{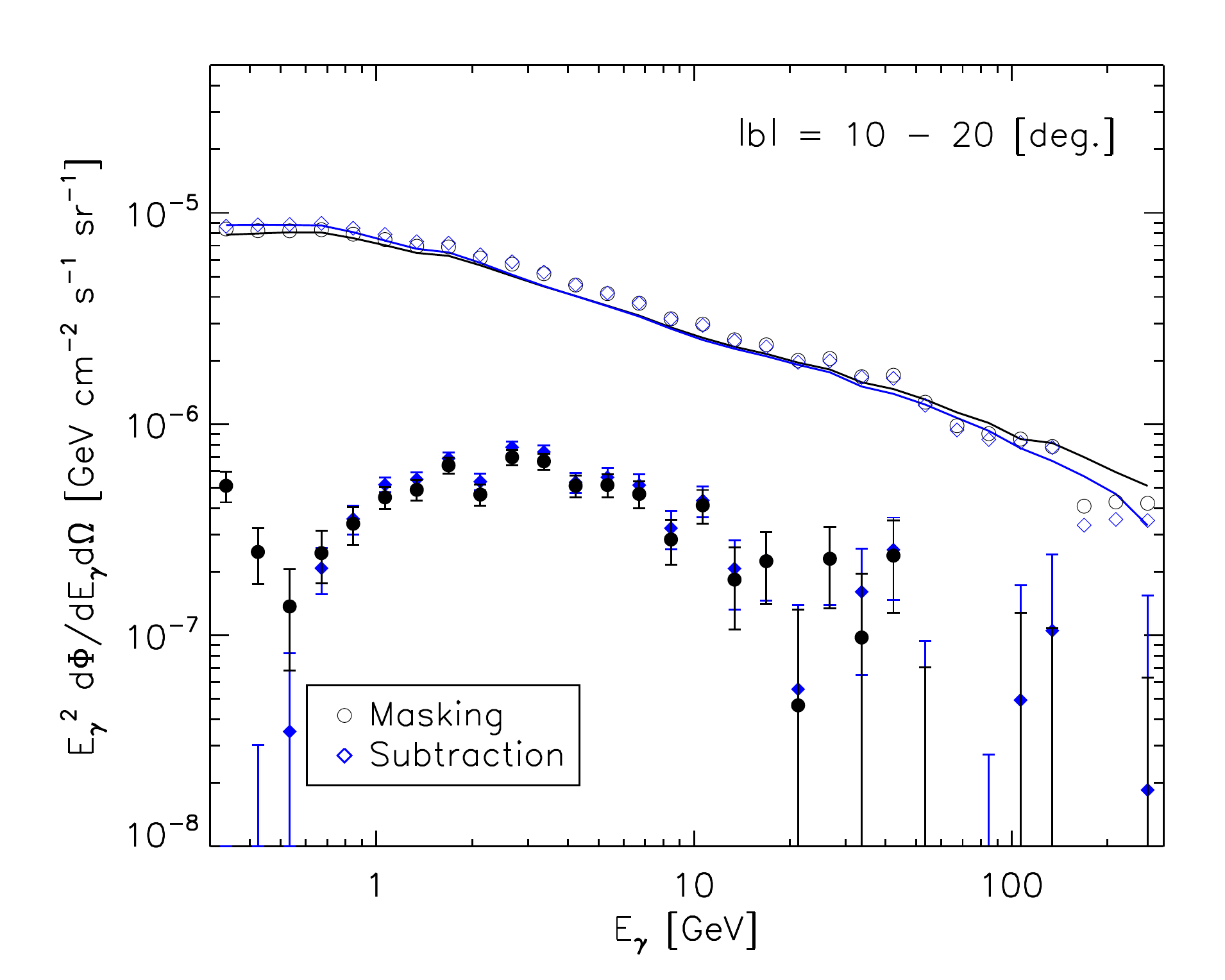}
   %\caption{\textit{Exposure Map}}\label{fig:ExposureMap}
    \end{minipage}\hspace{1.2 cm}
   \begin{minipage}{0.4\textwidth}
    \centering
    \includegraphics[scale=0.42]{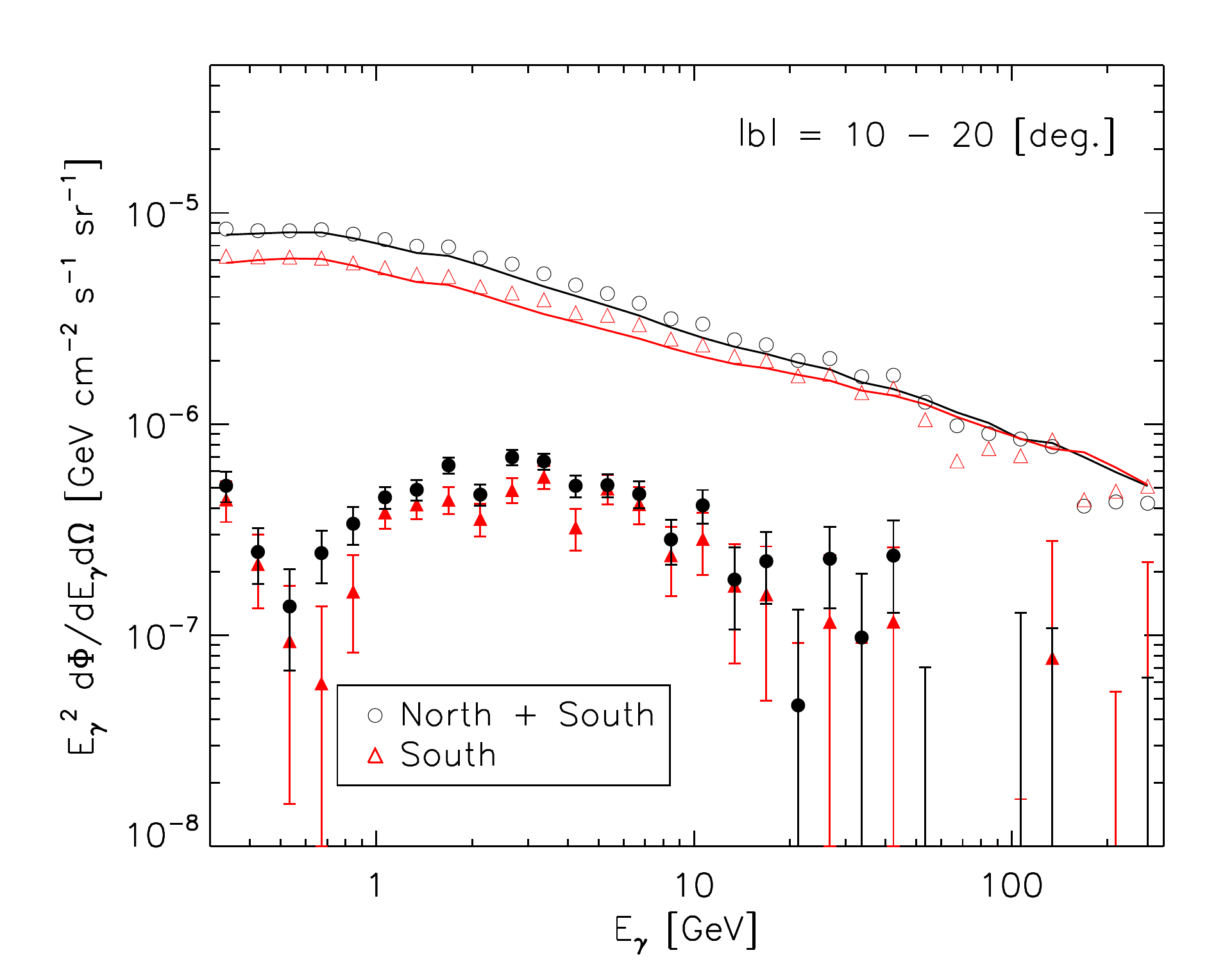}
    \end{minipage}
    \caption{\textit{
     Fermi bubbles energy spectrum in the second slice $|b|=1^{\circ}-10^{\circ}$. Details are given in the caption of Fig.~\ref{fig:MAIN2}.
    }}\label{fig:MAIN3}
\end{figure}
\begin{enumerate}

\item $|b|=1^{\circ} - 10^{\circ}$. The spectrum is flat up to energies $E_{\gamma} \sim 5$ GeV, thereafter it starts to decrease; at energies larger than $10$ GeV the signal is swamped by large statistical uncertainties. We report in this region a mild discrepancy compared to the results in Ref.~\cite{Hooper:2013rwa}, where the energy spectrum at $E_{\gamma}
 < 1$ GeV goes down.

\item $|b| = 10^{\circ}-20^{\circ}$. The spectrum clearly shows a bump peaked around $E_{\gamma}\sim 1-4$ GeV. This spectral features is consistent with Ref.~\cite{Hooper:2013rwa}.

\item $|b| = 20^{\circ}-50^{\circ}$. The spectrum presents a flattish behavior, in agreement with the result in Ref.~\cite{Hooper:2013rwa}.

\end{enumerate}

%Let us discuss in more detail these results.\\
In order to study the latitude-dependence of the Fermi bubbles energy spectrum in more details, and to check the stability of
our results in Fig.~\ref{fig:MAIN}, we perform the analysis in different setups.
In particular we use a different mask for the inner Galactic disk, namely $|b|<5^{\circ}$, we subtract the point source contribution instead of using the masking method (see Appendix~\ref{App:masking}), and we compare the whole Fermi bubbles region with the part lying in the Southern hemisphere of the sky.
As we shall see, these tests lead to a consistent and stable spectrum in all the slices of the Fermi bubbles.\vspace{0.2cm} \\
\textbf{1})~The first slice of the Fermi bubbles is close to the Galactic center, and has a latitude-dependent flux.
Although the first slice has some uncertainties, such as point source contamination, we obtain a consistent spectrum without North-South asymmetry.

First of all, the spectrum reveals the feature of latitude-dependence by comparing the results obtained using  the $1^\circ$ and the $5^\circ$ inner disk masks. In the upper left panel of Fig.~\ref{fig:MAIN2}, where we focus on the first slice $|b|<10^{\circ}$, there is no difference between the two inner disk masking methods
at $E_{\gamma}<1$ GeV. This is due to the fact that at these energies the point source masking radius is large, and thus almost the whole bubbles region in the interval $1^{\circ}-5^{\circ}$ is masked to remove the point sources. On the contrary at high energies, where the masking radius is significantly smaller, the contribution from $1^{\circ}-5^{\circ}$ is larger. To understand this difference, it is instructive to compare in these two cases the observed gamma-ray flux and the best-fit theoretical prediction from the Galactic diffuse model and the isotropic extragalactic component as done in Fig.~\ref{fig:MAIN2}. Going from $1^{\circ}$ mask to $5^{\circ}$ mask, the observed flux decreases, because a bright fraction of the emission is removed.
The diffuse flux exhibits the same behavior. The relative change, however, is smaller, leading to smaller residual values. This means that the $5^{\circ}$ mask for the inner Galactic disk not only removes the diffuse emission but also removes some extra contribution possibly related to Galactic center contamination or smoothing effects. The latitude-dependence is much clear using the point source subtraction method, as done in the upper right panel of Fig.~\ref{fig:MAIN2}. Further evidence in favor of this argument comes from the analysis of the energy spectrum as a function of the latitude. We present the latter in Fig.~\ref{fig:LAT}, where the increase of the energy spectrum at low latitudes, $|b|< 10^{\circ}$, is particularly evident.

Secondly, the subtraction and the masking methods are consistent. Let us remind that these two methods involve a different approach to remove the point source contribution, i.e. we subtract the point sources instead of masking them.
Some differences between masking and subtraction may arise at low energy, where the masking with large radius cover a considerable fraction of the analyzed area, and in low latitudes nearby the Galactic center where the concentration of point sources is high. We show the comparison
between point source subtraction and masking in the bottom left panel of Fig~\ref{fig:MAIN2}. As expected the largest discrepancy arises at low energy in the first slice, comparing the observed gamma-ray flux and
the best-fit prediction obtained from the combination of the diffuse model and the isotropic component. In particular they increase going from the masking to the subtraction method. This happens because in a masked region we are forced to remove not only the contribution corresponding to the point source but also the underlying diffuse emission. Using the subtraction, on the contrary, we include the latter in the
analysis.\footnote{To be more precise, using the masking method the Fermi bubbles region in the interval $|b|=1^{\circ}-5^{\circ}$ is masked at low energies because of the large point source contribution. Therefore, with this method we basically analyze  in the first slice only the region $|b|=5^{\circ}-10^{\circ}$. Using the subtraction method, on the contrary, the whole region $|b|= 1^{\circ}-10^{\circ}$ is analyzed which of course has larger averaged flux since it  involves the diffuse flux at low latitudes. The same argument can explain the opposite behavior at high energies.}
Notice, however, that this extra contributions cancel out in the computation of the residual values, leading to consistent results. Only in the first few bins the subtraction method gives
a smaller residual flux. Different aspects conspire to produce this distinctive feature.
The masking method, for instance, is characterized by a
residual point source contamination (see Appendix~\ref{App:maskingdetails}) that may become relevant at low energy (where the point sources are brighter) and low latitudes (where the point sources are larger in number). The flux itself, on the other hand, is strongly latitude-dependent thus being particularly sensitive to the masked region that in the first slice is concentrated at low latitudes. Finally, also small differences in the best-fit coefficients can affect the computation of the residual values.

Thirdly, the Fermi bubbles energy spectrum is North-South symmetric.
To reach this conclusion we start comparing the whole spectrum with the one obtained analyzing only the Southern hemisphere. We show this result in the bottom right panel of Fig.~\ref{fig:MAIN2}, where we report a difference between the two spectra. This is because the first slice of the Southern bubble covers only the latitude range from $5^\circ$ to $10^\circ$ rather than from $1^\circ$. The spectrum from the Southern bubble, as a consequence, has to be similar to the one obtained from the whole bubbles region but using the $5^\circ$ inner disk mask. This behavior can be observed comparing the Southern spectrum with the result of the $5^\circ$ disk analysis,  previously discussed with reference to the upper right panel of Fig.~\ref{fig:MAIN2}. A more clear evidence in favor of the North-South symmetry comes from the analysis of the energy spectrum as a function of the latitude in Fig.~\ref{fig:LAT}, where we report a good agreement in the comparison between the Southern hemisphere and the whole Fermi bubbles region.

Finally, as a caveat, all the uncertainties in the residual spectrum are statistical. The first slice of Fermi bubbles suffers from more systematic uncertainties than other slices.
These uncertainties result  from the fact that it is difficult, because of the finite resolution of the LAT, to distinguish between point sources and diffuse emission. Moreover, considering the diffuse emission model, cosmic rays propagation is not well known, and its uncertainties involve spectra
injection, transport parameters, magnetic fields and halo size. Finally, the interstellar radiation field and the gas distribution, crucial for the computation of ICS and Bremsstrahlung, suffer from large uncertainties near the Galactic center. As a consequence we will exclude the slice $|b|=1^{\circ}-10^{\circ}$ from the fit that will be performed in the next Section.\vspace{0.2cm} \\
\textbf{2})~The second slice of the Fermi bubbles, $|b| = 10^{\circ}-20^{\circ}$,
reveals a bump around $E_{\gamma}\sim 1-4$ GeV. Taking into account the statistical errors, this feature remains stable using different inner disk masks, opting for the point source subtraction method, and restricting the analysis to the Southern hemisphere. These results are shown in Fig.~\ref{fig:MAIN3}. As expected, the inner disk mask has little impact on the energy spectrum of the Fermi bubbles at high latitudes. Away from the Galactic center, the number of point sources significantly decreases; even at low energy, therefore, point source subtraction and masking agree very well. The only significant difference arises in the first few bins, as already pointed out and explained analyzing the spectrum in the first slice. In the bottom right panel of Fig.~\ref{fig:MAIN3} we report a good agreement in the comparison between the Southern hemisphere and the whole Fermi bubbles region; in particular the bump feature at $E_{\gamma}\sim 1-4$ GeV is still present in the energy spectrum. In particular we notice that the asymmetry in the observed gamma-ray flux is exactly counterbalanced by the diffuse emission, thus leading to consistent residual values.\vspace{0.2cm} \\
\textbf{3})~At higher latitudes, $|b| = 20^{\circ}-50^{\circ}$, the energy spectrum is almost flat. Moreover it remains stable if compared with the Southern hemisphere or with the results obtained using different disk masks and point source subtraction (see Appendix~\ref{App:B}). As mentioned in the Introduction, this result points towards the possibility that at these latitude the most prominent component of the Fermi bubbles spectrum comes from the existence of an extra population of electrons producing ICS photons. We will explore this hypothesis in the next Section.\vspace{0.2cm} \\
In conclusion, as pointed out in Ref.~\cite{Hooper:2013rwa}, two different components seem to emerge from the qualitative analysis of the Fermi bubbles spectrum.
The first component dominates at low latitudes, especially for $|b|=10^{\circ}-20^{\circ}$, producing a bump in the spectral shape at $E_{\gamma}\sim 1-4$ GeV. The second component, on the contrary, is responsible for the flat spectrum at higher latitudes. In the next Section we will verify the DM explanation for the bump feature together with the ICS photons for the flat spectrum.

%It is difficult, for instance, to disuncertaintiestinguish between point sources and diffuse emission.

%%%%%%%%%%%%%%%%%%%%%%%%%%%%%%%%%%%%%
\section{Fermi bubbles spectrum from Inverse Compton Scattering and Dark Matter}
\label{sec:ICS}
%%%%%%%%%%%%%%%%%%%%%%%%%%%%%%%%%%%%%

In this Section we fit the energy spectrum of the Fermi bubbles combining the gamma-ray photons
produced by an additional population of electrons via ICS on the ambient light, and the photons produced by DM annihilation via FSR. In Section~\ref{sec:ICStheory} we briefly review the ICS formalism. In Section~\ref{sec:Fermibubblesfit} we outline our fitting strategy and discuss our results.

%%%%%%%%%%%%%%%%%%%%%%%%%%%%%%%%%%%%%
\subsection{Gamma rays from Inverse Compton Scattering}\label{sec:ICStheory}
%%%%%%%%%%%%%%%%%%%%%%%%%%%%%%%%%%%%%

Given the energy spectrum and the density distribution of an electron population,
the differential photon flux produced by ICS on the photons of the InterStellar Radiation Field (ISFR, including CMB, infrared, and starlight), and detected on Earth within an angular region $d\Omega$ and energy $E_\g$  is~\cite{Cirelli:2009vg}\footnote{We follow Ref.~\cite{Cirelli:2009vg}, and refer readers to Ref.~\cite{Cirelli:2009vg,Cirelli:2010xx} and references therein for more details.}
 \be\label{eq:ICSFlux}
 \frac{d\Phi}{dE_\g d\Omega} =  \frac{1}{E_\g}  \int_{\rm{l.o.s.}} ds~\frac{j[E_\g, r(s)]}{4\pi}~,
  \ee
where $r$ is the distance between an emission cell, at which ICS photons are produced by electrons colliding with ISRF, and the Galactic center, $s$ is the distance between the Earth and the emission cell, $1/4\pi$ results from the isotropy of the ICS photon emission, and $j[E_\g, r(s)]$ is defined as
\be
j[E_\g, r(s)] =  \int^{E_{\rm cut}}_{m_e}  dE_e\,\, \mathcal{P}(E_\g,E_e,r)  \,\,  n_e(r,E_e)~.
\label{eq:power emission}
\ee
In Eq.~(\ref{eq:power emission}) $E_e$ is the initial energy of an electron before scattering on ISRF, $n_e(r,E_e)$, in units of cm$^{-3}$ GeV$^{-1}$, is the electron number density per unit energy at location $r$ with energy $E_e$,  and $\mathcal{P}(E_\g,E_e,r)$, in units of s$^{-1}$, is the differential power emitted into photons of energy $E_\g$ from electrons of energy $E_e$. The integration range is from the electron mass, $m_e$, to the highest energy of electrons, $E_{\rm cut}$.

Electrons move around the Galactic diffusion zone and lose energy via synchrotron radiation, Bremsstrahlung, ionization and ICS, and the energy loss is governed by the cosmic ray propagation equation~\cite{Strong:2007nh,Regis:2008ij}. Therefore, $n_e$ in Eq.~(\ref{eq:power emission}) should be the convoluted number density function and is different from the original injection spectrum. It is, however, reasonable to assume a power-law spectrum for $n_e$, regardless of the details of the propagation and associated
uncertainties, i.e., $n_{e}(r,E_e) \propto E^{\gamma}$, where the spectral shape $\gamma$ and the
 normalization will be determined by best-fits to the Fermi bubbles, as we will discuss in the next Section.

\begin{comment}
For the electron number density per unit energy we assume a simple power-law
$n_{e}(r,E_e) \propto E^{\gamma}$, where the spectral shape $\gamma$ and the
 normalization will be determined by best-fits to the Fermi bubbles energy spectrum, as we will discuss in the next Section.
In principle, electrons move around the Galactic diffusion zone and lose energy via synchrotron radiation, Bremsstrahlung, ionization and ICS, and the energy loss is governed by the cosmic ray propagation equation~\cite{Strong:2007nh,Regis:2008ij}. Therefore, $n_e$ in Eq.~(\ref{eq:power emission}) should be the convoluted number density function and is different from the original injection spectrum. However it is reasonable to assume that the power-law behavior remains
untouched, regardless of the details of the propagation and its
uncertainties. Given that normalization and spectral index are free parameters, therefore, this simply means that in our analysis we consider the electrons \textit{after} the propagation.
\end{comment}

The detailed derivation of $\mathcal{P}(E_\g,E_e,r)$ can be found in Ref.~\cite{Cirelli:2009vg}, and we outline here only the main points. Given an electron of energy $E_e$ and ISRF at location $r$, the energy loss rate of the electron into a photon of energy $E_\g$, in units of s$^{-1}$, is proportional to $\int dE_{\g^{\prime}} \,\, (E_{\g} - E^{\prime}_{\g} ) n_\g(E^{\prime}_{\g},r) \,\,  \frac{d\s}{dE_{\g}}(E_e, E_\g^{\prime}, E_\g) \,\, |v_e-v_\g|  $, where $E_\g$ $(E_{\g^{\prime}})$ is the photon energy after (before) ICS, $(E_{\g} - E_{\g^{\prime}} ) \simeq E_\g$ for ICS photons of interest, $|v_e-v_\g|$ is the initial electron-photon relative velocity, $n_{\g}(E_\g^{\prime},r)$ is the number density of photons of $E^{\prime}_{\g}$ in units of cm$^{-3}$ GeV$^{-1}$, and $\frac{d\s}{dE_{\g}}(E_e, E_\g^{\prime}, E_\g)$ is the differential Compton cross section with energy denoted by arguments for incoming and outgoing electron and photon. We then boost the system into the rest frame of the initial electron where the Compton cross section is in a simple form. Finally, we obtain
\bea
&&\mathcal{P}(E_\g,E_e,r) = \notag\\
&& \frac{3\s_T}{4\g^2} E_\g \int^1_{1/4\g^2} dq \left[ 1- \frac{1}{4 q \g^2 (1-\tilde{E_\g} ) }  \right]  \frac{n_\g(E^{\prime}_{\g},r)}{q} \left(  2q \log q + q +1 - 2q^2 + \frac{1-q}{2} \frac{ \tilde{E_\g}^2 }{1-\tilde{E_\g}}  \right)~,\notag\\
&&
\label{eq: power emission P}
\eea
where $\s_T=0.6652$ barn, the total Thomson cross section, $\g=E_e/m_e$, $ \tilde{E_\g}= E_\g / (\g m_e) $, $q=  \tilde{E_\g}/ \left[  \Gamma_{\g^{\prime}} (1-\tilde{E_\g})  \right]$, and $\Gamma_{\g^{\prime}} =4 E^{\prime}_\g \g / m_e $.

%%%%%%%%%%%%%%%%%%%%%%%%%%%%%%%%%%%%%
\subsection{Chi-square analysis and fitting results}\label{sec:Fermibubblesfit}
%%%%%%%%%%%%%%%%%%%%%%%%%%%%%%%%%%%%%

The flat behavior of the Fermi bubbles energy spectrum at high latitudes can be reproduced by means of ICS photons generated by an additional population of electrons with a power-law energy spectrum. On the qualitative level, in light of the results shown in Fig.~\ref{fig:MAIN}, this is certainly true in particular in the region $|b|=20^{\circ}-50^{\circ}$. The bump eminent especially at the slice $|b|=10^{\circ}-20^{\circ}$, however, suggests the existence of an extra component at low latitudes. In the following we will identify the latter with the FSR from DM annihilation into the Standard Model (SM) fermions. Combining FSR and ICS through a chi-square analysis, we will test the possibility  to realize the whole Fermi bubbles spectrum.

This approach has been already investigated in Ref.~\cite{Hooper:2013rwa}, where the spectrum was found to be compatible with a $\mathcal{O}(10)$ GeV DM particle annihilating into leptons or quarks, with a thermal averaged cross section $\langle \sigma v\rangle \sim 10^{-27}~{\rm cm}^{3}\,{\rm s}^{-1}$. In the rest of this Section we will explain our method and present our results.

For simplicity, we focus on DM annihilation into $b$-quarks only but the procedure is actually independent on the final state. Moreover, we fit the data from all the Fermi bubbles slices but the first one, $|b|=1^{\circ}-10^{\circ}$, because of the large astrophysical uncertainties mentioned in Section~\ref{Sec:ResultsAndComments}. We perform a $\chi^2$ analysis, and the procedure goes as follows.

First, we fit the data considering both ICS photons and FSR from DM annihilation. The former is given by Eq.~(\ref{eq:ICSFlux}) as previously discussed, while the latter follows from Eq.~(\ref{eq:FSR}). We use the generalized NFW profile in Eq.~(\ref{eq:genNFW}). We keep the ICS spectral shape universal within the region of the whole Fermi bubbles, where we assume a power-law spectrum with a cut-off energy, $E_{\rm cut}$, at 1.2 TeV. We vary the individual normalization of the electron density in each slice. The DM mass $M_{\rm DM}$ and the annihilation cross section $\langle \s v \rangle$ are fitting parameters as well. To sum, we use 7 parameters to fit the data.

%For the computation of the residual values, we neglect the masked points.
\begin{figure}[!htb!]
 \centering
  \begin{minipage}{0.4\textwidth}
   \centering
   \includegraphics[scale=0.42]{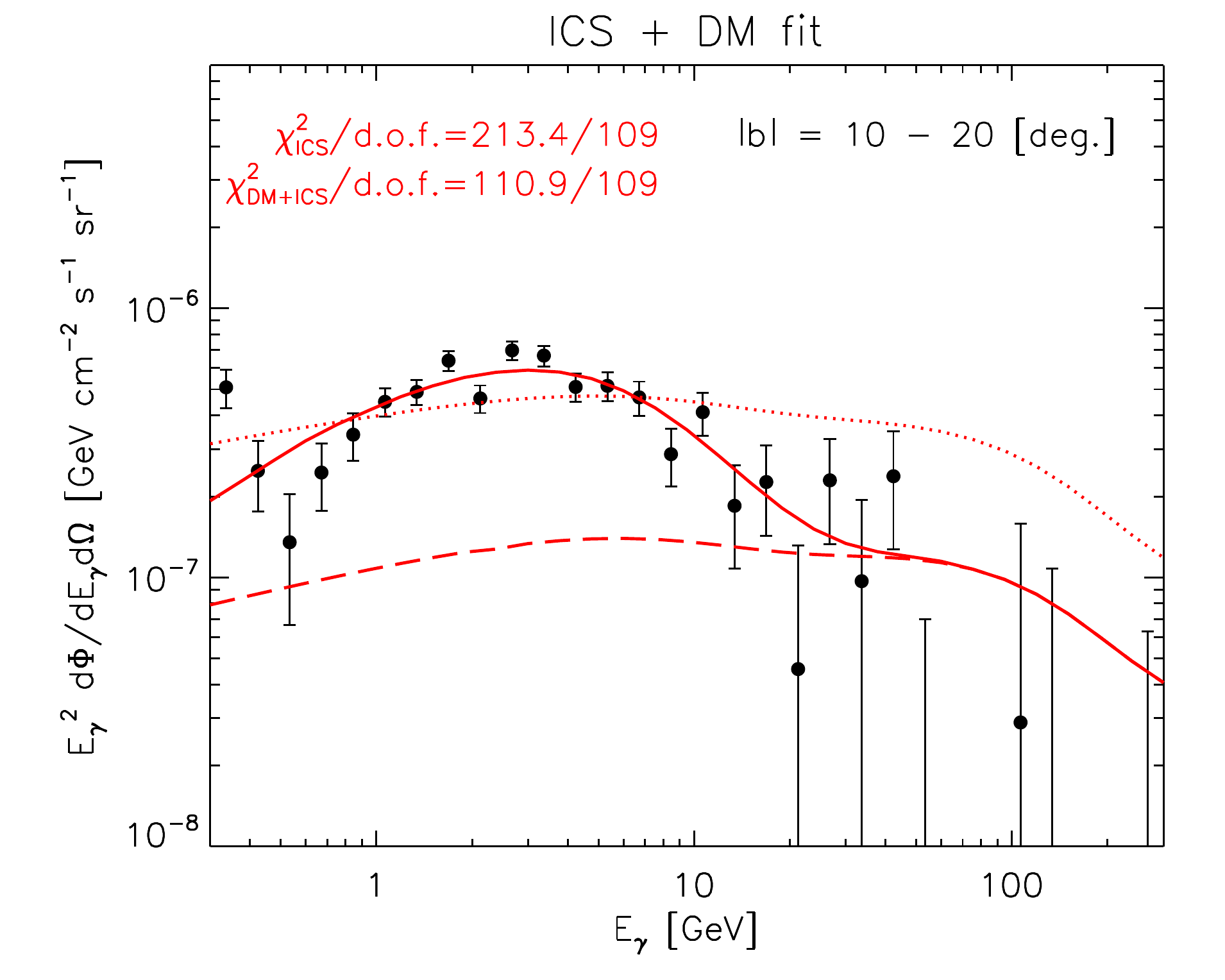}
   %\caption{\textit{Count Map}}\label{fig:CountMap}
    \end{minipage}\hspace{1.2 cm}
   \begin{minipage}{0.4\textwidth}
    \centering
    \includegraphics[scale=0.42]{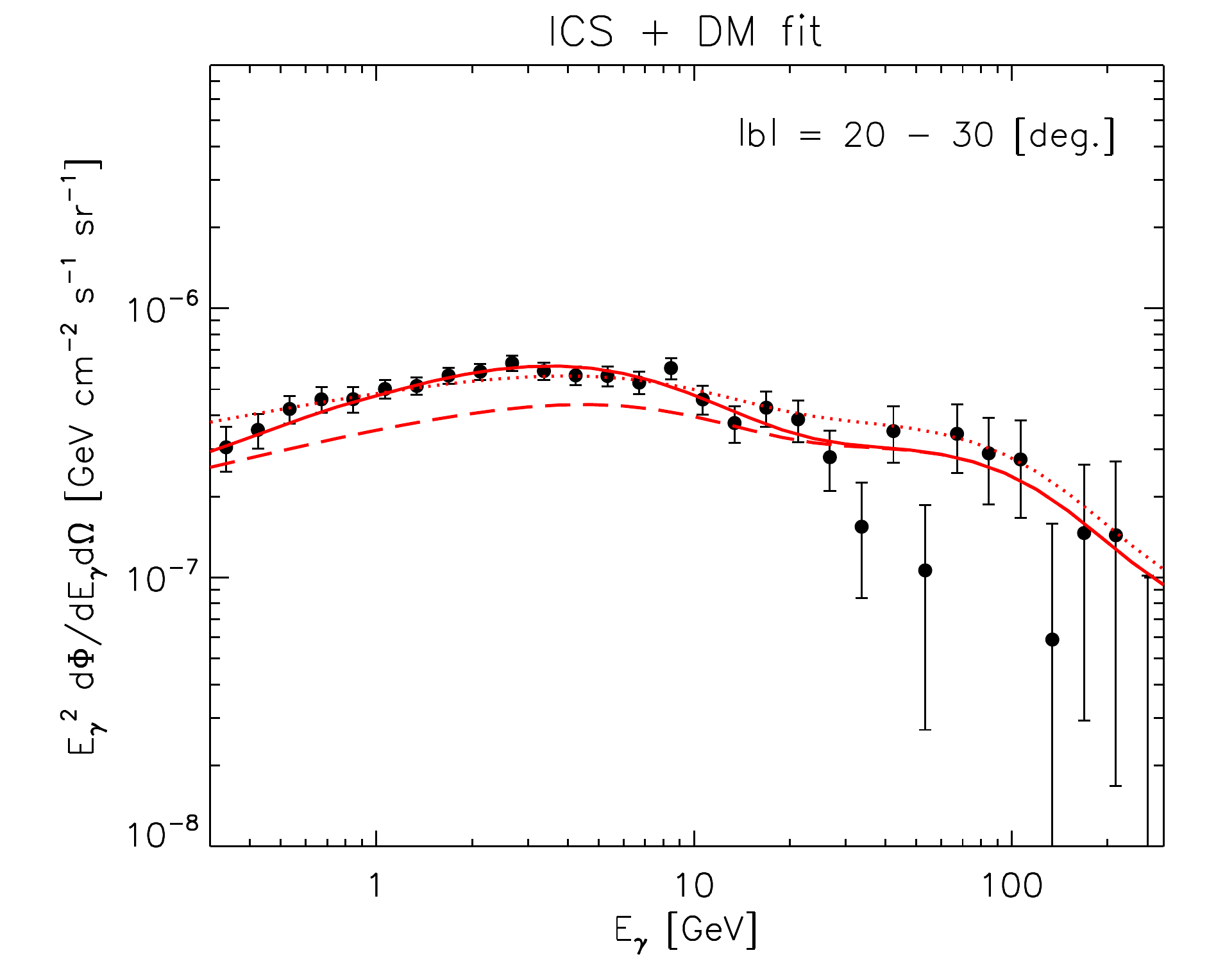}
    \end{minipage}\\
    \vspace{0.5 cm}
   \begin{minipage}{0.4\textwidth}
    \centering
   \includegraphics[scale=0.42]{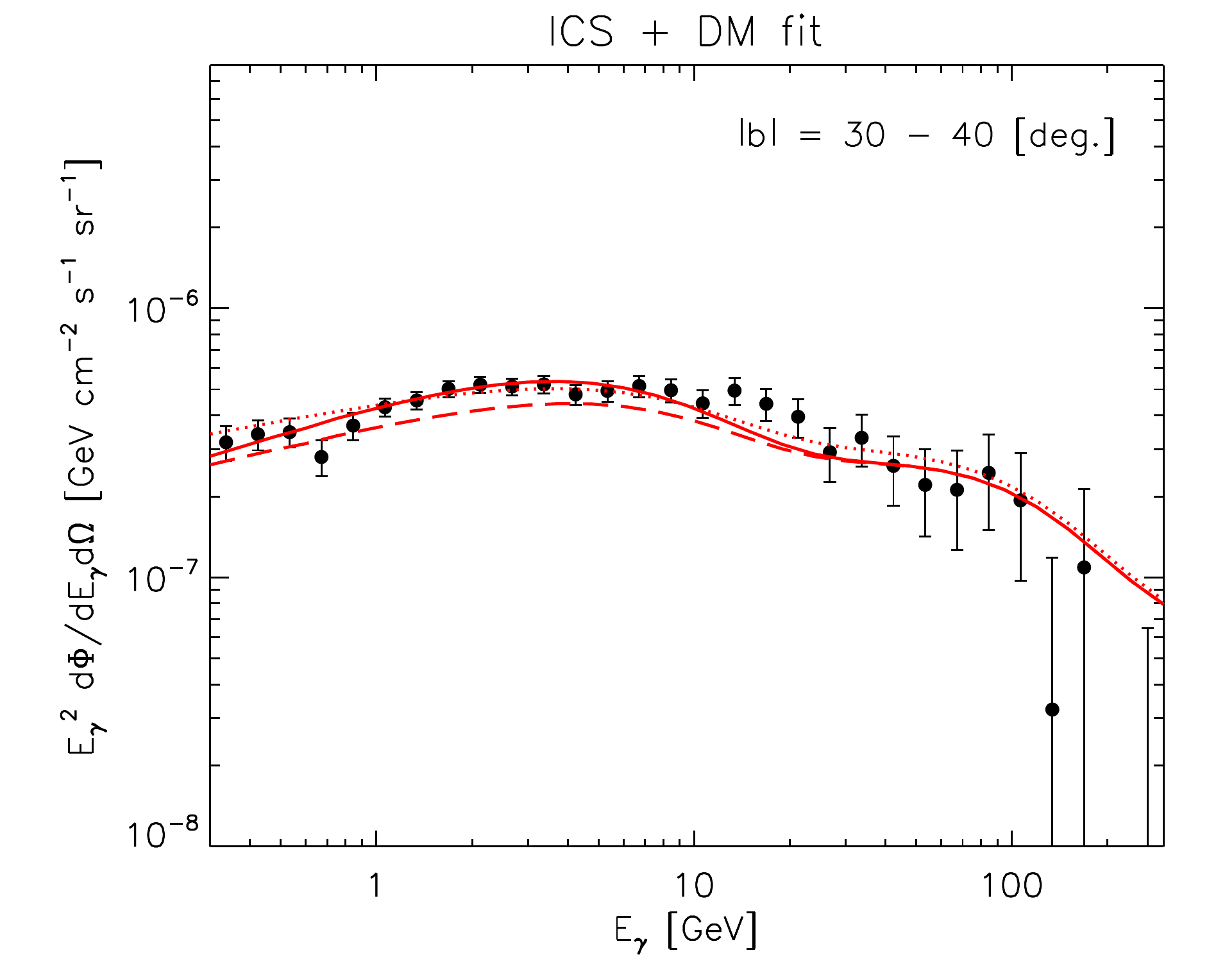}
   %\caption{\textit{Exposure Map}}\label{fig:ExposureMap}
    \end{minipage}\hspace{1.2 cm}
   \begin{minipage}{0.4\textwidth}
    \centering
    \includegraphics[scale=0.42]{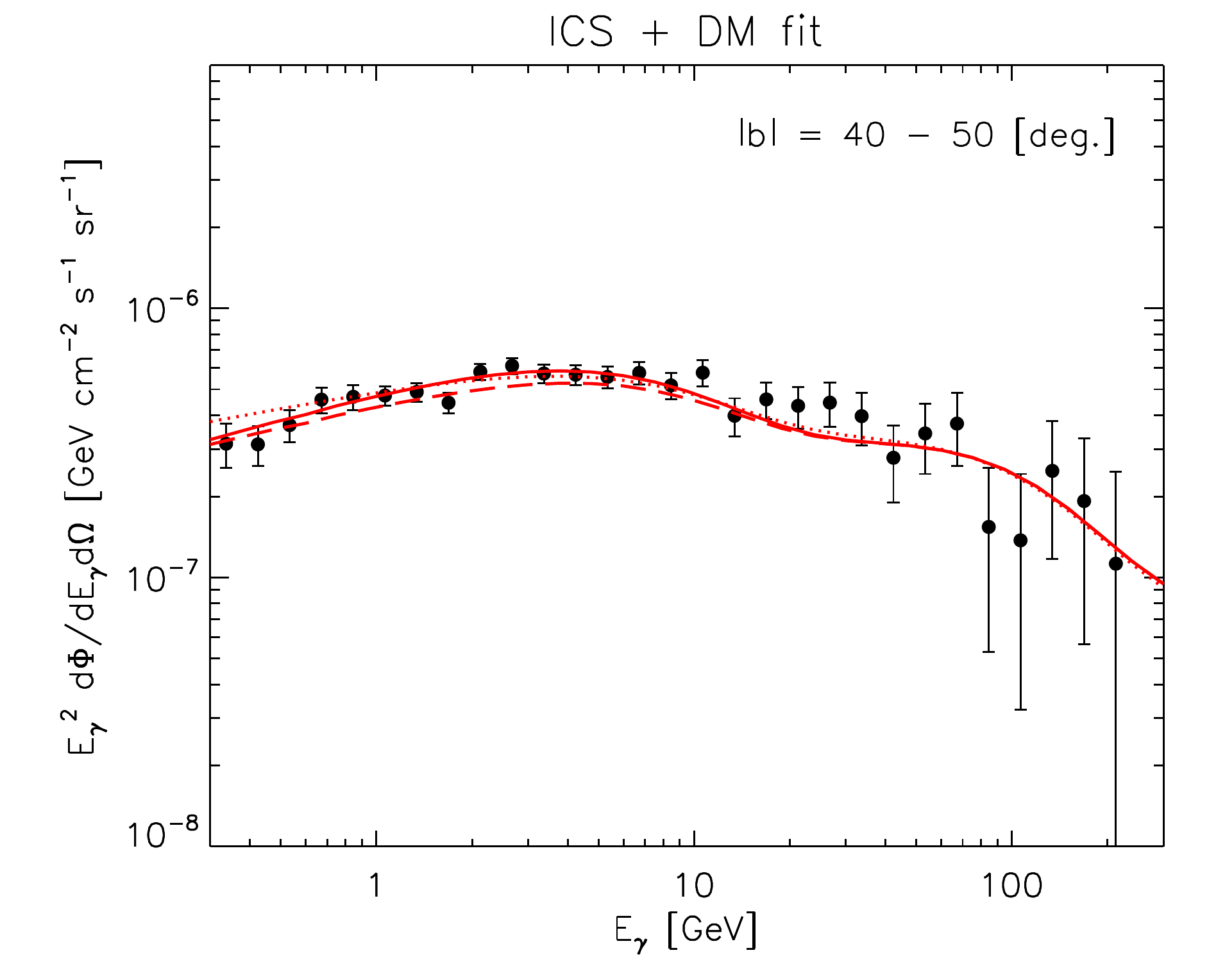}
    \end{minipage}
    \caption{\textit{
  Analysis of the energy spectrum of the Fermi bubbles in the four slices from $|b|=10^{\circ}-20^{\circ}$ to $|b|=40^{\circ}-50^{\circ}$. The solid line represents the best-fit result obtained combining ICS and FSR from DM annihilation into $b\overline{b}$. The dashed line retraces the ICS component, highlighting the role of the DM contribution in particular in the first slice, $|b|=10^{\circ}-20^{\circ}$, where a bump at $E_{\gamma}\sim 1-4$ GeV clearly arises.
  We also show the best-fit result obtained considering only ICS without DM (dotted line).
    }}\label{fig:1050ICSDM}
\end{figure}

Second, we repeat the same procedure but considering only the ICS photons. By comparing the results of the two $\chi^2$ analysis, we shall see that in the second case the fit is much worse, thus confirming the the reliability of our assumptions.

In Fig.~\ref{fig:1050ICSDM} we show the fitting results for each slice. We find $\chi_{\rm min}^2/{\rm d.o.f.} = 110.9/109$ for the combination of DM and ICS, and $ \chi_{\rm min}^2/{\rm d.o.f.} = 213.4/111$ for ICS only.\footnote{In addition, considering $|b| = 10^\circ - 30^\circ$ where the DM contribution is relatively important, we have $\chi_{\rm min}^2/{\rm d.o.f.} = 64.1/47$ for DM plus ICS, and $ \chi_{\rm min}^2/{\rm d.o.f.} = 154.8/49$ for ICS only. Furthermore, we have $\chi_{\rm min}^2/{\rm d.o.f.} = 49.5/46$ $(154.8/48)$ for DM-ICS (ICS), if we exclude the first energy bin which is subject to large contamination because of the poor angular resolution of the LAT at low energy.} It is therefore very clear that the combination of ICS and DM can account for the whole energy spectrum of the Fermi bubbles much better than ICS only. In particular at high latitudes, where the DM contribution is small, the ICS component is dominant and can fit the flattish spectrum of the Fermi bubbles. At low latitudes, especially for $|b| =10^{\circ} - 20^{\circ}$, ICS can not reproduce the bump at $E_{\gamma}\sim 1 - 4$ GeV. Notice, moreover, that our best-fit value for the spectral index of the power-law describing the spectrum of the electron population generating ICS photons is $\gamma = -2.39$. This number is in agreement with the typical values able to explain the WMAP haze observed in the microwave  \cite{Finkbeiner:2003im, Dobler:2007wv}.

\begin{figure}[!htb!]
 \centering
  \begin{minipage}{0.4\textwidth}
   \centering
   \includegraphics[scale=0.55]{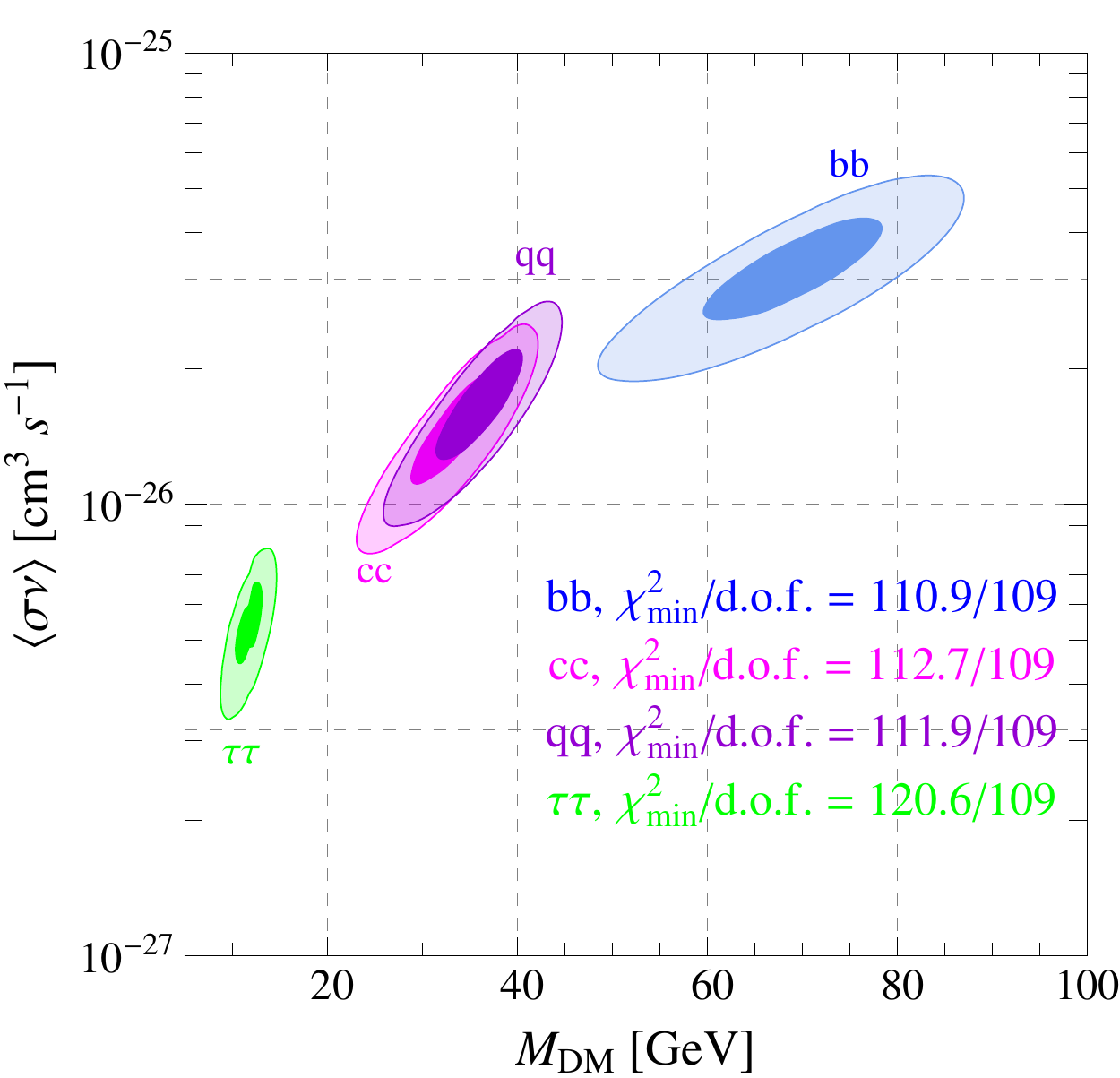}
    \end{minipage}\hspace{1 cm}
   \begin{minipage}{0.4\textwidth}
    \centering
    \includegraphics[scale=0.55]{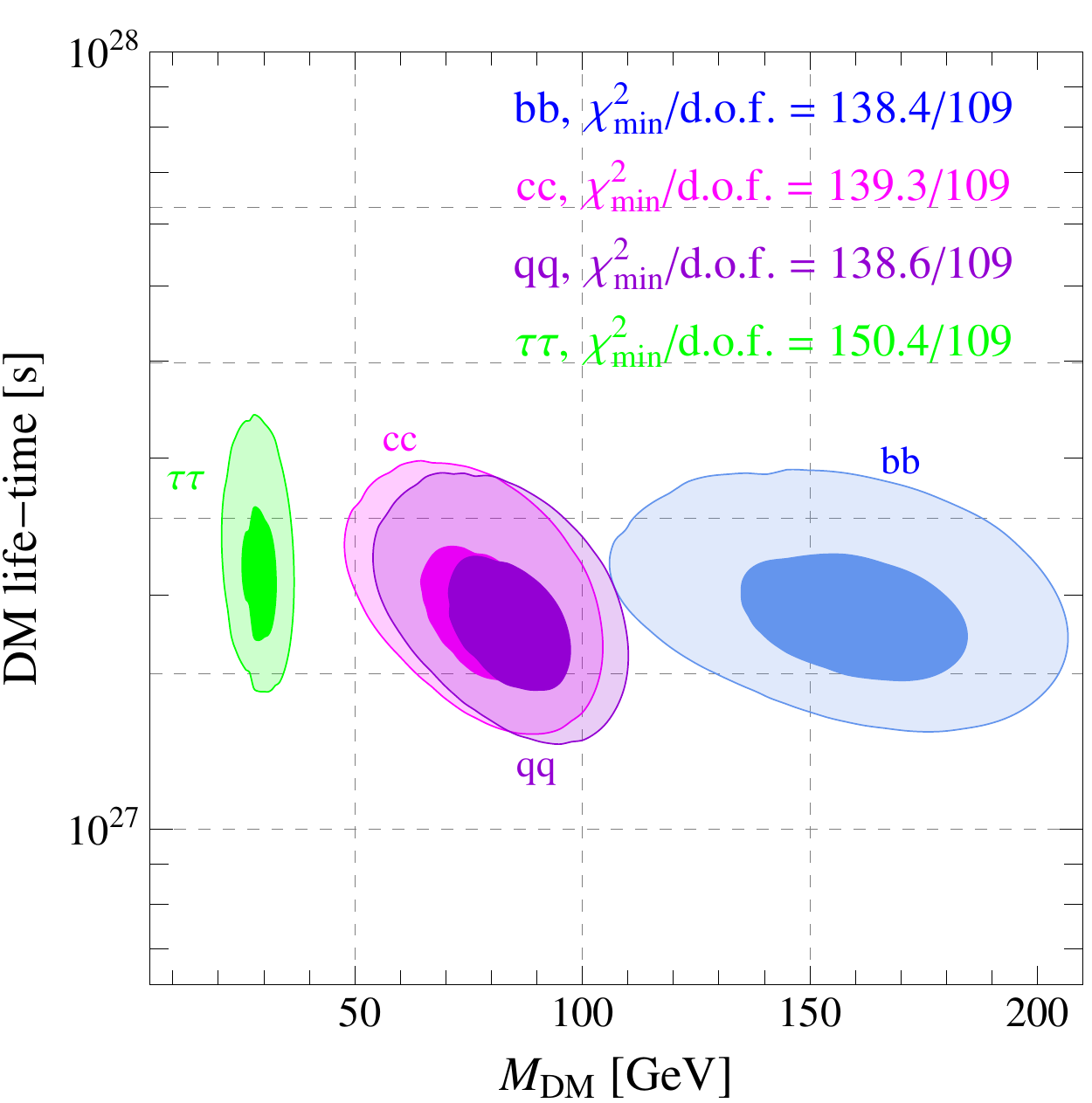}
    \end{minipage}\\
 \caption{\emph{
 Confidence regions (99\%  C.L. and 68\% C.L.) for the annihilating (left panel) and decaying (right panel) DM component in the analysis of the Fermi bubbles spectrum (see text for details).
 }}
 \label{fig:contour_AN_DEC}
\end{figure}

Generalizing the procedure described above, we study the interplay between ICS and FSR considering different final state. In Fig.~\ref{fig:contour_AN_DEC} we focus on the DM component, showing the 65\% and 99\% confidence regions for annihilating DM (left panel) and decaying DM (right panel). We perform a two-dimensional fit in the plane $(M_{\rm DM},\langle \sigma v\rangle)$,
marginalizing over the remaining parameters. Final states involving $\tau^{+}\tau^{-}$ have a harder FSR photon spectrum and in turn prefer a lower DM mass and smaller annihilation cross section for the annihilation DM and a smaller decay width for the decaying DM. The $\chi^2$s are similar among different final states.  Besides, by virtue of the feature of concentration of the gamma ray excess toward the Galactic center, the annihilation DM is by far preferred over the decaying DM; for example, in terms of the $b$-quark final states, $\chi_{\rm min}^2/{\rm d.o.f.}=110.9/109$ for annihilation but $\chi_{\rm min}^2/{\rm d.o.f.}=138.4/109$  for decay.

In Table~\ref{tab:DMbestfit} we summarize the best-fit values for $M_{\rm DM}$ and $\langle \sigma v\rangle$, together with the corresponding 1-$\sigma$ errors, considering DM annihilation into $b\overline{b}$, $c\overline{c}$, $q\overline{q}$ and $\tau^+\tau^-$. Our best-fit candidate corresponds to annihilation into $b\overline{b}$ with mass $M_{\rm DM}= 61.8^{+6.9}_{-4.9}$ GeV and cross section $\langle \sigma v\rangle =  3.30^{+0.69}_{-0.49}\times  10^{-26}~{\rm cm}^{3}\,{\rm s}^{-1}$.
Other final states, e.g. annihilation into $W^{+}W^{-}$, $e^+e^-$, $\mu^+\mu^-$, give worse results.

Before proceeding, it is important to address the following question. One may wonder if the bump characterizing the residual spectrum in the region \emph{within} the Fermi bubbles 
 at latitudes $|b|=10^{\circ}-20^{\circ}$ can be observed, in the same slice, also in
  the complementary region \emph{outside} the Fermi bubbles
   (but inside the rectangular mask, see Fig.~\ref{fig:BubbleTemplate}). 
   We have analyzed the complementary region, 
   comparing the observed energy spectrum, obtained after subtraction of the diffuse model, 
   with the gamma-ray flux produced by the annihilation of DM into $b\bar{b}$ 
   as predicted by our best fit candidate. We have found that the 
gamma-ray flux from DM annihilation
 never exceeds the residual flux of the bubbles 
 complement. 
However, for the same reason, the larger value of the residual flux (in particular at low energy, possibly related to a leakage of
 bright emission from the edges of the bubbles) disfavors the possibility to highlight in the complementary region the presence of the observed GeV bump.

Let us close this Section with a discussion of the DM profile dependence. Throughout this analysis, in fact, we made use of the gNFW profile as a benchmark model for the DM density distribution. It is interesting to notice that our results do not show a strong dependence on this choice. This happens because different profiles are actually  similar at latitudes $|b|>10^{\circ}$ (see Fig.~\ref{fig:Jfactor}), thus leading to mild quantitative differences. To be more precise, the best-fit candidate for the NFW profile is for annihilation into $b\overline{b}$ with $M_{\rm DM}=61.8$ GeV, $\langle \sigma v\rangle = 4.7 \times  10^{-26}~{\rm cm}^{3}\,{\rm s}^{-1}$ and $\chi_{\rm min}^2/{\rm d.o.f.} = 115.4/109$. The larger $\langle \sigma v\rangle$ for NFW results from a smaller $J$ factor for $|b|=10^{\circ}-20^{\circ}$, which is the most important region in terms of the DM component.

\begin{table}[!htb!]
\caption{\textit{DM contribution to the fit of the Fermi bubbles  energy spectrum. In correspondence of each channel we show
the best-fit values for mass and cross section together with the  corresponding 1-$\sigma$ errors and the ratio $\chi_{\rm min}^2/{\rm d.o.f.}$.}}
\begin{center}
\begin{tabular}{||c|c|c|c||}
\hline
\textbf{DM annihilation} & $M_{\rm DM}~~[GeV]$  & $\langle \sigma  v\rangle~~[cm^{3}s^{-1}]$ & $\chi_{\rm min}^2/{\rm d.o.f.}$ \\
\hline
$b\overline{b}$ & $61.8^{+6.9}_{-4.9}$  & $3.30^{+0.69}_{-0.49}\times  10^{-26}$ & $110.9/109$  \\
\hline
$c\overline{c}$ & $29.3^{+2.4}_{-3.4}$ & $1.54^{+0.26}_{-0.30}\times  10^{-26}$ & $112.7/109$  \\
\hline
$q\overline{q}$ & $32.0^{+2.6}_{-3.8}$ & $1.73^{+0.30}_{-0.30}\times  10^{-26}$ & $111.9/109$ \\
\hline
$\tau^{+}\tau^{-}$ & $10.6^{+0.5}_{-0.6}$  &  $5.63^{+0.58}_{-0.64}\times 10^{-27}$ & $120.6/109$ \\
\hline
\end{tabular}
\end{center}
\label{tab:DMbestfit}
\end{table}

%%%%%%%%%%%%%%%%%%%%%%%%%%%%%%%%%%%%%
\section{Dark Matter bounds from the Fermi bubbles}
\label{sec:bounds}
%%%%%%%%%%%%%%%%%%%%%%%%%%%%%%%%%%%%%

In this Section, we are not trying to explain the origin of the residual spectrum of Fermi bubbles, but
we would like to point out that the analysis of the Fermi bubbles energy spectrum also provide a competitive stringent bound on the cross section, especially for values of the DM mass away from the best-fit ones found in the previous Section. In the literature, indirect constraints on the DM thermally averaged annihilation cross section,  $\langle \sigma v\rangle $, have been derived from the LAT observations of the Galactic ridge~\cite{Ackermann:2012rg}, Galactic center~\cite{Hooper:2012sr, Gordon:2013vta}, Dwarf galaxies~\cite{GeringerSameth:2011iw,Ackermann:2011wa,Essig:2009jx},  isotropic diffuse gamma-ray background~\cite{Abdo:2010dk}, and galaxy clusters~\cite{Ackermann:2010rg}. We here derive upper limit on the DM cross section using the latitude-dependent energy spectrum
of the Fermi bubbles obtained in Section~\ref{fermi_s}.  We consider different annihilation channels as well as different DM density profiles,
comparing our results with those obtained from the Galactic center and Dwarf galaxies.

The annihilation of DM particles
can contribute to the gamma-ray flux both through FSR from SM  final states and through
ICS on the low energy background photons in the ISRF.
FSR has smaller uncertainty than ICS photon, since ICS not only relies on the DM density profile and annihilation channels, but also on the diffuse model and the ISRF. For this reason, we only focus on FSR constraints.
% and present the supplementary ICS constraints at the end of this Section.

The photon flux from DM FSR is proportional to the product of the injection spectrum times the squared of DM density integrated over the l.o.s., as shown in Eq.~(\ref{eq:FSR}).
For the local density of DM we take the value $\rho_\odot$ = 0.3 GeV /cm$^{-3}$, instead of $\rho_\odot = 0.4$ GeV /cm$^{-3}$ adopted in Section~\ref{sec:ICS}, in order to make a fair comparison with the results
available in the literature. Besides the gNFW defined in Eq.~(\ref{eq:genNFW}), we consider also the NFW and the isothermal profile. These profiles are written as
\bea
 \rho_{\rm NFW}(r) &\propto& \frac{1}{\left( r/R_s\right) \left[ 1+ \left( r/ R_s \right)\right]^{2} }~ ,\label{eq:NFW} \\
 \rho_{\rm gNFW}(r) &\propto& \frac{1}{\left( r/R_s\right)^\gamma \left[ 1+ \left( r/ R_s \right)\right]^{3-\gamma}} ~ ,\label{eq:gNFW} \\
 %\rho_{Ein} (r) &\propto& \exe the PSF's simple analytical fit to describe the radius of $68\%$ flux containment p \left\lbrace-  \frac{2}{\alpha} \left[ \left( \frac{r}{R_s}\right)^\alpha -1 \right]   \right\rbrace \ , \\
  \rho_{\rm ISO} (r)  &\propto& \frac{1}{1 + \left( r/r_c \right)^2} ~,\label{eq:ISO}
\eea
where for the scale radius $R_s$ we use $R_s = 24.42$ kpc for the NFW and $R_s = 20$ kpc for the gNFW, while we use $r_c = 4.38$ kpc for the isothermal core radius. There are some general comments to be made. In particular, due to the fact that the FSR photon flux is proportional to the square of the DM density through the $J$ factor, it is expected that the constraint  on $\langle \sigma v\rangle$ mainly comes from the slices of the Fermi bubbles closest to the Galactic center exhibiting, as a consequence, a sharp dependence on the slope of the DM profile.

 \begin{figure}[!htb!]
   \centering
       \includegraphics[scale=0.7]{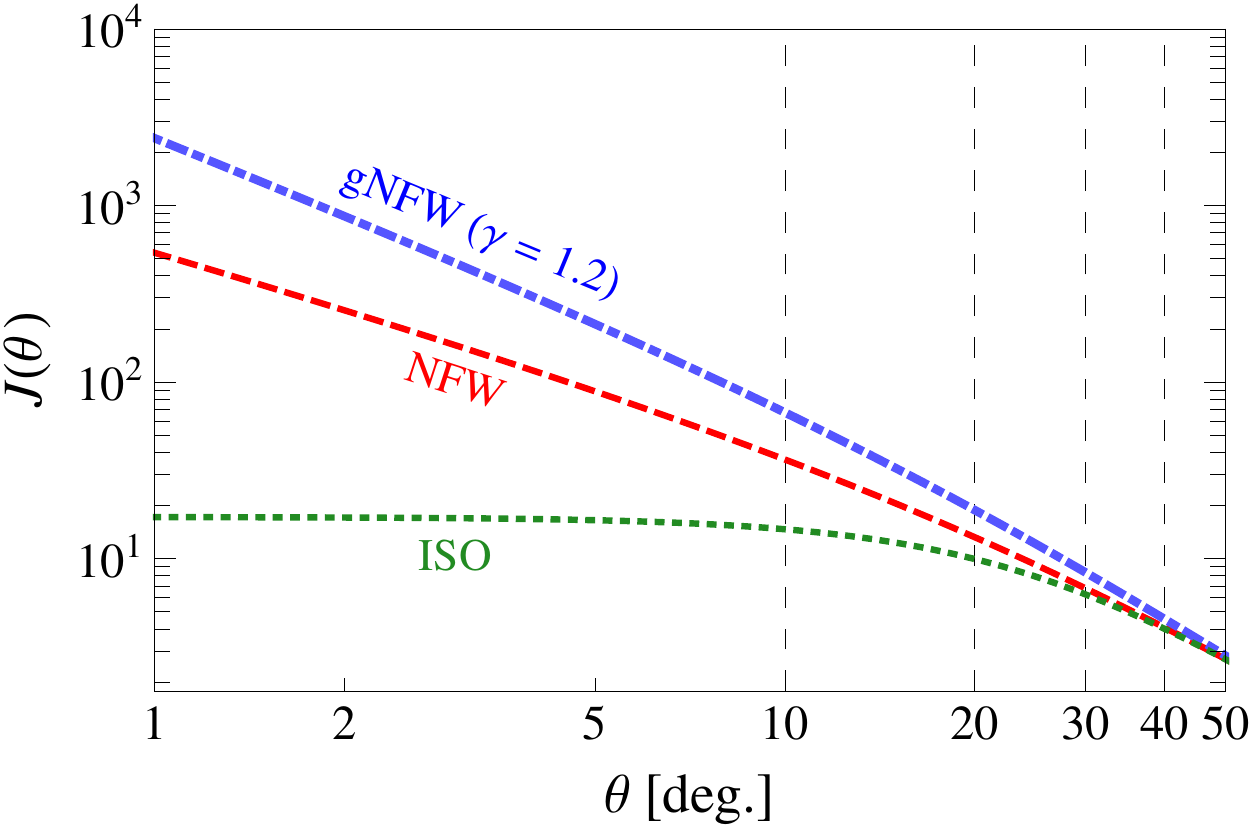}
 \caption{\emph{
 J factor as a function of the angle $\theta$ from the Galactic center for the DM density profiles in Eqs.~(\ref{eq:NFW}-\ref{eq:ISO}).
 For the gNFW we use $\gamma=1.2$. The vertical dashed lines correspond to the latitudes of the
 five slices of the Fermi bubbles as defined in Fig.~\ref{fig:BubbleTemplate}.
  }}
 \label{fig:Jfactor}
\end{figure}

More quantitatively, we show in Fig.~\ref{fig:Jfactor} the $J$ factor, defined in Eq.~(\ref{eq:FSR}), as a function of the angle from the Galactic center for the three DM density profiles analyzed in this Section.
  From this plot it is clear that the gNFW profile will give rise to the most stringent constraint and the isothermal one to the loosest; besides, the constraints based on the  gNFW are dominated by the first slice of the Fermi bubbles but for the isothermal profile, the second slice also plays an important role. The difference between these two extreme
 cases quantifies the uncertainties resulting from different DM profiles.

%\begin{table}
 %\centering
%\begin{tabular}{|c|c|c|c|c|}
%\hline
% J & ~~~~~ NFW~~~~~  & gNFW ( $\gamma = 1.2$ ) & gNFW ($\gamma = 1.4$) & Isothermal \\
%\hline
%\hline
%$1^\circ <|b|< 10 ^\circ$ &  53.3 & 122.0 &  256.9 & 14.9  \\
%\hline
%$5^\circ <|b|< 10 ^\circ$ &  40.0 & 77.8 & 133.8  & 14.6 \\
%\hline
%$10^\circ <|b|< 20 ^\circ$ &  17.1 & 26.5 & 37.2 &  11.0 \\
%\hline
%$20^\circ <|b|< 30 ^\circ$ & 8.1  & 10.5 & 12.6 & 7.1 \\
%\hline
%$30^\circ <|b|< 40 ^\circ$ & 4.9  & 5.6 & 6.2 & 4.7 \\
%\hline
%$40^\circ <|b|< 50 ^\circ$ &  3.3 &  3.6 & 3.8  & 3.3 \\
%\hline
%\end{tabular}
%\caption{J factor for various density profiles and for different slices of Fermi Bubbles}
%\label{table:Jfactor}
%\end{table}

As previously anticipated, the data we use in order to derive the bounds on the DM annihilation cross section come from the latitude-dependent energy spectrum of the Fermi bubbles shown in Fig.~\ref{fig:MAIN}.
We use the data obtained with the $1^\circ$ inner disk mask and the point source masking method.
 Even if we use all the five slices in the interval $|b|=1^{\circ}-50^{\circ}$, let us stress once again  that the most relevant ones are the first two, $|b|=1^{\circ}-10^{\circ}$ and $|b|=10^{\circ}-20^{\circ}$.
%Notice, moreover, that the data obtained using  the $5^\circ$ inner disk mask (upper panel in Fig.~\ref{fig:MAIN2}) or the South hemisphere
%(Fig.~\ref{fig:MAIN3}) give even stronger constraints. To be conservative, however, we use the data from  the whole sky analysis with $1^\circ $ inner disk mask.

%In the table~\ref{table:Jfactor}, we list J factor for different slices of Fermi bubbles including $1^\circ %$ inner disk mask and $5^\circ $ inner disk mask, and for various DM density profiles.
\begin{figure}[!htb!]
 \centering
%\vspace{-2cm}
%\hspace{-1cm}
\includegraphics[scale=0.85]{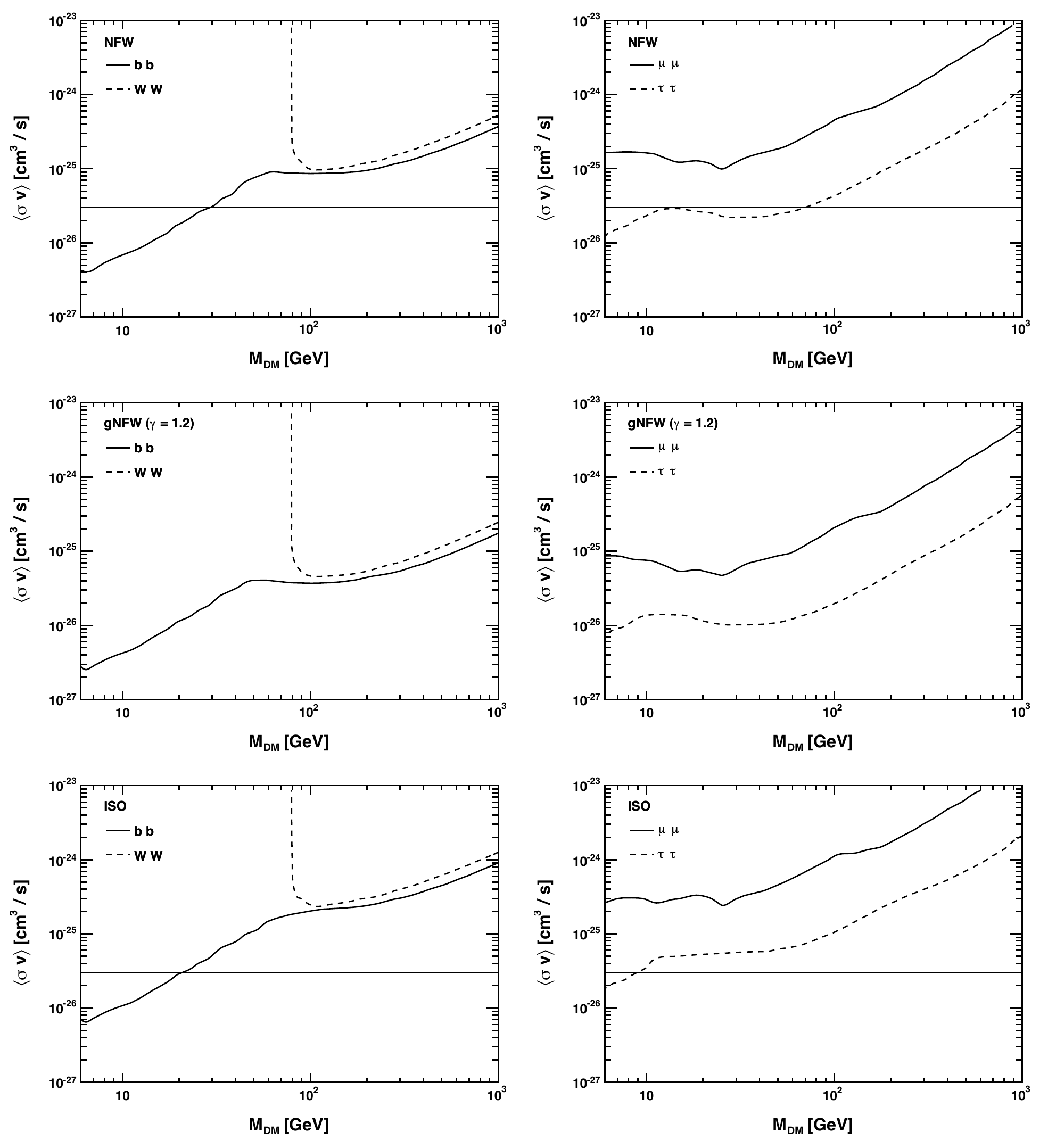}
\caption{
\textit{95\% C.L. upper bound on DM annihilation cross section for various final states and DM density profiles. The normalized density at the Sun is $\rho_\odot = 0.3~\GeV / cm^3$. The horizontal line corresponds to the simple thermal relic $\langle \sigma v \rangle= 3.0 \times 10^{-26} cm^3 /s $} } \label{fig:quarkletpon_FSR}
\end{figure}

Our constraints are shown in Fig.~\ref{fig:quarkletpon_FSR} in correspondence of four different final state: $b\overline{b}$, $W^{+}W^-$,
$\tau^+\tau^-$ and $\mu^+\mu^-$.
We present the 95\% (2-$\sigma$) C.L. upper bound in the plane $(M_{\rm DM}, \langle \sigma v \rangle)$.
 The isothermal density profile, as a cored profile, does not give constraints much different compared with the NFW profile, since the second slice of Fermi Bubble, $|b| = 10^{\circ}-20^\circ$, plays the most important role for both these profiles. For the gNFW profile with inner slope $\gamma = 1.2$, on the contrary, the constraints are a factor of few stronger than the others, since the first slice of Fermi bubbles, $|b| = 1^{\circ}-10^\circ$, contributes most.

 \begin{figure}[!htb!]
\centering
%\vspace{-2cm}
%\hspace{-1cm}
% \centering
\includegraphics[scale=0.85]{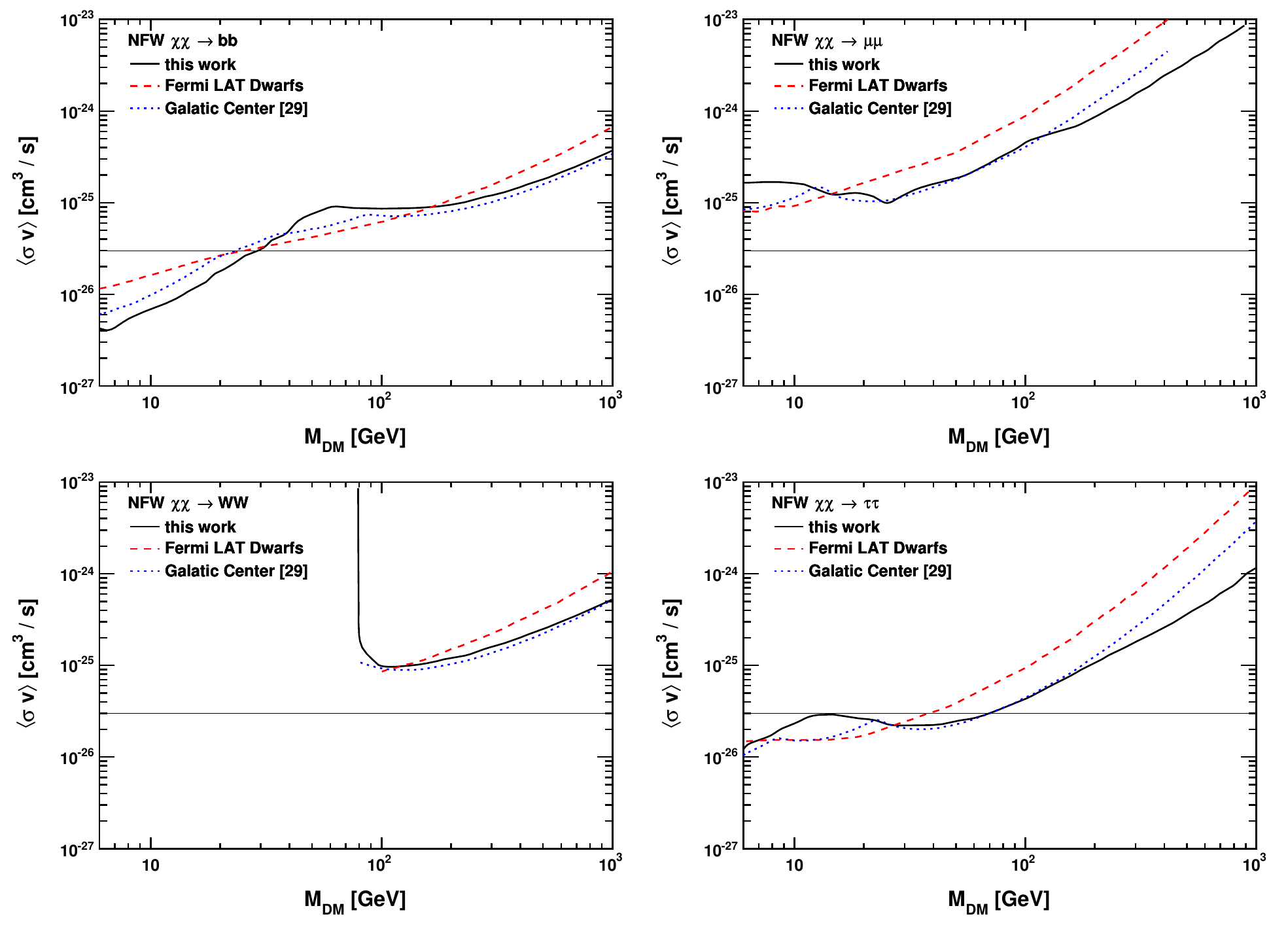}
\caption{\textit{Comparison between the bounds on the DM annihilation cross section obtained in this work analyzing the energy spectrum of the Fermi bubbles and the existing literature. The solid black line corresponds to this work. The constraints from Dwarf galaxies~\cite{Ackermann:2011wa} are shown in dashed red line while
 the Galactic center results~\cite{Hooper:2012sr} are in dotted blue line.}} \label{fig:NFWC_FSR}
\end{figure}

In Fig.~\ref{fig:NFWC_FSR}, we compare our results obtained using the NFW profile with the existing literature. We focus on the constraints from Dwarf galaxies~\cite{Ackermann:2011wa} and Galactic center~\cite{Hooper:2012sr} since they are relatively stronger than the other observations. In all the final states and almost all the mass range, our constraints from the Fermi bubbles are stronger than those from the Dwarf galaxies in Ref.~\cite{Ackermann:2011wa}.
In the low mass region for the $b\overline{b}$ final state ($M_{\rm DM} < 30$ GeV) and
in the high mass region for the $\tau^+ \tau^-$ final state ($M_{\rm DM} > 100$ GeV), the bounds of this work are stronger than those coming from the Galactic centre in Ref.~\cite{Hooper:2012sr}.

%%%%%%%%%%%%%%%%%%%%%%%%%%%%%%%%%%%%%
\section{Conclusions}
\label{sec:Conclusions}
%%%%%%%%%%%%%%%%%%%%%%%%%%%%%%%%%%%%%

Notwithstanding its elusive and photophobic nature, the possibility to hunt for DM in the sky via gamma rays is promising.

 In this paper we have analyzed the energy spectrum of the Fermi  bubbles, two  giant gamma-ray lobes extending northward and southward  from the Galactic center, at different latitudes. Our  main results can be summarized as follows.

\begin{enumerate}[label=\roman{*}., ref=(\roman{*})]

\item  Confirming the results in Ref.~\cite{Hooper:2013rwa}, we have  found that
 the energy spectrum of the Fermi bubbles, going from high to low  latitudes, presents different characteristics.
 
  At high latitude, $|b| = 20^{\circ} - 50^{\circ}$, the Fermi  bubbles spectrum
 is flattish (in
 $E_{\gamma}^2d\Phi/dE_{\gamma}d\Omega$), and can be reproduced by ICS photons generated by the the additional electron population with cut-off energy $E_{\rm cut}=1.2$ TeV and the power-law spectrum. Our best-fit value for the corresponding spectral index
  is $\gamma =-2.39$, in agreement with the typical values capable of explaining the WMAP haze in the microwave.
  
  Al low latitude, $|b| = 10^{\circ} - 20^{\circ}$, a
  bump at $E_{\gamma}=1-4$ GeV stands out in the spectral shape. This
  feature can not be explained with the only existence of the ICS  photons previously discussed. On the contrary, including in the fit  the FSR from DM annihilation the whole spectrum of the Fermi bubbles  can be realized. By virtue of the concentration of DM near the Galactic  center, in fact, the DM component leaves untouched the energy spectrum  at high latitudes, providing at the same time an excellent fit for the  bump at low latitudes.  Our best-fit candidate corresponds to  annihilation into $b\overline{b}$ with mass $M_{\rm DM}= 61.8^{+6.9}_{-4.9}$ GeV and cross section $\langle \sigma v\rangle =  3.30^{+0.69}_{-0.49}\times  10^{-26}~{\rm cm}^{3}\,{\rm s}^{-1}$.
  
  To study the interplay between ICS and DM, we have used a different  strategy compared to the one adopted in Ref.~\cite{Hooper:2013rwa}. In  order to check the stability of our results, moreover, we have  performed a number of checks concerning in particular the size of the  Galactic disk mask, the North-South symmetry of the signal, the event  categories used in the analysis, and the method employed to handle the  point source emission.

\item From a more general perspective,  we have used the energy  spectrum of the Fermi bubbles in the region $|b| =  1^{\circ}-20^{\circ}$ to derive conservative but stringent upper  limits on the DM annihilation cross section for different  final states. By comparing with the existing literature, we have found  that the bounds from the Fermi bubbles are, in most part of the  analyzed parameter space, stronger than those obtained from the Dwarf  galaxies and compatible to those obtained from the Galactic center. In  particular in the low mass region ($M_{\rm DM}< 30$ GeV) for  $b\overline{b}$ final state and in the high mass region ($M_{\rm DM}>  100$ GeV) for $\tau^+\tau^-$ final state we are able to place the most  stringent  bounds.

\end{enumerate}

In conclusion, two distinct but complementary directions arise from  this analysis. On the one hand, it is interesting to examine  from the particle  physics perspective the DM signal proposed in  Ref.~\cite{Hooper:2013rwa} and confirmed in this work. This issue will be studied in-depth in the second part of this paper. On the other hand, as suggested by the analysis of the Fermi bubbles  in the
slice $|b|=10^{\circ}-20^{\circ}$, a more careful investigation in the  region around the Galactic center is mandatory. The presence of  smaller uncertainties compared to those affecting the inner emission,  in fact, may play a crucial role to definitively highlight the first  non-gravitational evidence of DM.
Also this direction will be pursued in a forthcoming work.

A final comment is mandatory. In this paper, we have considered only statistical uncertainties.
As a caveat, the systematic uncertainties have the possibility to tune the DM
residual spectrum. The \texttt{PASS6(V11)} diffuse model template, used in this paper to study the Galactic diffuse emission, was developed by the Fermi collaboration as a result of a complicated fit. In summary, the procedure goes as follows. First, using spectral line surveys of atomic hydrogen and carbon monoxide, it is possible to derive the distribution of interstellar gas in Galactocentric rings. Second, the infrared emission from cold interstellar dust is used to correct the density distribution of  atomic hydrogen in directions where its optical depth was over or under-estimated. Gamma-ray photons produced by Bremsstrahlung and $\pi^0$ decay are obtained from these maps by means of gas-correlated method. Separately, models of magnetic fields, ISRF, and injection of cosmic rays (protons and electrons) are assumed. Finally, using the numerical code GALPROP,\footnote{\href{http://galprop.stanford.edu/}{GALPROP website.}} it is possible to propagate these cosmic rays through the gas in order to obtain the gamma-ray photons produced by ICS. As a final step, the skymaps obtained by means of this procedure are fitted using the observed gamma-ray data. Throughout this analysis, there are numerous sources of systematic errors. 
The diffuse model is generated by separating the sky into many Gaussian patches; the size of the patches will bring systematic uncertainties.
To derive the gas correlated templates, the atomic and molecular hydrogen density assume uniform spin temperature, which is another source of uncertainties.
Besides, the presence of non-gas correlated components, such as the Loop I structure, will contribute with further uncertainties. As far as this last point is concerned, note that the study of the residual spectrum in the South hemisphere, as well as the comparison with the spectrum obtained from the whole Fermi bubbles region, represents a way to estimate the uncertainties associated with the Loop I structure. Finally, unresolved Point sources and instrumental effects are additional sources of systematic uncertainties. As implied from this discussion, a detailed analysis of these effects needs the use of a numerical code, such as DRAGON\footnote{\href{http://www.dragonproject.org/Home.html}{DRAGON website.}} or GALPROP, in order to properly estimate the size of these uncertainties.  Given its importance, this topic deserves further study.

\subsection*{Aknowledgement}

We are especially thankful to Ilias Cholis, Alessandro Cuoco, Tracy Slatyer, Piero Ullio, Aaron Vincent and Gabrijela Zaharijas for many precious and enlightening discussions.
We also thank Zhen-Yi Cai, Marco Cirelli, Andrea De Simone, Gregory Dobler, Michele Frigerio, Daniele Gaggero, Farinaldo Queiroz, Meng Su and Daniele Vetrugno for useful advices.\\
The work of A.U. is supported by the ERC Advanced Grant n$^{\circ}$ $267985$, ``Electroweak Symmetry Breaking, Flavour and Dark Matter: One Solution for Three Mysteries" (DaMeSyFla). W.-C.H. would like to thank the hospitality of Northwestern University, where part of this work was performed.

\appendix

%%%%%%%%%%%%%%%%%%%%%%%%%%%%%%%%%%%%%
\section{Fermi-LAT gamma-ray full sky analysis}\label{app:A}
%%%%%%%%%%%%%%%%%%%%%%%%%%%%%%%%%%%%%
In this Appendix we describe in detail our procedure for the analysis of the Fermi-LAT data. 
\subsection{Event selection and counts maps}\label{App:Count}
%The Fermi Gamma Ray Space Telescope spacecraft \cite{Fermi_website}  - launched on 11 June 2008 - is a %space observatory
%devoted to the gamma-ray analysis of the Milky Way galaxy. The main instrument aboard is the
%Large Area Telescope (LAT) \cite{LAT_website}, able to detect photons in the energy range from about 20 MeV %to more than 300 GeV. 
The \texttt{PASS7(V6)} dataset \cite{Fermi_Data} used in this analysis contains two different types of files. The  \href{http://heasarc.gsfc.nasa.gov/FTP/fermi/data/lat/weekly/p7v6/photon/}{event data files} provide all the informations describing the collected photons, e.g. their energy and their reconstructed arrival direction;  the \href{http://heasarc.gsfc.nasa.gov/FTP/fermi/data/lat/weekly/p7v6/spacecraft/}{spacecraft files}, on the contrary, contain all the information regarding the spacecraft, e.g. position and orientation for the typical time interval of 30 seconds.\footnote{The interested reader can find a more detailed description of the gamma-ray data in  the correspondent section of the Fermi Science Tools manual, aka  ``\href{http://fermi.gsfc.nasa.gov/ssc/data/analysis/documentation/Cicerone/}{\textit{Cicerone}}".} 
The events are classified in four classes denoted as \texttt{TRANSIENT}, \texttt{SOURCE}, \texttt{CLEAN} and \texttt{ULTRACLEAN}. We use the \texttt{ULTRACLEAN} event class to reduce the cosmic-ray background contribution.
For completeness in Appendix~\ref{App:eventclass} we repeat our analysis using \texttt{SOURCE} and \texttt{CLEAN} categories.

Lastly, each photon is further labelled as \texttt{front} or \texttt{back} according to region of the LAT detector in which - interacting with a tungsten atom - the photon is converted into an electron-positron pair. We analyze both front- and back-converting events, generating two different sets of counts maps. However for energies smaller than $1$ GeV, as explained in Section~\ref{App:masking}, we use only front-converting events.

To analyze the data we use the \href{http://fermi.gsfc.nasa.gov/ssc/data/analysis/}{Fermi Science Tools}.
In order to generate the counts maps we use \texttt{gtselect} and \texttt{gtmktime} to create a filtered FITS file according  to our selection criteria. In particular we impose the zenith angle cut $z_{cut}=100^{\circ}$ to reduce the contamination from the Earth limb, while 
we use recommended cuts on data quality, nominal science configuration and rocking angle, i.e. \texttt{DATA\_QUAL = 1}, \texttt{LAT\_CONFIG = 1}, \texttt{ABS(ROCK\_ANGLE) < 52}. 
We specify \texttt{ROIcut = no}, as recommended for the full-sky analysis.\footnote{We have checked that different choices for the zenith angle cut, namely  $z_{cut}=90^{\circ},\,105^{\circ}$ do not change our results.} We bin the selected data in the range between $0.3$ GeV and $300$ GeV in 30 log-spaced energy bins,  while for the galactic coordinates $l$ and $b$ we use a spatial grid $l\times b= 720\times 360$ pixels, with resolution $0.5$ degrees/pixels. 
To manipulate and analyze the generated FITS files, we use a \texttt{NSIDE} $= 256$ \href{http://healpix.jpl.nasa.gov/}{HEALPix} grid; in this way we can fully benefit from 
the iso-latitude equal-area pixelization  algorithm used by HEALPix in the analysis of the sky maps. We generate the counts maps in HEALPix format using the \href{http://www.exelisvis.com/ProductsServices/IDL.aspx}{IDL} software.  For this purpose - as well as for any other IDL analysis throughout this paper - we use our own code.\footnote{The interested reader can find on the webpage \href{http://www.sdss3.org/dr8/software/idlutils.php}{idlutils} a collection of IDL routines for astronomical applications.}  
 
\subsection{Livetime cubes and exposure maps}\label{App:Exposure}

Considering the detection of a photon from a given point $(l,b)$ of the sky, the response of the LAT crucially depends on the  angle between the incident direction of the photon and the orientation of the instrument. The latter is defined by the LAT boresight, i.e.  the line normal to the top surface of the LAT. Since this angle changes during the orbital motion of the spacecraft, the measured number of photons depends on the amount of time spent by the instrument at each angle w.r.t. the position in the sky of the analyzed source. This information is stored in a three-dimensional grid, called ``livetime cube''. We generate the livetime cube using 
\texttt{gtltcube}. As recommended by the Fermi \textit{Cicerone}, we carefully specify the \texttt{zmax} cut in order to match the zenith cut imposed in \texttt{gtselect};  in this way we are able to exclude time intervals in which the condition on the zenith angle is not respected. For a given source position in the sky and a given energy, the exposure map accounts for the correspondent total exposure measured in cm$^2$s. Combining the livetime cube with the response function of the instrument using \texttt{gtexpcube2}, we generate the exposure maps in FITS format. For the LAT instrument response function we use \texttt{P7ULTRACLEAN\_V6::FRONT} and \texttt{P7ULTRACLEAN\_V6::BACK} for, respectively, front- and back-converting \texttt{ULTRACLEAN} events. As done before for counts maps, we interpolate the exposure FITS files on a \texttt{NSIDE} $= 256$ HEALPix grid using the IDL software.
%Counts map divided by the exposure and pixel solid angle produce the correspondent observed intensity map, 
%measured in [$photons/cm^{2}/s/sr$]. 

%\subsection{The Galactic diffuse model}\label{app:diffuse}
%The Galactic diffuse gamma-ray emission is produced by interactions of cosmic rays 
%with interstellar gas and low-energy radiation fields. 
%In more detail cosmic ray electrons produce synchrotron radiation interacting with magnetic fields due to their %spiral motion. Furthermore they produce Bremsstrahlung radiation via interactions with the matter in the %interstellar medium.
%Another contribution is the inverse Compton scattering between cosmic ray electrons and photons of the low-%energy
%interstellar radiation field. Moreover cosmic ray protons interacting with the interstellar medium produce gamma %rays via neutral $\pi^0$ decay.\\
%The Galactic diffuse emission is a background for our analysis. In Section~\ref{App:Residual} it will be %subtracted from the observed counts maps in order to highlight the residual gamma-ray contribution that we are %interested in. \\
%We use the \texttt{PASS6(V11)} \href{http://fermi.gsfc.nasa.gov/ssc/data/access/lat/BackgroundModels.html}%{diffuse model template} provided by the Fermi collaboration in order to model the Galactic diffuse emission.\footnote{We use the \texttt{PASS6(V11)} diffuse model template instead of the more recent  \texttt{PASS7(V6)} because the latter already contains a template for the Fermi bubbles.}

\subsection{Point and extended sources: masking vs. subtraction}\label{App:masking}

%The number of photons per unit solid angles and unit energy predicted by the Galactic diffuse model is given by %the correspondent diffuse template times the exposure.\\
Galactic gamma-rays are not only produced by
the interactions of cosmic rays with the interstellar gas and low-energy radiation fields but also come from point sources, e.g. pulsars such as Crab, Geminga and Vela, as well as extended sources, e.g. galaxies 
such as Centaurus A.

Before comparing the observed gamma-ray sky with the Galactic diffuse model prediction,
% in order to get the residual maps
 therefore,
we need to take into account the contribution of the point and extended sources.
% two more operations are mandatory - first we mask all the point and the extended sources, then we smooth all %the sky maps.\\
The \href{http://fermi.gsfc.nasa.gov/ssc/data/access/lat/2yr_catalog/}{LAT 2-year Point Source Catalog} (L2PSC) contains
all the available information about  $1862$ point sources and $11$ extended sources, e.g. the position in the sky and the energy spectrum. The LAT has finite angular resolution as well as finite precision; as a consequence the point sources are broadened in the data and appear to be wider than they actually are. The smaller the energy of the detected photons the larger the broadening effect. The information about the angular resolution of the instrument is encoded in the PSF. Indicating as $\theta$ the angle between the apparent photon direction and a given position in the sky, the PSF, measured in sr$^{-1}$, is the probability density to reconstruct such direction for a gamma-ray photon with energy $E_{\gamma}$. As a result, the PSF is normalized according to
\be
2\pi \int d\theta\,\sin\theta\,{\rm PSF}( \theta, E_{\gamma}) = 1~,
\ee
and it also depends on the event class. We use two methods to analyze the data. One is the point source masking, based on an analytical fit for the PSF. The other method is the point source subtraction. In the second method, we use \texttt{gtpsf} to generate the in-flight PSF for all the point sources. In the following we discuss in detail both these methods.

\subsubsection{Masking}\label{App:maskingdetails}
Following Ref.~\cite{Ackermann:2012kna} we use for the PSF a simple analytical fit describing the radius of $68\%$ flux containment 
\begin{equation}
r_{68}(E_{\gamma}) = \sqrt{\left[c_0 \left(\frac{E_{\gamma}}{100\,{\rm MeV}}\right)^{-\beta}\right]^2+c_1^2}~,
\end{equation}\label{eq:PSF68rad}
where the values of the coefficients are summarized in Table~\ref{tab:PSFcoeff}.\\
\begin{table}[!htb!]
\caption{\textit{Coefficient for the analytical fit in Eq.~(\ref{eq:PSF68rad}).}}
\begin{center}
\begin{tabular}{|c|c|c|c|}
\hline
Conversion type & $c_0~~[{\rm deg.}]$  & $c_1~~[{\rm deg.}]$ & $\beta$\\
\hline
\texttt{Front} & 3.3 & 0.1 & 0.78 \\
\hline
\texttt{Back} & 6.6 & 0.2 & 0.78 \\
\hline
\end{tabular}
\end{center}
\label{tab:PSFcoeff}
\end{table}
We associate each point source of the catalogue a disk mask with radius $r_{95}(E_{\gamma})$. As described in Ref.~\cite{Ackermann:2012kna}, the PSF is not a Gaussian distribution and the ratio 
$r_{95}(E_{\gamma})/r_{68}(E_{\gamma})$  is wider 
than Gaussian. We consider $r_{95}(E_{\gamma})=3\times r_{68}(E_{\gamma})$,  that is a reliable value in the energy range relevant for our analysis.
For the extended 
sources, on the contrary, we use a fixed masking radius according to the correspondent Extended Source template provided by the Fermi collaboration. Moreover we mask the inner disk in the region $|b| < 1^{\circ}$, $ |l| < 60^{\circ}$. In order to test the presence of further residual contaminations from disk-correlated emission, in Appendix~\ref{App:Galacticdisk} we use a larger disk mask  $|b| < 5^{\circ}$, $ |l| < 60^{\circ}$. We generate, for both front- and back-converting events, two sets of masks. They differ by the presence of an extra rectangular mask in correspondence of the Fermi bubbles region ($-30^{\circ} < b < 30^{\circ}$, $-53.5^{\circ} < l < 50^{\circ}$). Since the Galactic diffuse model does not include the bubble template, in fact, we need to exclude this region in the comparison with the observed counts maps. Masked maps are obtained, for each energy bin, by simply multiplying the analyzed map by the correspondent mask.

\subsubsection{Subtraction}

\begin{figure}[!htb!]
   \centering
       \includegraphics[scale=0.45]{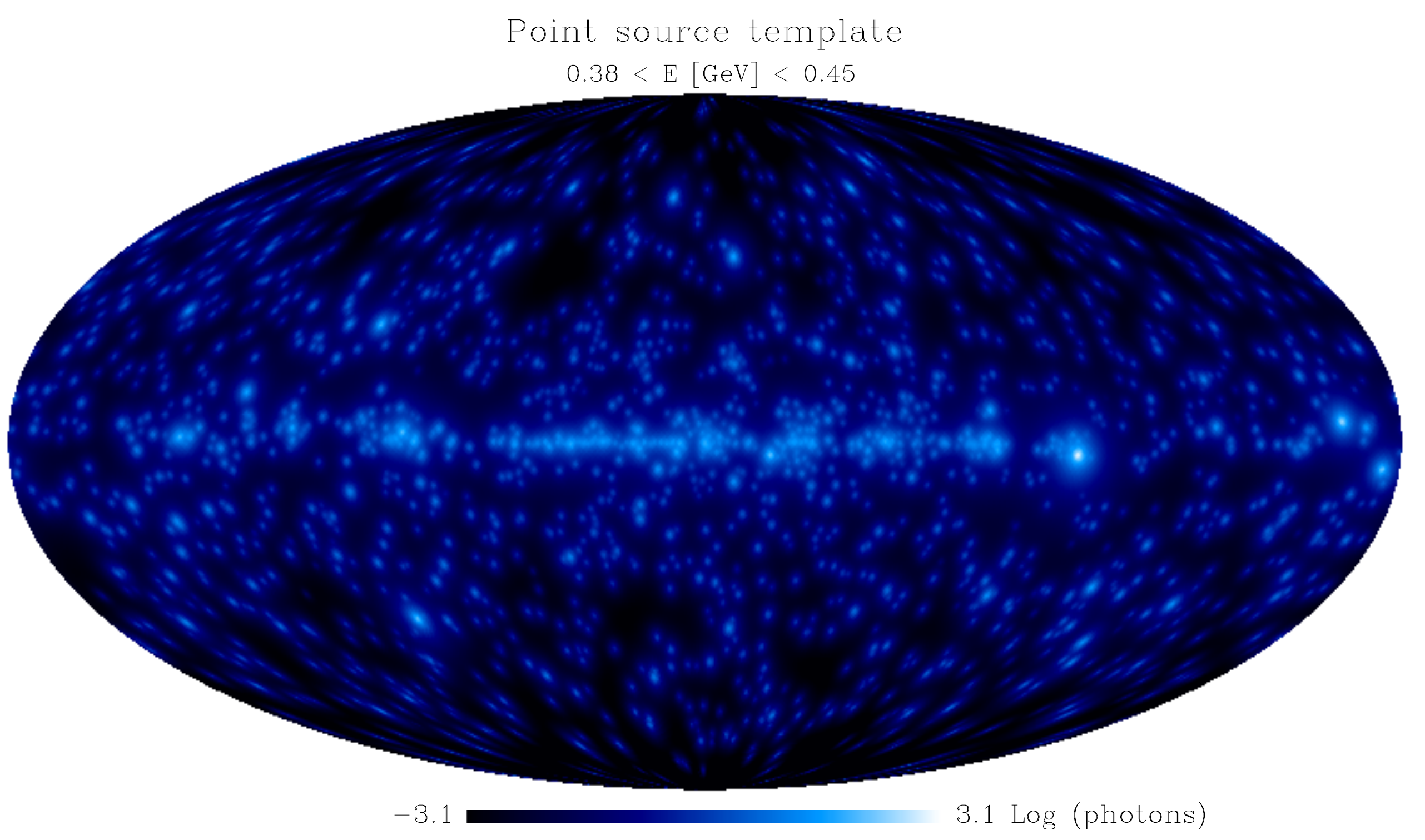}
 \caption{\emph{
 Point source template for front-converting photons in the energy interval  $E_{\gamma} = 0.38 - 0.45$ GeV.
 }}
 \label{Fig:PSMap}
\end{figure}

Point source subtraction consists in subtracting from the observed counts maps the photons coming from the point sources. This information can be obtained generating a point source template.
The first step to generate the point source template is to read the point source information, such as the position in the sky and the energy spectrum, from the L2PSC. 
By specifying the exposure livetime cube and the instrument response function, \texttt{gtpsf} calculates for each point source the PSF as a function of energy and angle $\theta$. The observed gamma-ray flux at energy $E_{\gamma}$ and position $(l,b)$ is therefore given by
\be\label{eq:PSFlux}
\frac{d \Phi}{d E_{\gamma} d \Omega} ( E_{\gamma}, l, b) 
= \sum_{i={\rm point\,sources}} \mathrm{Flux}_{i}( E_{\gamma})\times \mathrm{PSF}_{i}(\theta_i , E_{\gamma})
\ee
where the angle $\theta_i$ is quantitatively defined as the angular distance, projected in the Galactic plane, between the observation point $(l,b)$ and the source position, 
$\theta_i = \sqrt{( l - l_i)^2+ ( b - b_i)^2}$. For each point source we compute the corresponding
 flux assuming for the spectral parameters the central values reported by the L2PSC. In Eq.~(\ref{eq:PSFlux}) we sum over all the point and extended sources with the exception of the pulsar wind nebulae Vela X and MSH 15-52, that we mask. In addition we mask the inner Galactic disk
and the Fermi bubbles region as explained in Appendix~\ref{App:maskingdetails}. Multiplying 
the flux in Eq.~(\ref{eq:PSFlux}) by the exposure, the pixel solid angle and the energy width, we find the total number of photons associated with the point source emission for a given energy interval.  In
 Fig.~\ref{Fig:PSMap} we show our point source template for front-converting events in the
 representative interval $E_{\gamma} = 0.38 - 0.45$ GeV.

\subsection{Smoothing}\label{App:smoothing}

Because of the PSF each sky map has a resolution that, in the Gaussian  approximation, is given by a Gaussian distribution with full width at half maximum (FWHM) $f_{\rm raw} = 2\,r_{50} \approx 1.56\,r_{68}$. In order to properly compare different maps, therefore,  we first need to smooth each of them to a common value; the latter is defined, following Ref.~\cite{Dobler:2009xz}, by a
Gaussian distribution with FWHM $f_{{\rm target}}= 2^{\circ}$ for $E_{\gamma} > 1$ GeV ($f_{{\rm target}}= 3^{\circ}$ for $E_{\gamma} < 1$ GeV). This means that we need to smooth each map by the kernel $f_{\rm kernel}=\sqrt{f_{\rm target}^2 - f_{\rm raw}^2}$. Since for back-converting events $f_{\rm raw}$ is large at low energy, for $E_{\gamma} < 1$ GeV we use only front-converting ones. All the counts maps and exposure maps are masked and smoothed following this prescription. For the diffuse model skymaps, on the contrary, we use the kernel $f_{\rm kernel}=\sqrt{f_{\rm target}^2 - f_{0}^2}$, where $f_0 = 0.25^{\circ}$ is the spatial resolution of the template provided by the Fermi collaboration.

We also checked that our results are stable choosing $f_{\rm target} = f_{\rm raw}$, i.e. using unsmoothed counts and exposure maps, and smoothing the diffuse model by the corresponding kernel.

\section{Exploring alternative setups}\label{App:B}

In this Appendix we explore different setups for the Fermi bubbles analysis.
In particular in Section~\ref{App:eventclass} we use different event categories, comparing the energy spectra w.r.t. those  obtained in the main part of the paper using the  \texttt{ULTRACLEAN} events.  In Section~\ref{App:Galacticdisk}, on the contrary, we use a different mask for the Galactic disk and the Fermi bubbles region. We conclude in Section~\ref{App:NS} with the analysis of the residual energy spectra obtained considering separately the North and the South hemisphere. The purpose of the Appendix~\ref{App:B} is to illustrate qualitatively to what extent the observed spectral features depend on these selection criteria.

\subsection{Event class}\label{App:eventclass}

\begin{figure}[!htb!]
 \centering
  \begin{minipage}{0.4\textwidth}
   \centering
   \includegraphics[scale=0.42]{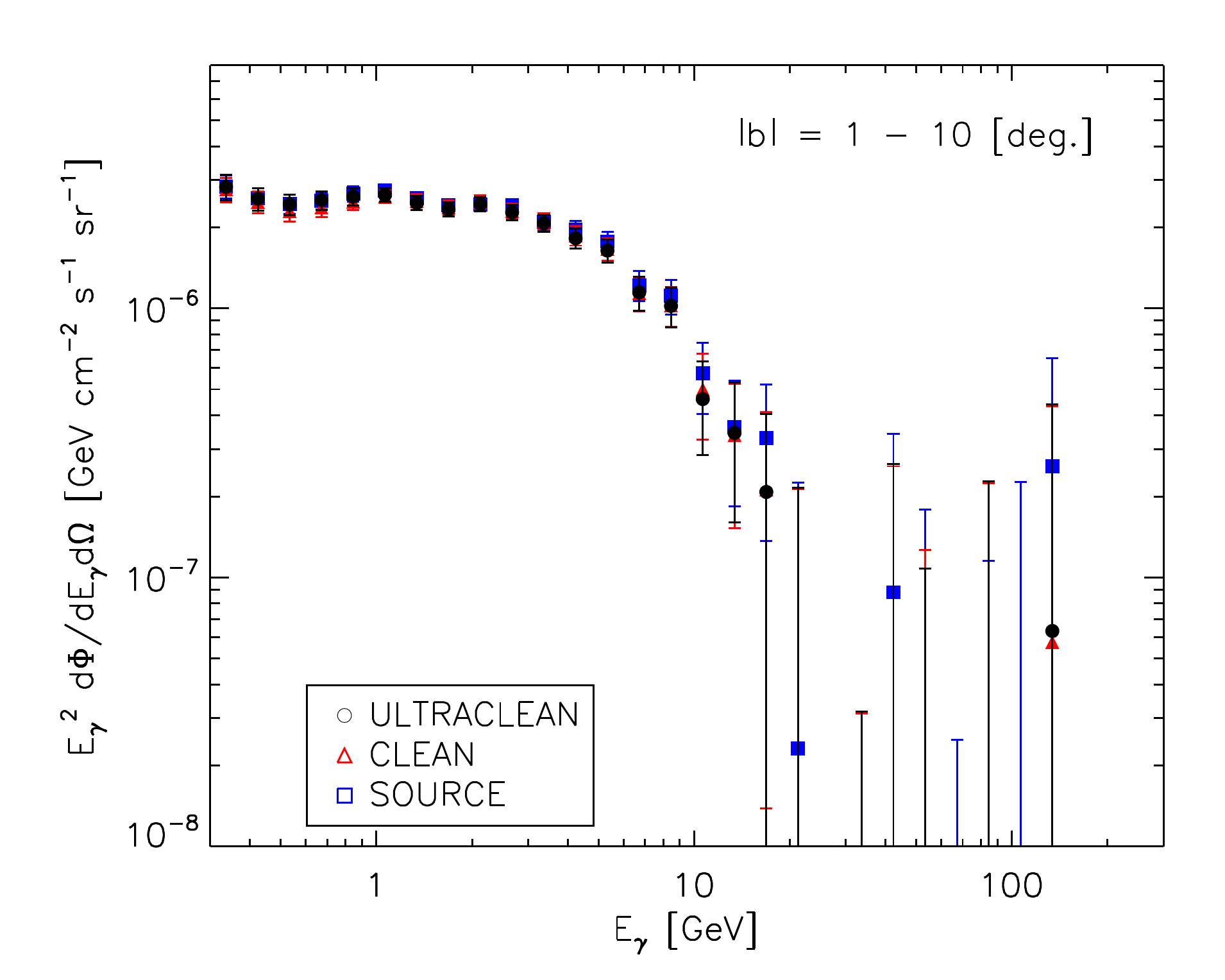}
   %\caption{\textit{Count Map}}\label{fig:CountMap}
    \end{minipage}\hspace{1.2 cm}
   \begin{minipage}{0.4\textwidth}
    \centering
    \includegraphics[scale=0.42]{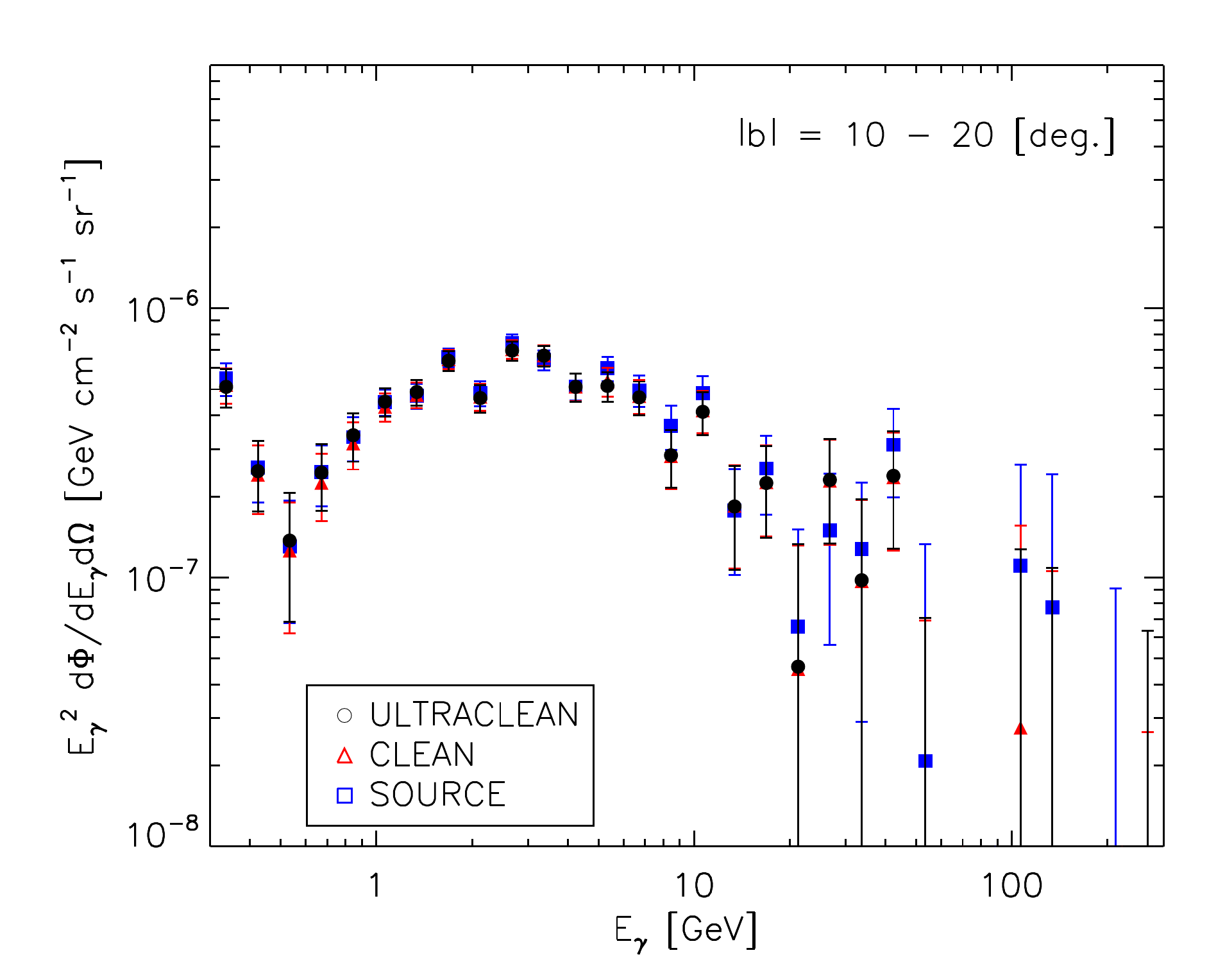}
    \end{minipage}\\
    \vspace{0.5 cm}
   \begin{minipage}{0.4\textwidth}
    \centering
   \includegraphics[scale=0.42]{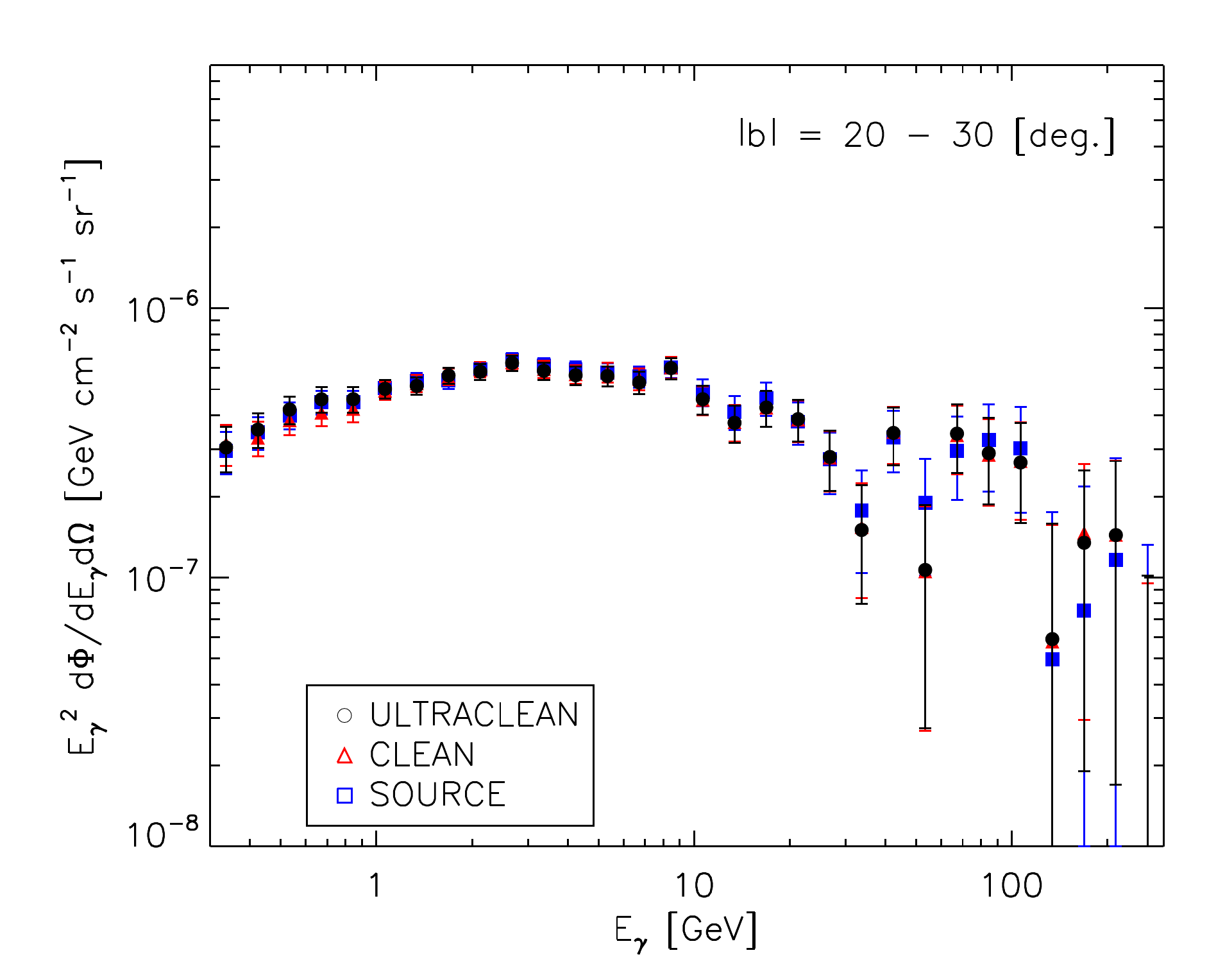}
   %\caption{\textit{Exposure Map}}\label{fig:ExposureMap}
    \end{minipage}\hspace{1.2 cm}
   \begin{minipage}{0.4\textwidth}
    \centering
    \includegraphics[scale=0.42]{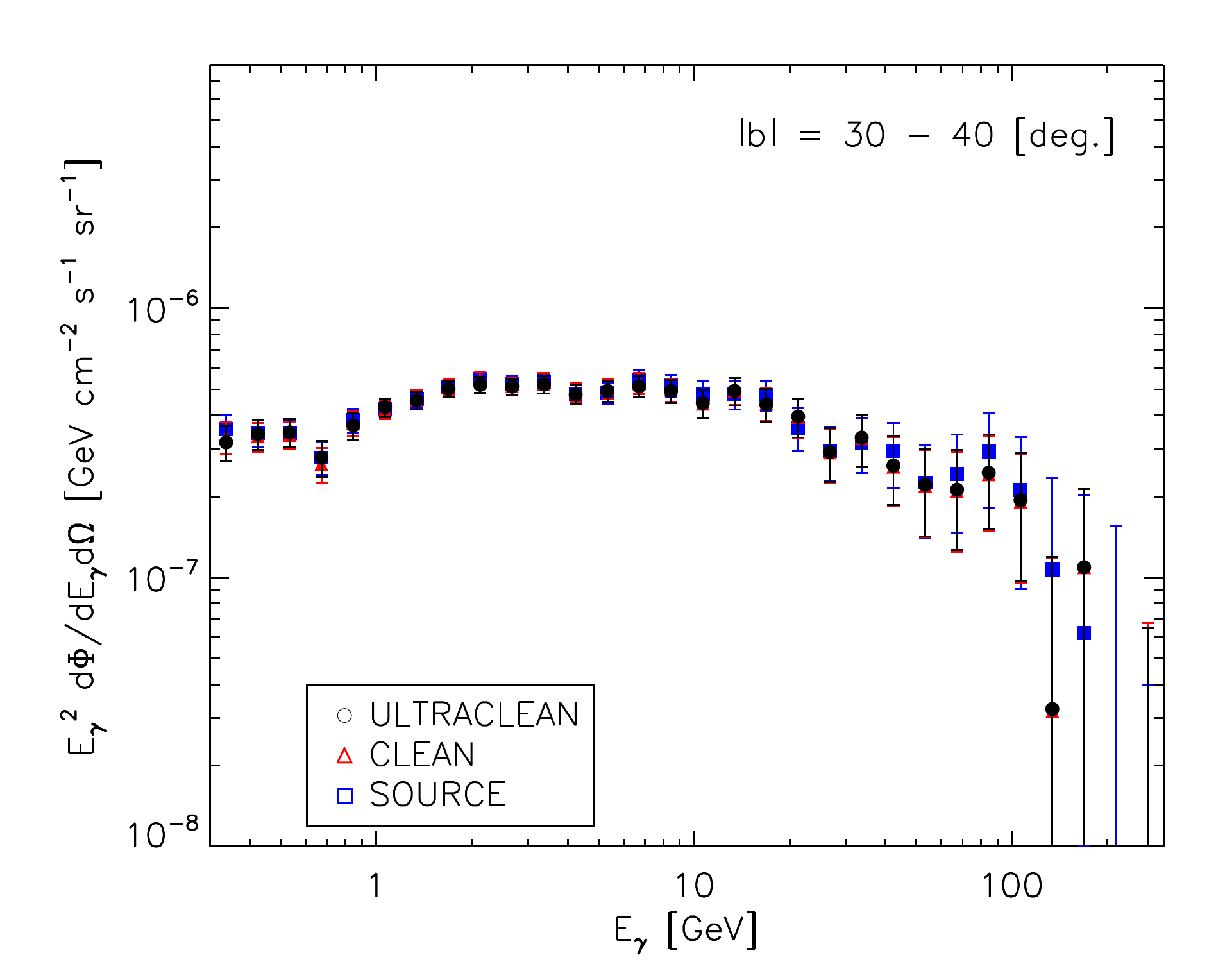}
    \end{minipage}\\
     \vspace{0.5 cm}
       \begin{minipage}{1\textwidth}
    \centering
   \includegraphics[scale=0.42]{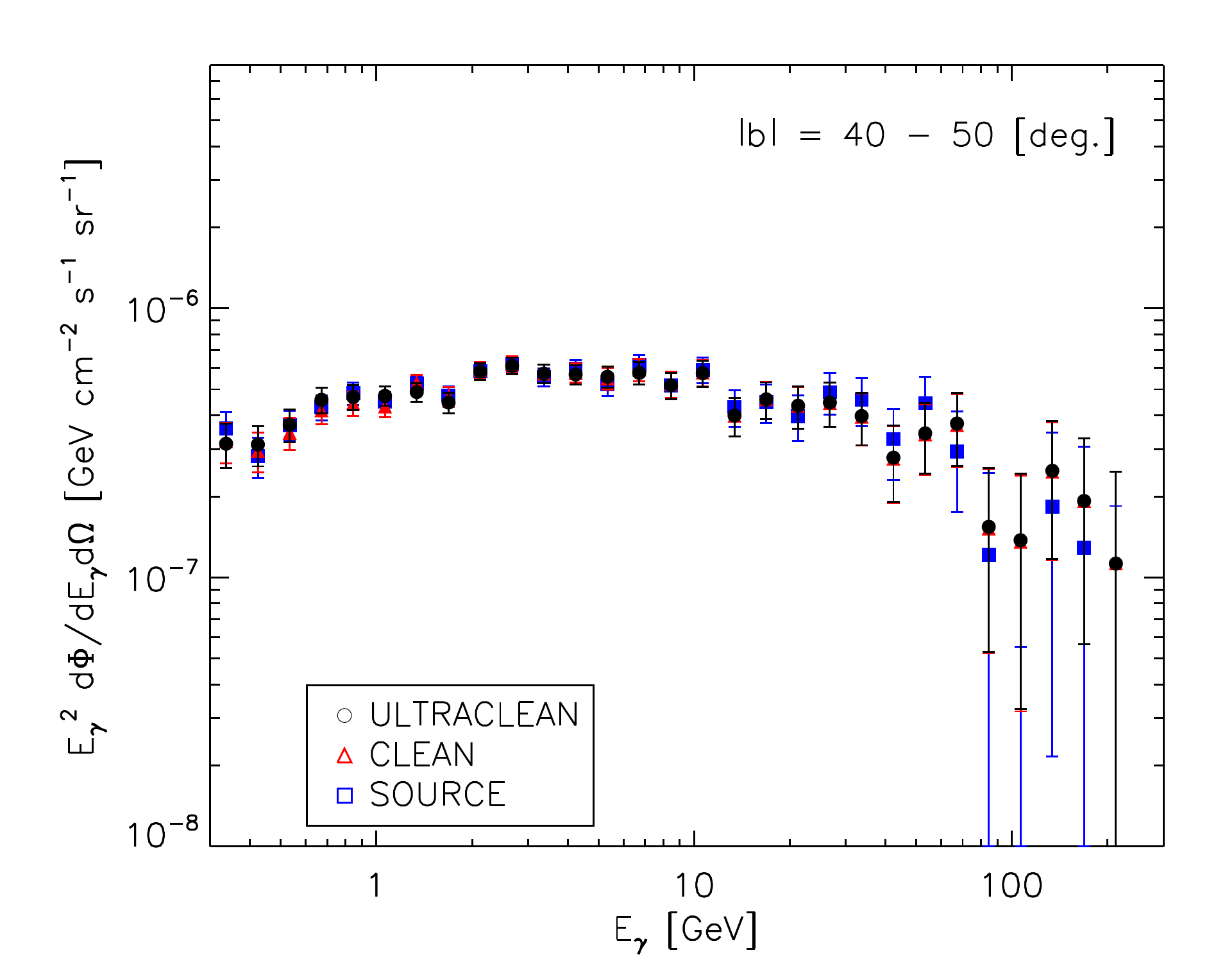}
  % \caption{\textit{Galactic diffuse model}}\label{fig:DiffuseMap}
    \end{minipage}
    \caption{\textit{
    {\underline{EVENT CLASS ANALYSIS}}.
   Fermi bubbles  energy spectrum obtained using \texttt{ULTRACLEAN} (black dots), \texttt{CLEAN} (red triangles) and \texttt{SOURCE} (blue squares) events. 
   %We mask the inner disk in the region $|b|< 1^{\circ}$, $|l|<60^{\circ}$.
    }}\label{fig:class}
\end{figure}

In this Section we analyze the Fermi bubbles energy spectrum 
comparing the \texttt{ULTRACLEAN} events  with the \texttt{CLEAN} and \texttt{SOURCE} categories. 
They are characterized by a different residual contamination from cosmic rays. Roughly speaking,  from \texttt{SOURCE} to \texttt{ULTRACLEAN} through \texttt{CLEAN},  event  data are subjected to gradually tighter cuts to reduce contamination from primary cosmic ray protons, primary cosmic ray electrons and secondary cosmic rays \cite{Ackermann:2012kna}. We show our results in Fig.~\ref{fig:class}. As argued in Ref.~\cite{Hooper:2013rwa}, we find that the features of the Fermi bubbles spectrum remain intact without any significant change. 

\subsection{Galactic disk and bubbles mask}\label{App:Galacticdisk}

\begin{figure}[!htb!]
 \centering
  \begin{minipage}{0.4\textwidth}
   \centering
   \includegraphics[scale=0.42]{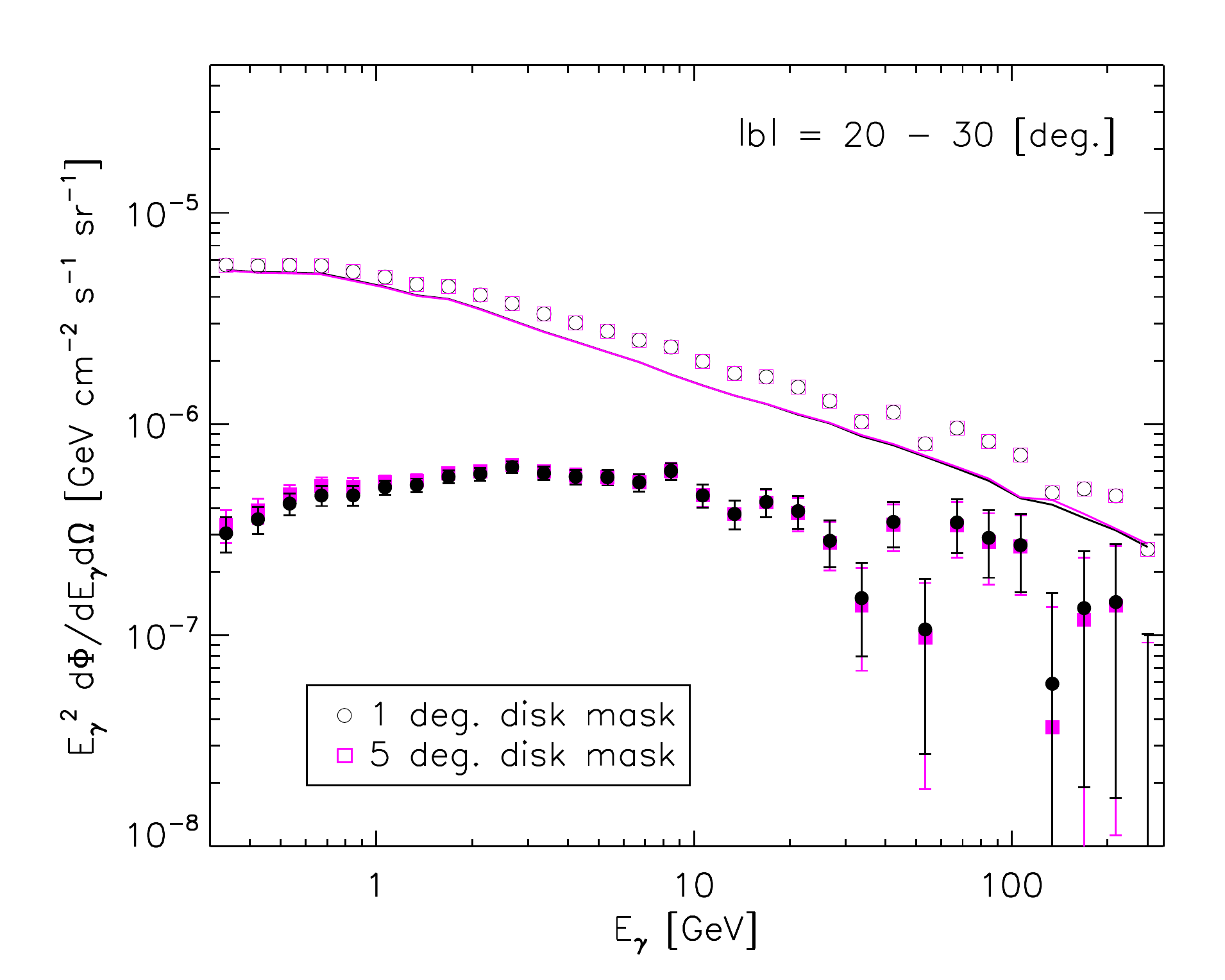}
   %\caption{\textit{Count Map}}\label{fig:CountMap}
    \end{minipage}\hspace{1.2 cm}
   \begin{minipage}{0.4\textwidth}
    \centering
    \includegraphics[scale=0.42]{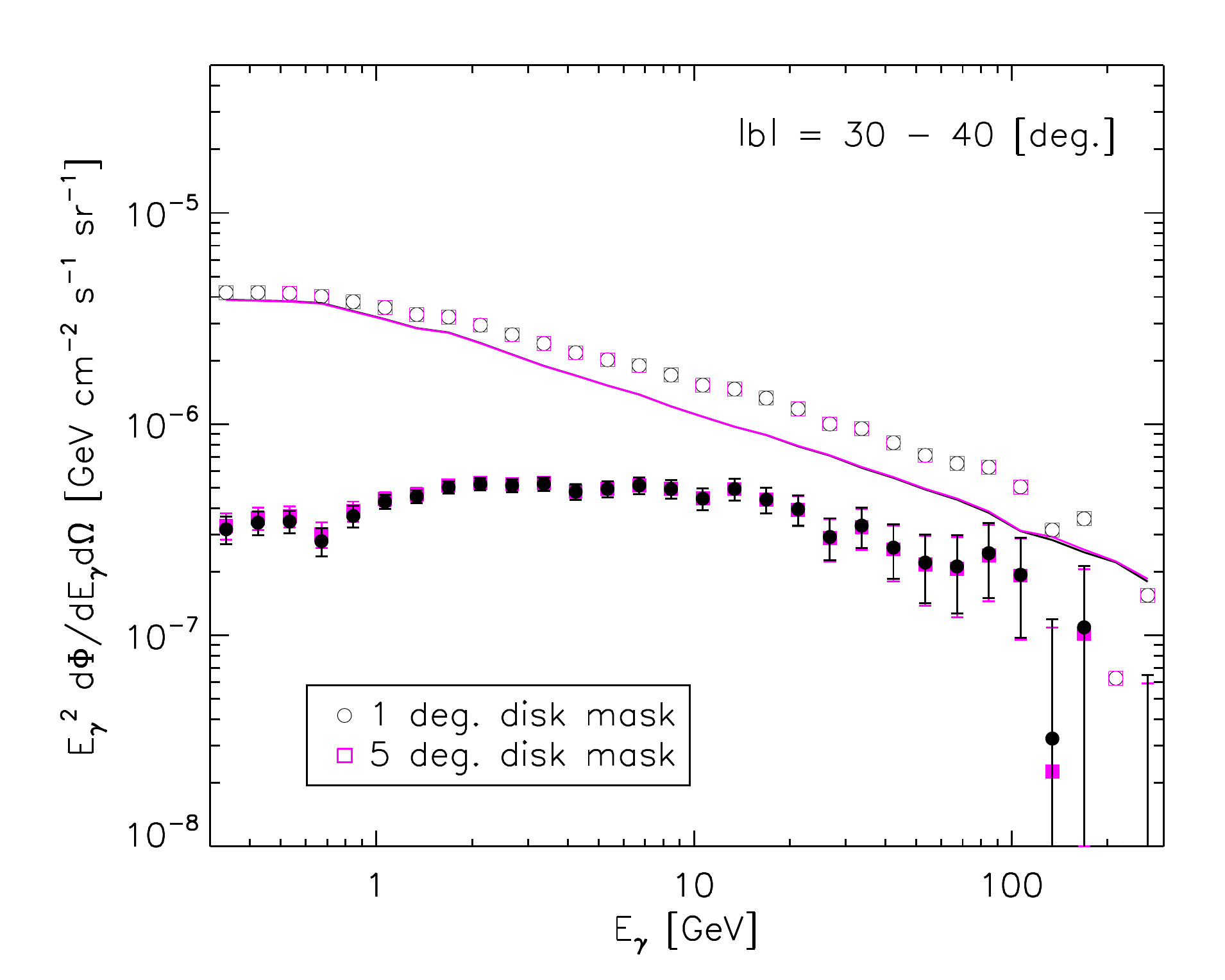}
    \end{minipage}\\
    \vspace{0.5 cm}
    \begin{minipage}{1\textwidth}
    \centering
   \includegraphics[scale=0.42]{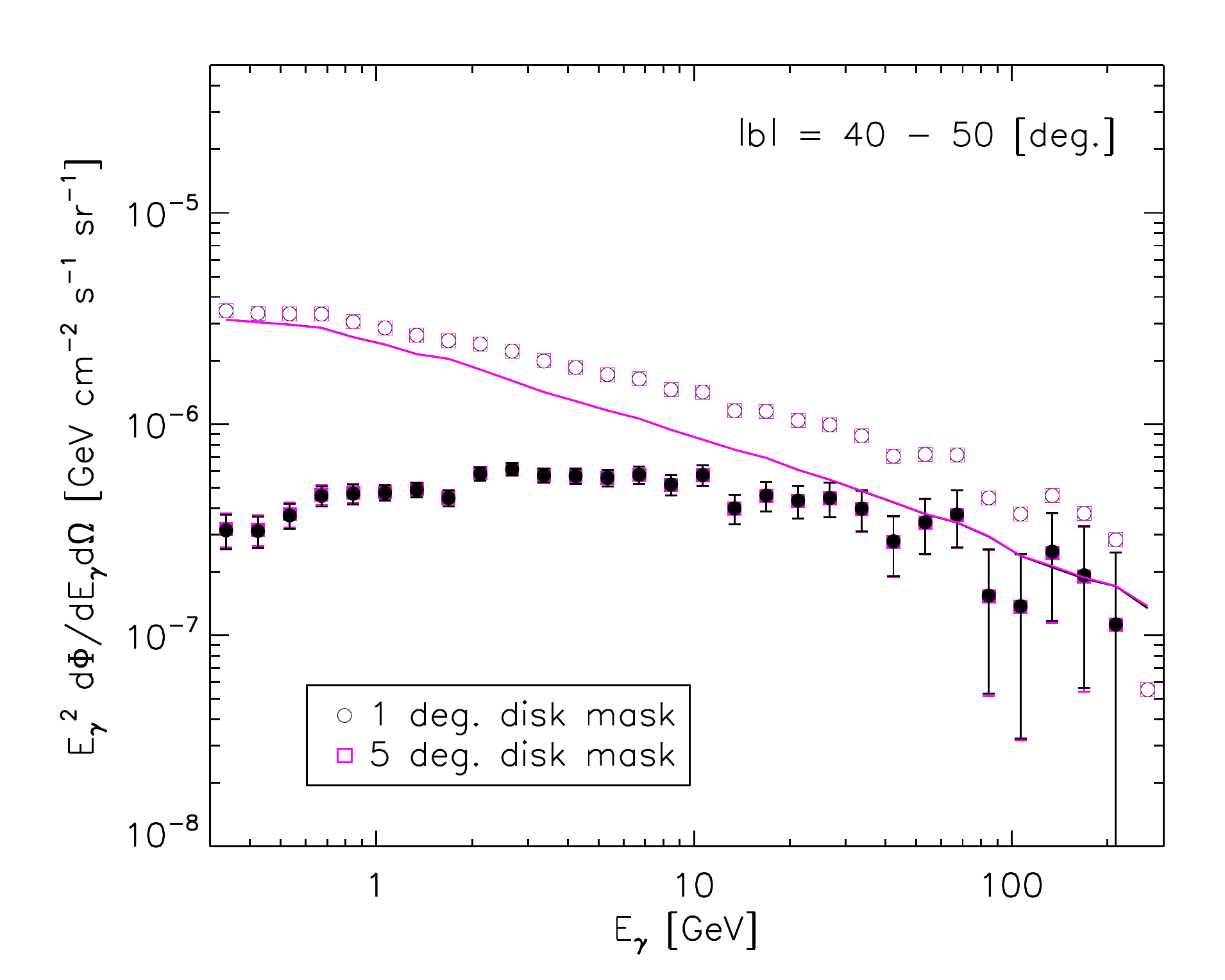}
  % \caption{\textit{Galactic diffuse model}}\label{fig:DiffuseMap}
    \end{minipage}
    \caption{\textit{
    {\underline{GALACTIC DISK ANALYSIS}}.
Fermi bubbles energy spectrum obtained masking the inner Galactic disk in the region 
$|b|< 1^{\circ}$, $|l|<60^{\circ}$ (black dots) and $|b|< 5^{\circ}$, $|l|<60^{\circ}$ (purple squares).
 }}\label{Fig:disk}
\end{figure}

\begin{figure}[!htb!]
 \centering
  \begin{minipage}{0.4\textwidth}
   \centering
   \includegraphics[scale=0.42]{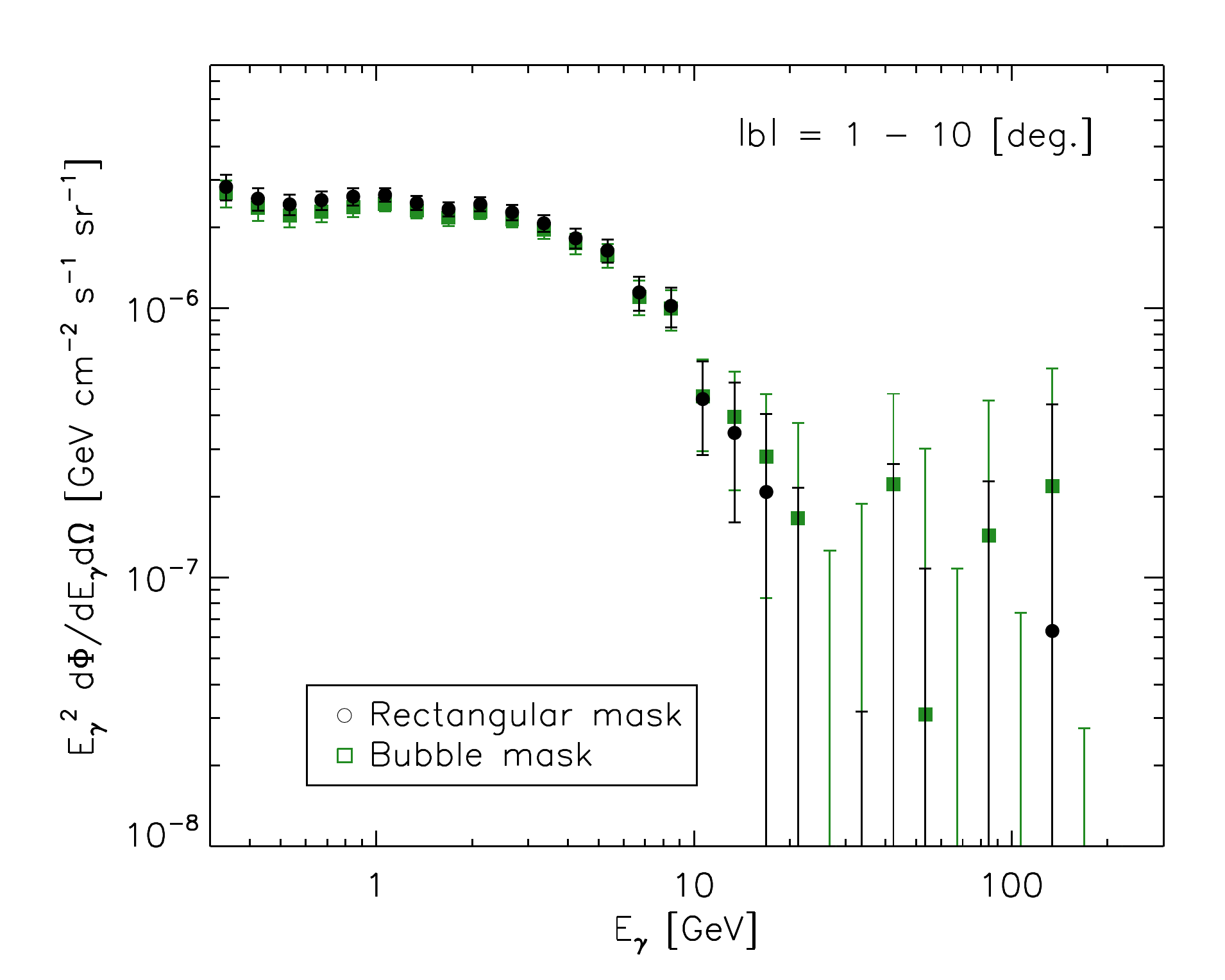}
   %\caption{\textit{Count Map}}\label{fig:CountMap}
    \end{minipage}\hspace{1.2 cm}
   \begin{minipage}{0.4\textwidth}
    \centering
    \includegraphics[scale=0.42]{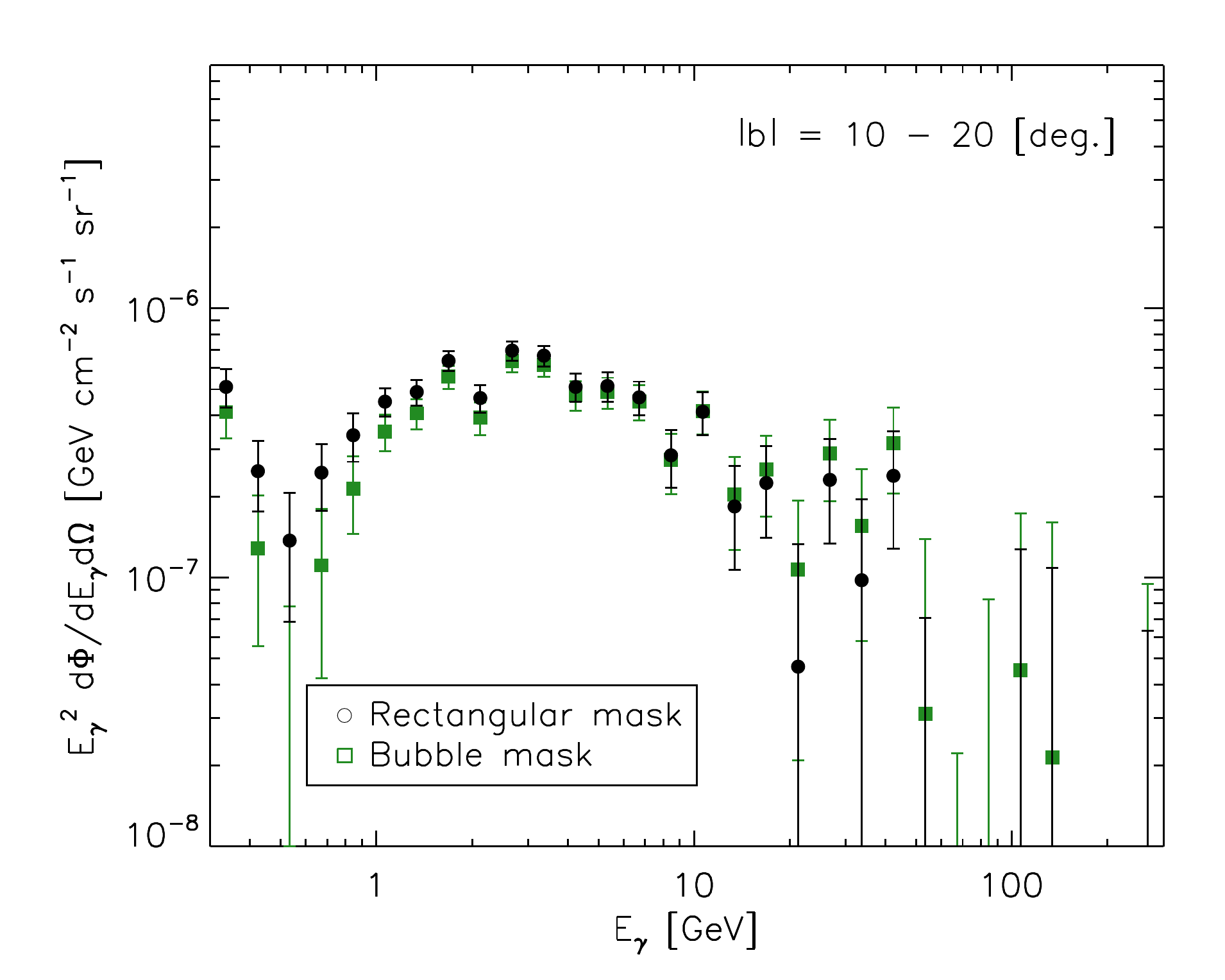}
    \end{minipage}\\
    \vspace{0.5 cm}
   \begin{minipage}{0.4\textwidth}
    \centering
   \includegraphics[scale=0.42]{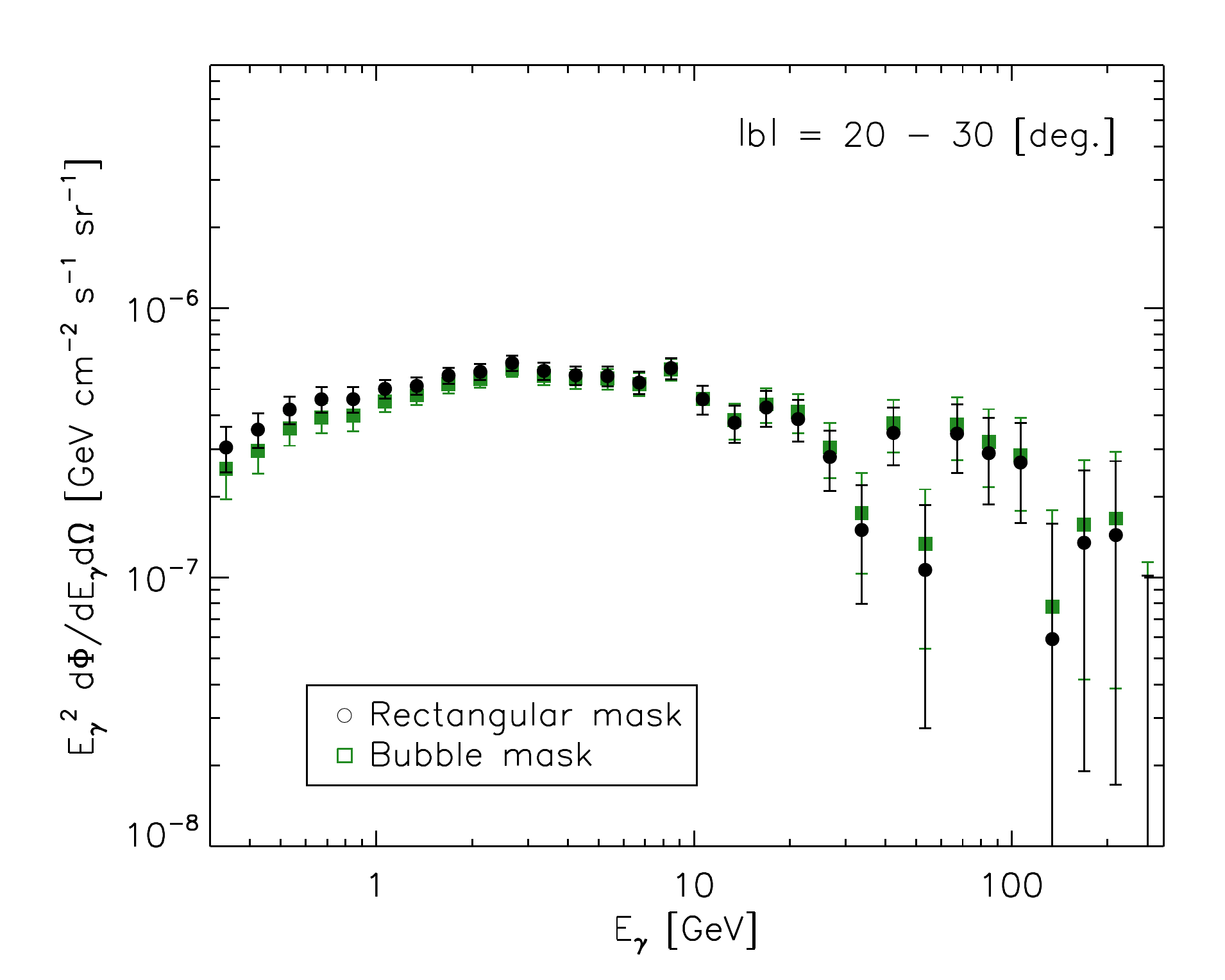}
   %\caption{\textit{Exposure Map}}\label{fig:ExposureMap}
    \end{minipage}\hspace{1.2 cm}
   \begin{minipage}{0.4\textwidth}
    \centering
    \includegraphics[scale=0.42]{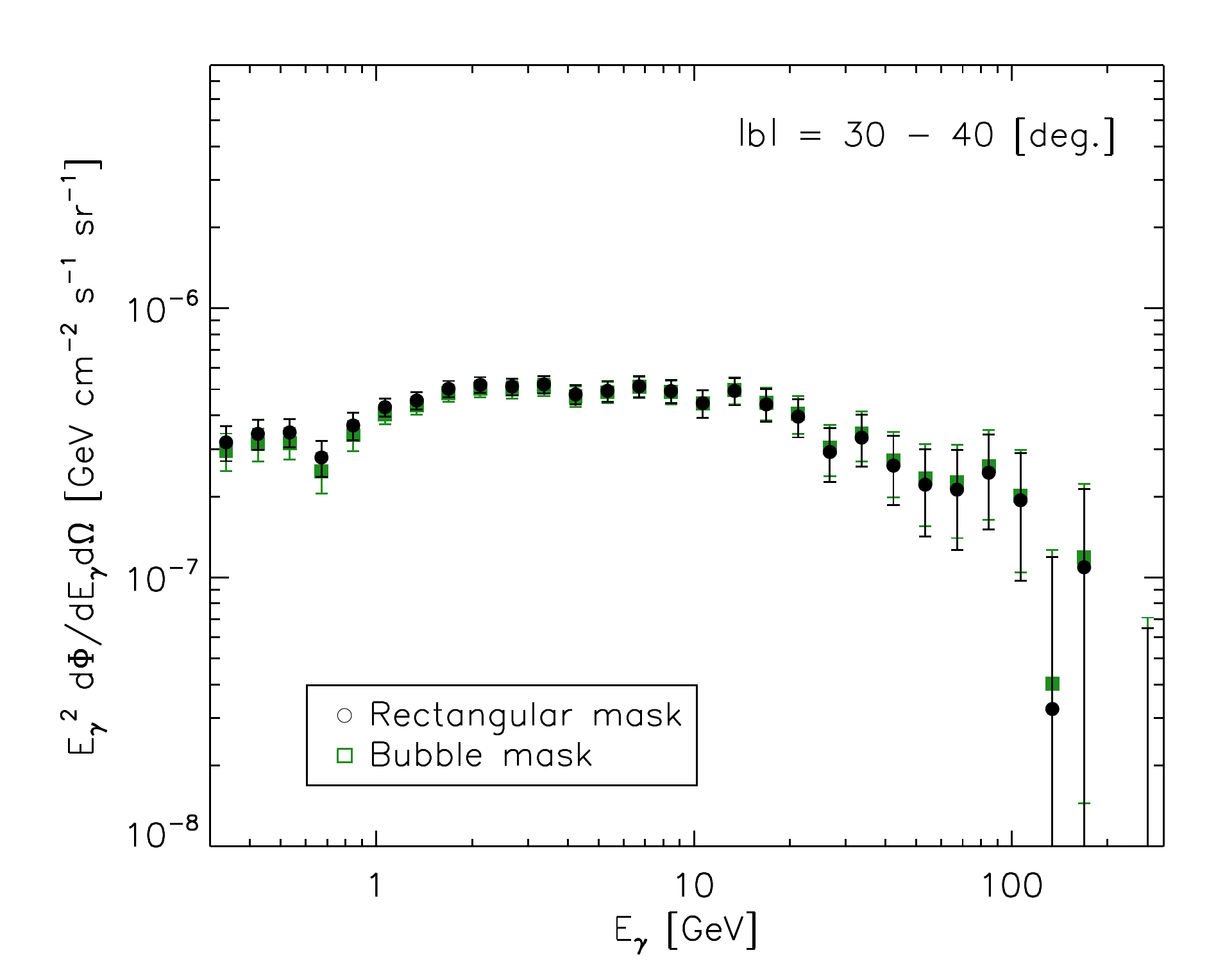}
    \end{minipage}\\
     \vspace{0.5 cm}
       \begin{minipage}{1\textwidth}
    \centering
   \includegraphics[scale=0.42]{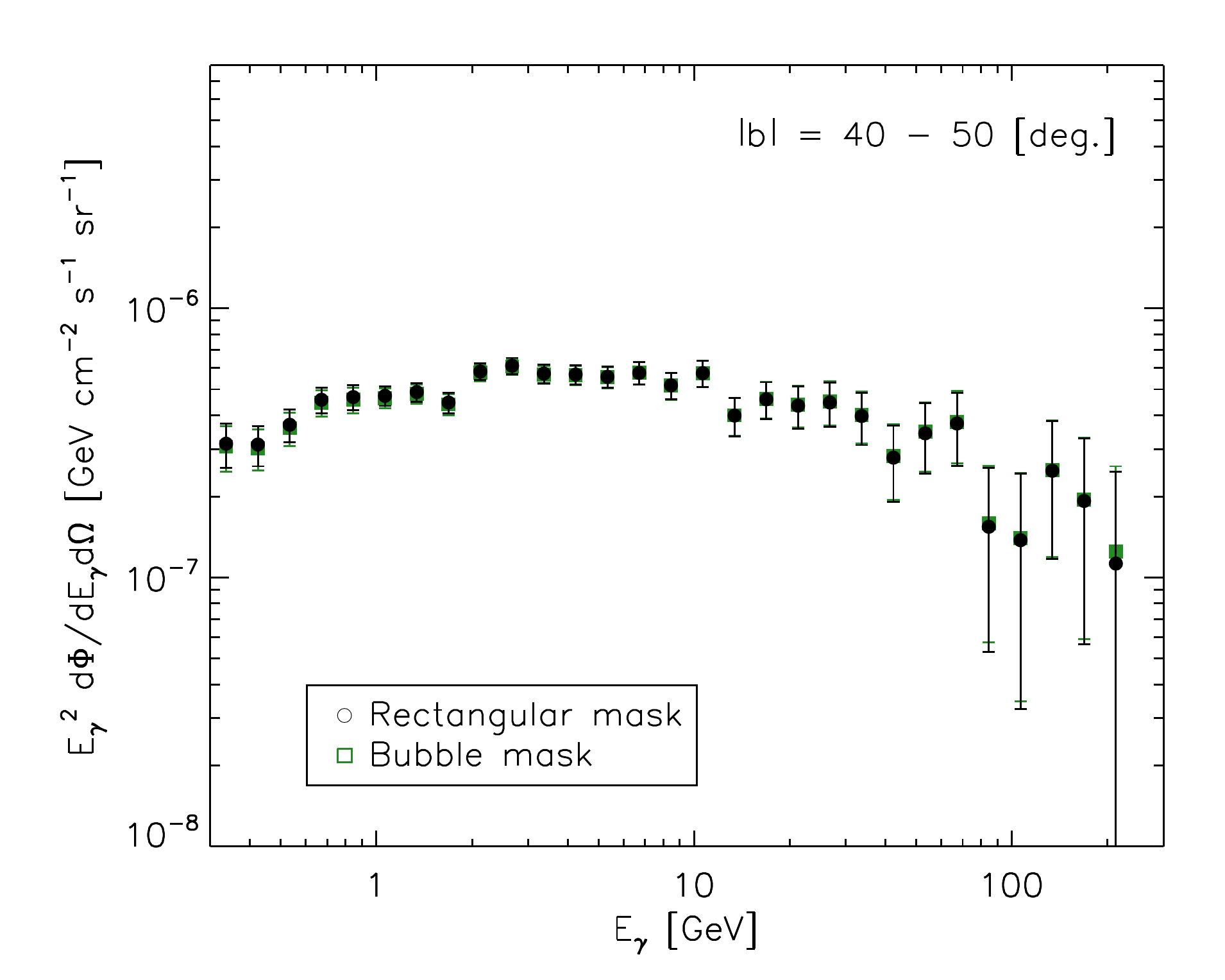}
  % \caption{\textit{Galactic diffuse model}}\label{fig:DiffuseMap}
    \end{minipage}
    \caption{\textit{   {\underline{FERMI BUBBLES MASK}}.
 Fermi bubbles   energy spectrum  obtained masking the Fermi bubbles as defined by the red region in Fig.~\ref{fig:BubbleTemplate}  (green squares). For comparison we also show the same energy spectra obtained using the rectangular mask, as defined in Appendix~\ref{App:masking} (black dots). 
    %We mask the inner disk in the region $|b|< 1^{\circ}$, $|l|<60^{\circ}$.
    }}\label{Fig:Mask}
\end{figure}

We here present the Fermi bubbles energy spectrum obtained 
masking the Galactic disk in the region  $|b| < 5^{\circ}$,  $|l| < 60^{\circ}$.  We compare this spectrum with the one 
obtained in the main part of the paper using the values $|b| < 1^{\circ}$,  $|l| < 60^{\circ}$. 
The Galactic disk cut involves directly only the first slice of the Fermi bubbles in the North hemisphere (see Fig.~\ref{fig:BubbleTemplate}); nevertheless the resulting energy spectrum is slightly modified, especially in the low-energy region. We show our results in Fig.~\ref{Fig:disk} for $|b|=20^{\circ}-
50^{\circ}$ (see the upper panel in Fig.~\ref{fig:MAIN2} for the regions $|b|< 10^{\circ}$, $|b|=10^{\circ}-20^{\circ}$). We find consistent results.\\
Additionally, we test a different masking method for the Fermi bubbles region. In particular, instead of the rectangular mask defined by the coordinates $-30^{\circ} < b < 30^{\circ}$, $-53.5^{\circ} < l < 50^{\circ}$ (see Appendix~\ref{App:masking}), we
use a mask reproducing the edges of the bubbles in Fig.~\ref{fig:BubbleTemplate}. In this way, concerning the bubbles, we exclude 
from the fitting procedure only the inner red region instead of the whole rectangle. We show the energy spectrum in Fig.~\ref{Fig:Mask}. We find consistent results in all the analyzed regions. 

\subsection{Point source subtraction}\label{App:PSS}

\begin{figure}[!htb!]
 \centering
  \begin{minipage}{0.4\textwidth}
   \centering
   \includegraphics[scale=0.42]{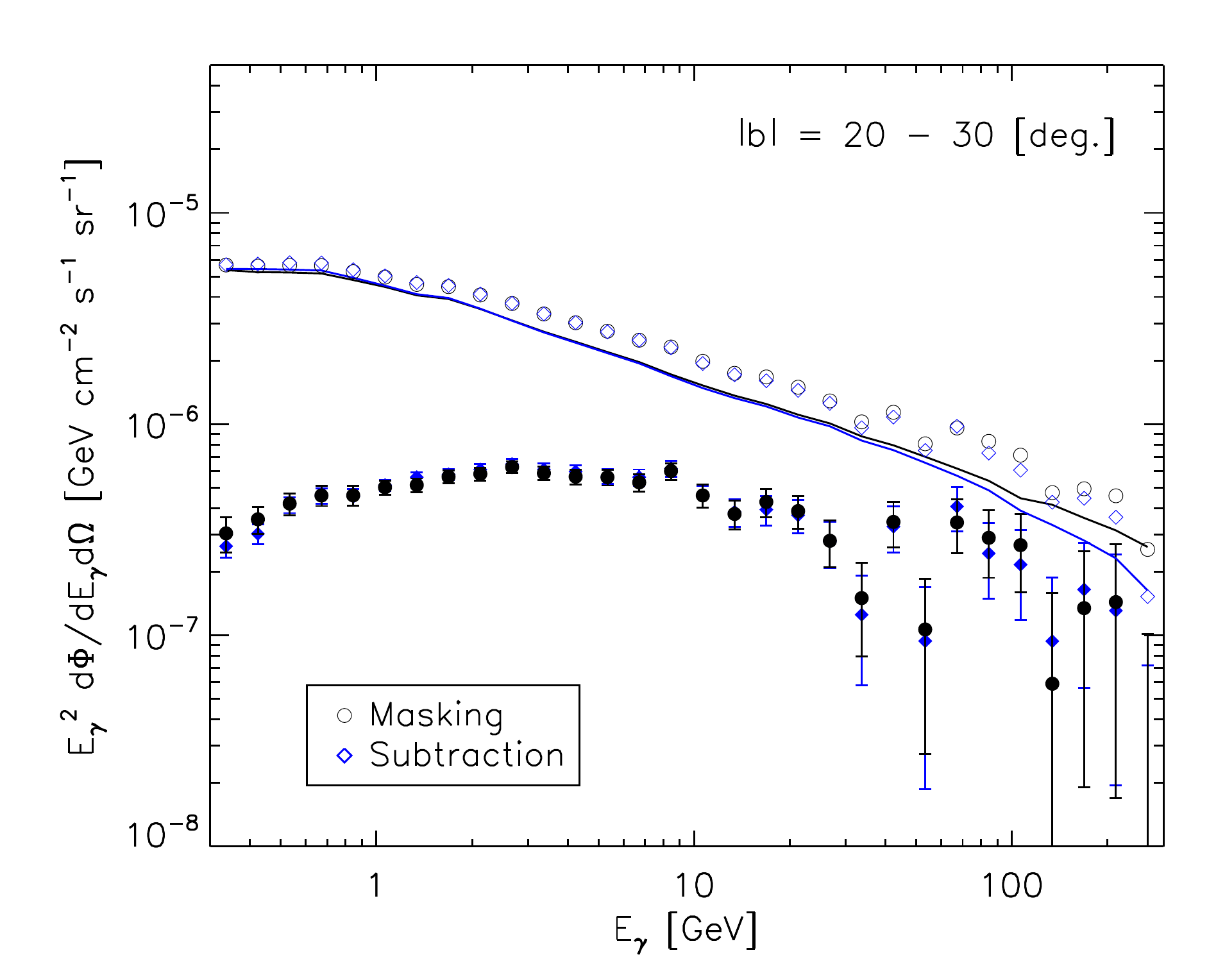}
   %\caption{\textit{Count Map}}\label{fig:CountMap}
    \end{minipage}\hspace{1.2 cm}
   \begin{minipage}{0.4\textwidth}
    \centering
    \includegraphics[scale=0.42]{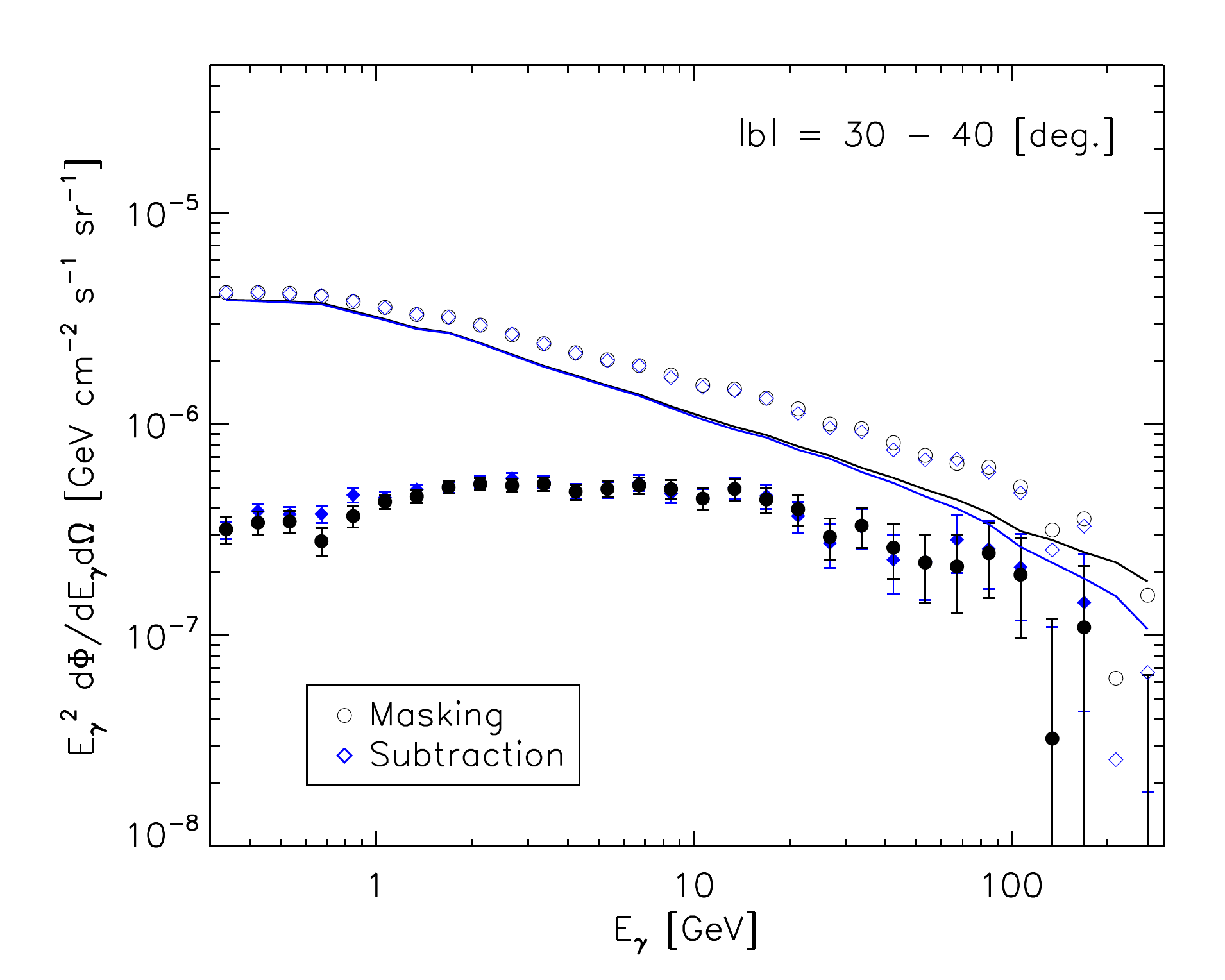}
    \end{minipage}\\
    \vspace{0.5 cm}
    \begin{minipage}{1\textwidth}
    \centering
   \includegraphics[scale=0.42]{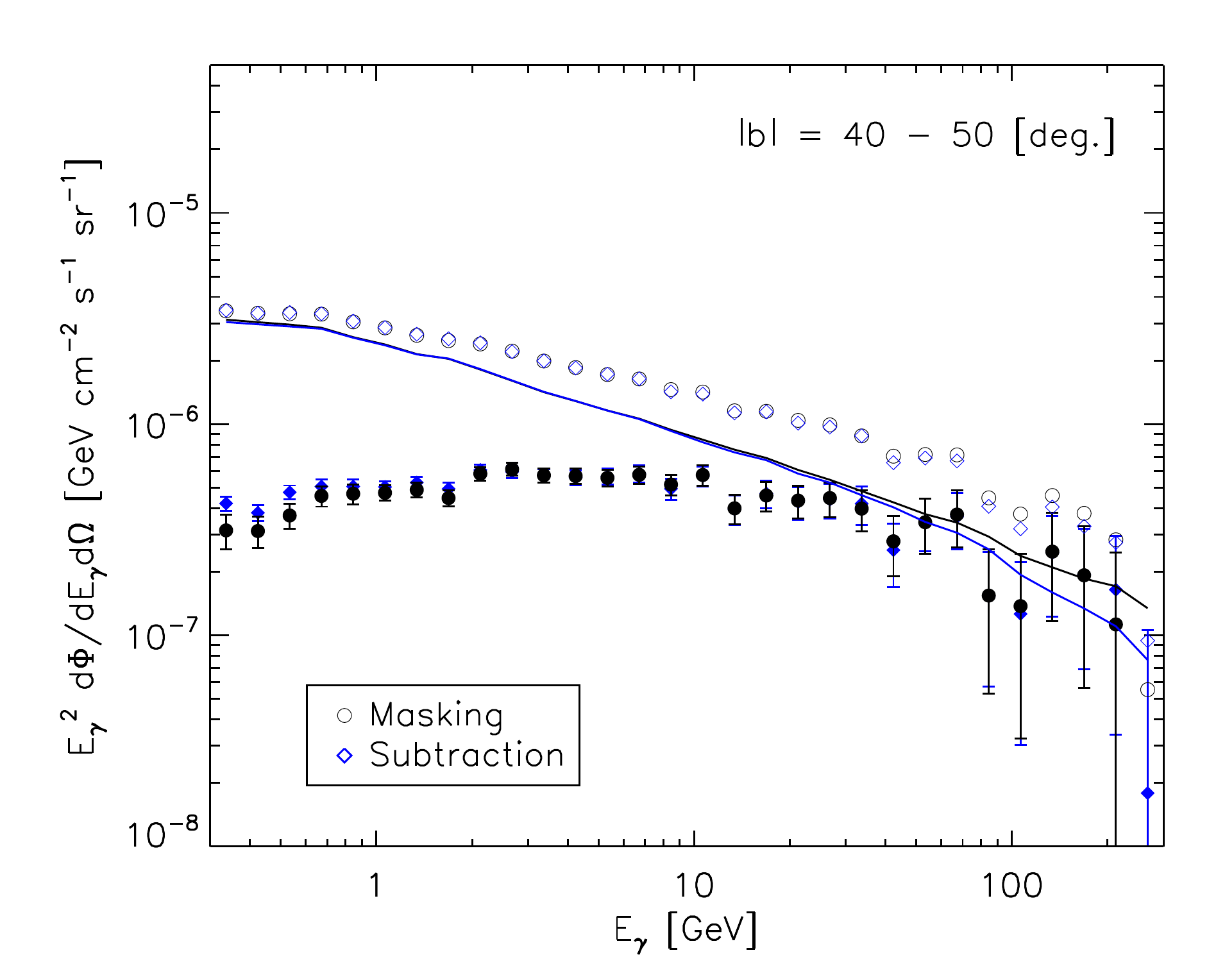}
  % \caption{\textit{Galactic diffuse model}}\label{fig:DiffuseMap}
    \end{minipage}
    \caption{\textit{
    {\underline{MASKING vs. SUBTRACTION}}.
   Fermi bubbles energy spectrum obtained considering point source masking (black filled dots) and point source subtraction (blue filled diamonds).    For comparison we also show  the observed flux and the best-fit theoretical prediction from the Galactic diffuse model and the isotropic extragalactic component (we use same color code w.r.t. the residual values but, respectively,  with empty symbols and solid lines). 
    }}\label{Fig:PSS}
\end{figure}

In this Section we show the energy spectrum of the Fermi bubbles comparing point source masking and subtraction. We show our results in Fig.~\ref{Fig:PSS} for $|b|=20^{\circ}-
50^{\circ}$ (see the lower panel in Fig.~\ref{fig:MAIN2} for the regions $|b|=1^{\circ}-10^{\circ}$, $|b|=10^{\circ}-20^{\circ}$). At high latitudes the two methods give the same results.

\subsection{North-South asymmetry}\label{App:NS}

\begin{figure}[!htb!]
 \centering
  \begin{minipage}{0.4\textwidth}
   \centering
   \includegraphics[scale=0.42]{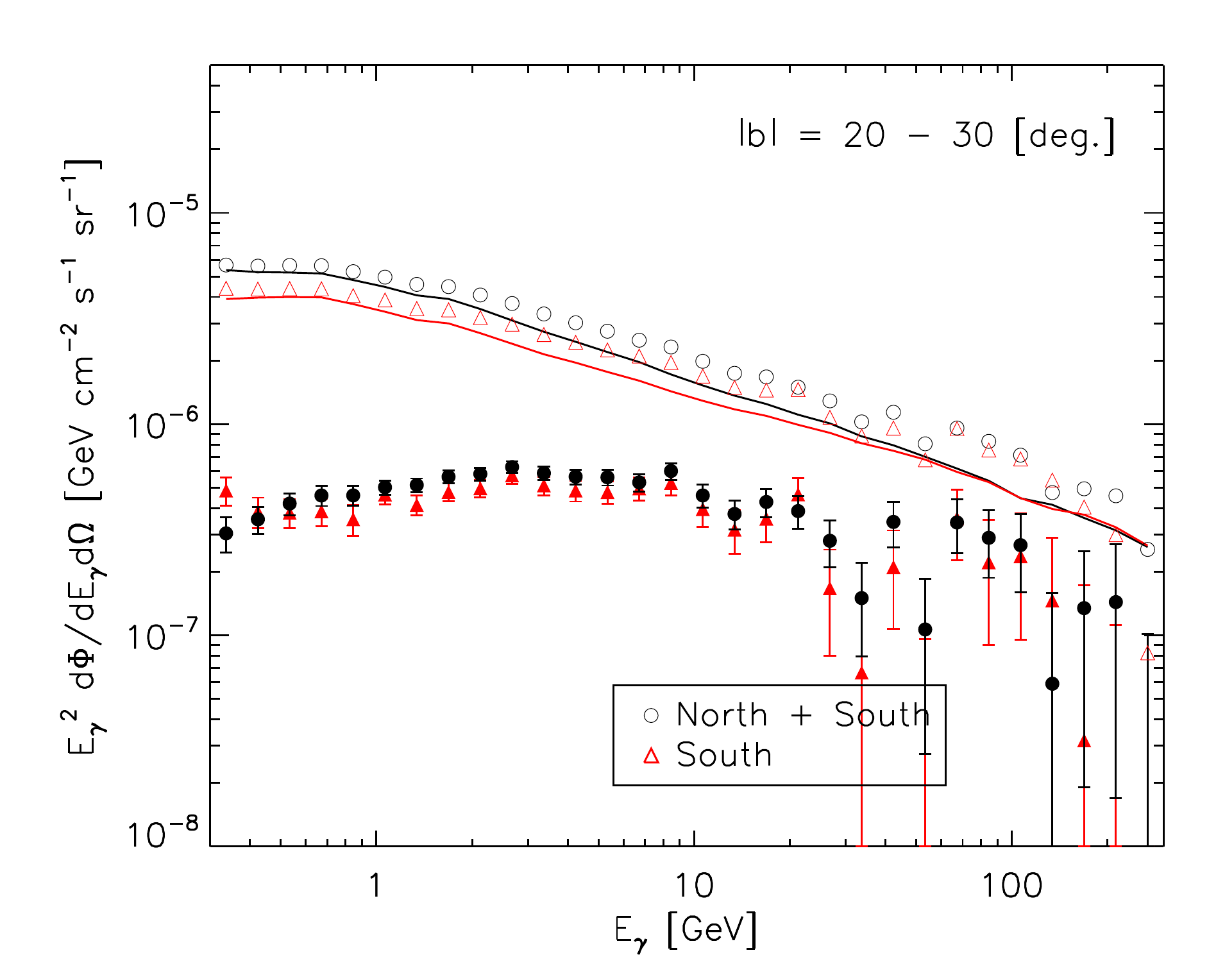}
   %\caption{\textit{Count Map}}\label{fig:CountMap}
    \end{minipage}\hspace{1.2 cm}
   \begin{minipage}{0.4\textwidth}
    \centering
    \includegraphics[scale=0.42]{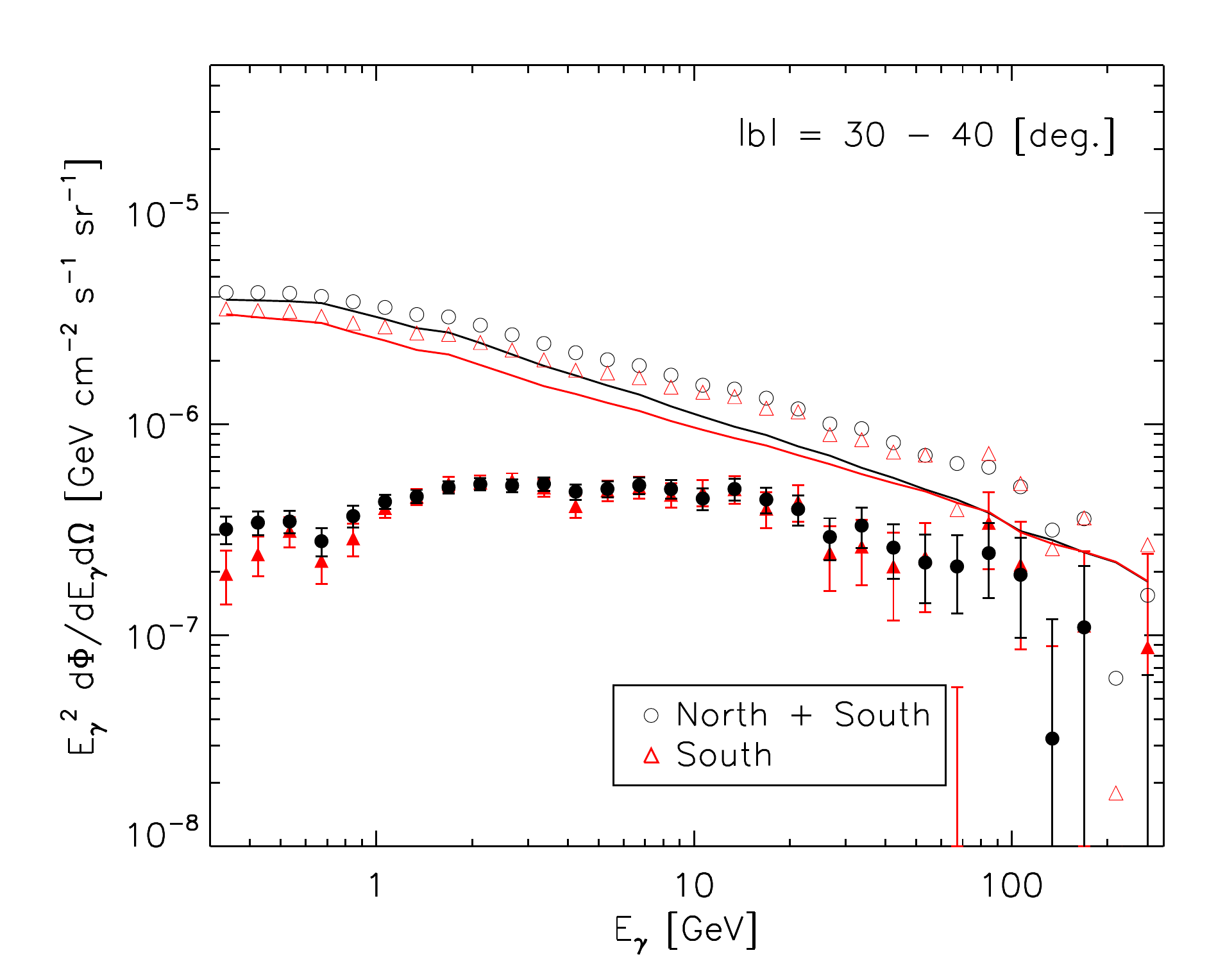}
    \end{minipage}\\
    \vspace{0.5 cm}
    \begin{minipage}{1\textwidth}
    \centering
   \includegraphics[scale=0.42]{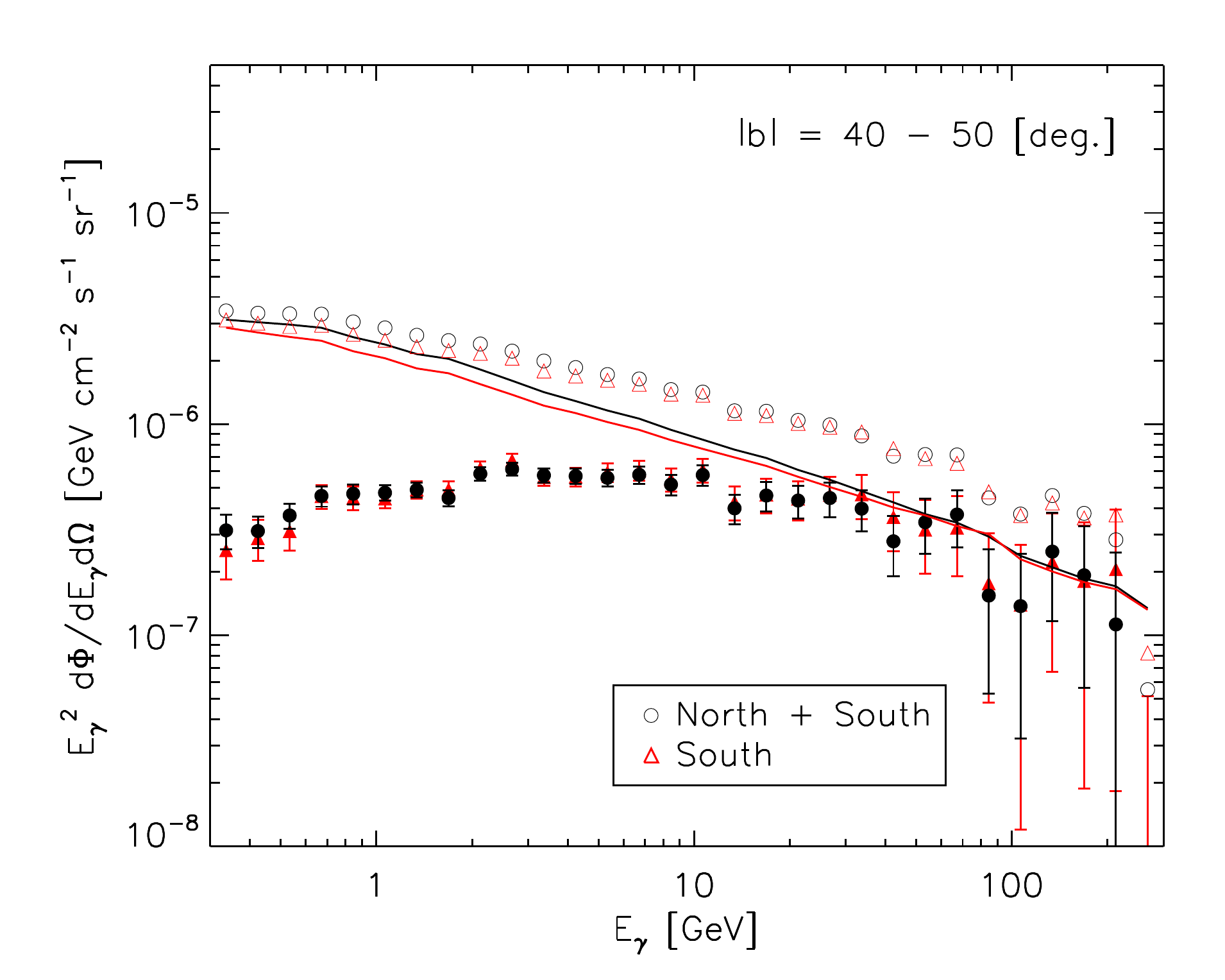}
  % \caption{\textit{Galactic diffuse model}}\label{fig:DiffuseMap}
    \end{minipage}
    \caption{\textit{
    {\underline{NORTH/SOUTH ASYMMETRY}}.
   Fermi bubbles energy spectrum obtained considering the South hemisphere (red filled triangles) and the North+South hemispheres (black filled dots).
   For comparison we also show  the observed flux and the best-fit theoretical prediction from the Galactic diffuse model and the isotropic extragalactic component (we use same color code w.r.t. the residual values but, respectively,  with empty symbols and solid lines). 
   %We use the \texttt{Ultraclean} events, masking the inner disk in the region $|b|< 1^{\circ}$, $|l|<60^{\circ}$.
    }}\label{Fig:South}
\end{figure}

In this Section we compare the energy spectrum of the Fermi bubbles between the North and the South hemisphere. We show our results in Fig.~\ref{Fig:South} for $|b|=20^{\circ}-
50^{\circ}$ (see Fig.~\ref{fig:MAIN3} for the regions $|b|=1^{\circ}-10^{\circ}$, $|b|=10^{\circ}-20^{\circ}$).  
At high latitudes we do not spot any significant asymmetry in the energy spectrum.

\newpage

\bibliography{Fermi_Bubbles}
\bibliographystyle{h-physrev}

\end{document}